\RequirePackage{ifpdf}
\documentclass[a4paper,11pt]{article} 
\usepackage[usenames,dvipsnames]{pstricks}
\usepackage{jheppub}
\usepackage[T1]{fontenc}
\usepackage[utf8]{inputenc}
\usepackage[english]{babel}
\usepackage[babel]{csquotes}
\usepackage{amsmath}
\usepackage{amsfonts}
\usepackage{amssymb}
\usepackage{mathtools}
\usepackage{mathrsfs}
\usepackage{bm}
\usepackage{bbold}
\usepackage{braket}
\usepackage{graphicx}
\usepackage{comment}
\usepackage{booktabs}
\usepackage{array}
\usepackage{color}
\usepackage{slashed}
\usepackage{tikz}
\usepackage{float} 
\usepackage{placeins}

\usepackage{epsfig,framed}
\usepackage{hyperref}
\input epsf.sty
\usepackage{hhline}
\usepackage{bm}
\usepackage{array, caption}
\usepackage{fmtcount} 
\usepackage{array,multirow}

\newcolumntype{C}[1]{>{\centering\arraybackslash}m{#1}}

\newcommand{\ei}{\end{itemize}}

\newcommand{\bse}{\begin{subequations}}
\newcommand{\ese}{\end{subequations}}

\newcolumntype{P}[1]{>{\centering\arraybackslash}p{#1}}

\usepackage[numbers]{natbib} 

\DeclarePairedDelimiter\ceil{\lceil}{\rceil}
\DeclarePairedDelimiter\floor{\lfloor}{\rfloor}

\author[a]{Josua Faller}
\author[a,b,c]{Sourav Sarkar}
\author[d,e]{Mritunjay Verma}

\affiliation[a]{\it Institut f{\"u}r Physik, Humboldt-Universit{\"a}t zu Berlin, IRIS-Adlershof, Zum Gro{\ss}en Windkanal 6, 12489 Berlin, Germany}
\affiliation[b]{\it Institut f{\"u}r Mathematik, Humboldt-Universit{\"a}t zu Berlin, IRIS-Adlershof, Zum Gro{\ss}en Windkanal 6, 12489 Berlin, Germany}
\affiliation[c]{\it Max-Planck-Institut f{\"u}r Gravitationsphysik, Albert-Einstein-Institut, Am M{\"u}hlenberg 1, 14476 Potsdam, Germany}
\affiliation[d]{\it Harish-Chandra Research Institute, HBNI, Chhatnag Road, Jhunsi, Allahabad-211019, India }
\affiliation[e] {\it International Centre for Theoretical Sciences, TIFR, Hesaraghatta, Hubli, Bengaluru-560089, India }
\emailAdd{josuafaller@physik.hu-berlin.de}
\emailAdd{sarkar@physik.hu-berlin.de}
\emailAdd{mritunjayverma@hri.res.in}

\preprint{HU-EP-17/25}
\preprint{HU-MATH-2017-08}
\preprint{HRI/ST/1704}
\preprint{\parbox{3cm}{HRI/ST/1704\\ HU-EP-17/25\\ HU-MATH-2017-08}}

\bigskip
\abstract{We define Mellin amplitudes for the fermion-scalar four point function and the fermion four point function. The Mellin amplitude thus defined has multiple components each associated with a tensor structure. In the case of three spacetime dimensions, we explicitly show that each component factorizes on dynamical poles onto components of the Mellin amplitudes for the corresponding three point functions. The novelty here is that for a given exchanged primary, each component of the Mellin amplitude may in general have more than one series of poles. We present a few examples of Mellin amplitudes for tree-level Witten diagrams and tree-level conformal Feynman integrals with fermionic legs, which illustrate the general properties. }

\title{Mellin Amplitudes for Fermionic Conformal Correlators}

\keywords{Conformal Field Theory, Mellin Amplitudes}

\newcommand{\abs}[1]{\lvert #1 \rvert}
\newcommand{\mdelta}[1]{\hat{\delta}\left( #1 \right)} 
\newcommand{\Ga}[1]{\Gamma\left(#1\right)} 
\newcommand{\Be}[2]{B\left(\frac{#1}{2},\frac{#2}{2} \right)}
\newcommand{\be}{\begin{eqnarray}}
\newcommand{\ee}{\end{eqnarray}}

\newcommand{\Cooint}[1]{\int \mathcal{D} #1} 

\newcommand{\Singprop}[3]{\frac{\Gamma\left( #3 \right)}{\abs{#1 #2}^{2 #3}} } 
\newcommand{\Singpro}[3]{\frac{1}{\abs{#1 #2}^{2 #3}} } 
\newcommand{\Singferprop}[5]{ \frac{\slashed{#1}_{#2} - \slashed{#3}_{#4}}{\abs{{#1}_{#2} - {#3}_{#4}}^{2 #5 + 1}} \Gamma\left( #5 + \frac{1}{2}\right)} 
\newcommand{\Singferpro}[5]{ \frac{\slashed{#1}_{#2} - \slashed{#3}_{#4}}{\abs{{#1}_{#2} - {#3}_{#4}}^{2 #5 + 1}}} 
\newcommand{\Multiprop}[4]{ \frac{1}{ \abs{#1 #2}^{2 #3}} \Gamma\left( #3 #4\right)} 

\newcommand{\Mellint}[2]{\prod_{#1 \leq i < l}^{#2} \int_{c_{il}- i \infty}^{c_{il} + i \infty} \left( ds_{il} \right) } 
\newcommand{\Mellintt}[2]{\prod_{#1 \leq i < l}^{#2} \int_{\tilde{c}_{il}- i \infty}^{\tilde{c}_{il} + i \infty} \left( d\tilde{s}_{il} \right) } 
\newcommand{\Mellintu}[2]{\prod_{i=#1}^{#2} \int_{c_{iu}- i \infty}^{c_{iu} + i \infty} \left( ds_{iu} \right) } 

\newcommand{\mellintb}[1]{\int_{\bar{c}_{#1}- i \infty}^{\bar{c}_{#1} + i \infty} \left( d\bar{s}_{#1} \right)}
\newcommand{\mellintt}[1]{\int_{\tilde{c}_{#1}- i \infty}^{\tilde{c}_{#1} + i \infty} \left( d\tilde{s}_{#1} \right)}

\newcommand{\Melldel}[2]{\prod_{i=#1}^{#2} \mdelta{\Delta_i - \sum_{k=#1, k\neq i}^{#2} s_{ik}}} 
\newcommand{\Melldeltwist}[2]{\prod_{i=#1}^{#2} \mdelta{\tau_i - \sum_{k=#1, k\neq i}^{#2} s_{ik}}} 

\newcommand{\ten}[2]{\frac{\slashed{#1}_{#2}}{\abs{#1_{#2}}}} 

\newcommand{\spinpr}[1]{\left[#1\right]} 

\newcommand{\f}{\frac}

\newcommand{\qeq}{{\hbox{=\kern-2.3mm ? \kern.5mm }}}
\renewcommand{\qeq}{=}

\newcommand{\non}{\nonumber}

\def\one{{\hbox{ 1\kern-.8mm l}}}
\def\zero{{\hbox{ 0\kern-1.5mm 0}}}

\begin{document}

\maketitle
\flushbottom

\section{Introduction}
\label{sec:intro}

From the pioneering work of Mack \cite{Mack:2009mi, Mack:2009gy}, followed up on later by Penedones and several others \cite{Penedones:2010ue, Costa:2012cb, Fitzpatrick:2011ia, Rastelli:2017ecj, Fitzpatrick:2011hu}, we now understand that Mellin space provides us with a natural representation to study conformal correlation functions. The Mellin representation of conformal correlation functions is analogous to the momentum space representation of scattering amplitudes. Information on operator dimensions and structure constants in a Conformal Field Theory (CFT) is encoded in the poles and residues of Mellin amplitudes. The amplitude factorizes on these poles onto lower point Mellin amplitudes. In the context of large $N$ theories, the Mellin amplitude gains special prominence as it is now a meromorphic function with its poles encoding information on only the exchanged single trace operators. Through the so-called ``flat space limit'', Mellin amplitudes of $d$ dimensional CFT are concretely related to scattering amplitudes in $d+1$ dimensional Quantum Field Theory (QFT). \\
\\
Complementary to the conceptual advances, there has also been significant progress in the application of the Mellin representation on various fronts. It has been shown that Witten diagrams, at least at tree level, are easy to calculate and take very simple forms in Mellin space \cite{Paulos:2011ie, Nandan:2011wc}. There has also been some progress on Mellin amplitudes of loop level Witten diagrams \cite{Aharony:2016dwx, Yuan:2017vgp, Cardona:2017tsw}. It may in fact be possible to bootstrap the full holographic correlator as shown in \cite{Rastelli:2016nze, Rastelli:2017udc} for the four point function of one-half BPS single trace operators in the context of IIB supergravity in $AdS_{5}\times S^{5}$. In the context of higher-spin holography, there have been efforts to understand the non-locality in the bulk interactions with Mellin amplitudes in the dual free CFT \cite{Ponomarev:2017qab, Bekaert:2016ezc, Taronna:2016ats}. Through the flat space limit, the conformal bootstrap has been related to to the S-matrix bootstrap \cite{Paulos:2016fap}. A new approach to the conformal bootstrap has been developed in Mellin space \cite{Gopakumar:2016wkt, Gopakumar:2016cpb}. In this method, conformal correlation functions are expanded in a manifestly crossing symmetric basis of functions provided by exchange Witten diagrams (in three channels). Demanding consistency with the Operator Product Expansion (OPE) one obtains constraints on operator dimensions and OPE coefficients. \\
\\
So far, the literature on Mellin amplitudes focusses almost exclusively on correlation functions of scalar operators. It is natural to ask if one can define Mellin amplitudes for correlation functions of operators with spin and make similar progress conceptually and with applications as for scalar correlation functions. Mellin amplitudes for correlation functions of scalars and one integer spin operator were defined in \cite{Goncalves:2014rfa} with the purpose of studying factorization of scalar Mellin amplitudes onto lower point Mellin amplitudes. However the Mellin amplitudes of the spinning correlators themselves have not been studied as such.\\
\\
This deficiency is especially significant in the context of fermionic conformal correlation functions. Fermionic operators do not appear in the OPE of scalar operators. Therefore, if one desires to access the fermionic sector of a CFT, it is necessary to consider correlation functions with at least spin-half operators. Moreover, spinning correlation functions in general can potentially provide us with more information on CFT data than scalar correlation functions.\\
\\
In this paper, we make an attempt at studying Mellin amplitudes for correlation functions with spin-half fermions. We define Mellin amplitudes for the four point function of two fermions and two scalars and the four point function of four fermions. For simplicity, we restrict our analysis of the analyticity properties of the Mellin amplitude to three dimensions. Defining the Mellin amplitude involves making choice of a basis of tensor structures and the Mellin amplitude has one component corresponding to each basis element. Generically, the separation of the tensorial part might introduce spurious singularities in the conformal blocks, as noted in \cite{Iliesiu:2015qra, Kravchuk:2016qvl}. Therefore not all bases are suitable for defining a Mellin amplitude with the desired analyticity properties. After defining the Mellin amplitude suitably, we proceed to examine the pole structure by looking at the behavior of the correlator in the OPE limit. This also makes the factorization of the four point Mellin amplitude manifest. \\
\\
The three point function of two fermions and a boson has multiple tensor structures. Generically, this results in each component of the Mellin amplitude having more than one distinct series (two in our case) of poles corresponding to each primary operator exchanged in the OPE in a given channel. We always choose tensor structures of definite parity for both the three point and four point functions as this choice leads to simplifications in the pole structure when the three point functions are of definite parity. It must be noted that the pole structure of the Mellin amplitude is related to the choice of basis and is tunable as such. \\
\\
After this preliminary analysis of the properties of the Mellin amplitude, we compute some Mellin amplitudes corresponding to tree level Witten diagrams and tree level conformal Feynman integrals. These examples illustrate the generic predictions on the pole structure considering the parity of the three point functions in each case.  It should be straightforward to generalize our study to four dimensions. The definition also trivially extends to $n$-point functions when supplemented by a concrete choice of tensor structures. \\
\\
The article has three main chapters. Chapter \ref{sec:generalities} starts with a basic review of Mellin amplitudes for scalar correlators in Section \ref{subsec:scalar}. We present the basis of tensor structures we would be using in each case in Section \ref{subsec:tens} and define the Mellin amplitude for fermionic correlation functions in Section \ref{subsec:defn}. We present the pole structure of the fermion scalar four point correlator and the four fermion correlator in Sections \ref{subsec:pole2f} and \ref{subsec:4ferm-pole} respectively. In Chapter \ref{sec:wittendiagrams}, we present results for Mellin amplitudes for a few tree level Witten diagrams and in Chapter \ref{sec:perturbative}, we present results for Mellin amplitudes for some conformal Feynman integrals. We end with a discussion of our results and future directions in Section \ref{sec:disco}. Calculations, methods and a short review of fermions in $AdS$ are provided in the Appendices.

\section{Mellin Amplitudes for Fermionic Correlators}
\label{sec:generalities}

\subsection{Review: Mellin representation of scalar correlators}
\label{subsec:scalar}

In this section, we shall briefly review the basics of Mellin amplitudes for scalar correlators. The Mellin amplitude for the connected part of a scalar correlator was defined by Mack \cite{Mack:2009mi} in the following manner (Euclidean signature):
\begin{eqnarray}
\Braket{O_{1}(x_{1})O_{2}(x_{2})\cdots O_{n}(x_{n})}_{c}=\int [ds_{ij}]\; \prod_{i<j}\Gamma\left(s_{ij}\right)x_{ij}^{-2s_{ij}}
\mathcal{M}\left(\{s_{ij}\}\right).            \label{scalardef}
\end{eqnarray}
The integral in ~\eqref{scalardef} is a Mellin-Barnes integral and the contours run parallel to the imaginary axis. The Mellin variables $s_{ij}$ are not all independent but satisfy the following constraints,
\be
\Delta_{i}-\sum_{j\neq i}s_{ij}=0  \;\;\;\; \forall i        \label{thiago}
\ee
These conformality constraints ~\eqref{thiago} ensure that the right hand side of ~\eqref{scalardef} transforms properly under conformal transformations. The number of independent Mellin variables $s_{ij}$ is $\frac{n(n-3)}{2}$ which is the same as the number of independent cross-ratios. For $n>d+2$, the dimension of the conformal moduli space is less than this (see \cite{Kravchuk:2016qvl}) and the associated Mellin amplitude is non-unique (see \cite{Penedones:2016voo}). \\
\\
The conformality constraints can be interpreted in terms of Mellin momenta $k_{i}$ with $k_{i}\cdot k_{j}=s_{ij}$ and an on-shell condition $k_{i}^{2}=-\Delta_{i}$ as the overall conservation of Mellin momentum $\sum_i k_{i}=0$. One can thus relate the Mellin variables to Mandelstam variables $S_{i_{1}\cdots i_{a}}$ as
\begin{eqnarray}
S_{i_{1}\cdots i_{a}}=-\left(k_{i_{1}}+\cdots+k_{i_{a}}\right)^{2}=-2\sum_{l<k\leq a}s_{i_{l}i_{k}}+\sum_{j=1}^{a} \Delta_{i_{j}}.          \label{mandelstam}
\end{eqnarray}
\\
The location of the poles in a given Mandelstam variable $S_{i_{1}i_{2}}$ is at the twists of the operators in the OPE of $O_{i_{1}}O_{i_{2}}$ that contribute to the correlator. The Mellin amplitude factorizes at these poles and the residue is proportional to the Mellin amplitudes of the corresponding lower point correlators as dictated by the OPE. \\
\\
As an example, let us look at the case of the four point function. 
\be
\Braket{O_{1}(x_{1})O_{2}(x_{2})O_{3}(x_{3})O_{4}(x_{4})}_{c}& =& \left(\frac{x_{24}^{2}}{x_{14}^{2}}\right)^{\frac{\Delta_{1}-\Delta_{2}}{2}}\left(\frac{x_{14}^{2}}{x_{13}^{2}}\right)^{\frac{\Delta_{3}-\Delta_{4}}{2}}\frac{\mathcal{A}(u,v)}{\left(x_{12}^{2}\right)^{\frac{\Delta_{1}+\Delta_{2}}{2}}\left(x_{34}^{2}\right)^{\frac{\Delta_{3}+\Delta_{4}}{2}},}      \non \\
u &=&  \frac{x_{12}^{2}x_{34}^{2}}{x_{13}^{2}x_{24}^{2}},     \;\;\;\;\;\;\;\;\;\;\;\;    v=  \frac{x_{14}^{2}x_{23}^{2}}{x_{13}^{2}x_{24}^{2}},                \label{fourpoint}
\ee
\be
\mathcal{A}(u,v)&=&\int_{c_{s}-i\infty}^{c_{s}+i\infty} \frac{ds}{2\pi i} \int_{c_{t}-i\infty}^{c_{t}+i\infty} \frac{dt}{2\pi i} \,  
\mathcal{M}(s,t)  u^{\frac{s}{2}}v^{-\frac{s+t-\Delta_{1}-\Delta_{4}}{2}}\Gamma\left(\frac{\Delta_{1}+\Delta_{2}-s}{2}\right) \Gamma\left(\frac{\Delta_{3}+\Delta_{4}-s}{2}\right)   \non    \\
  && \Gamma\left(\frac{\Delta_{1}+\Delta_{3}-t}{2}\right)   \Gamma\left(\frac{\Delta_{2}+\Delta_{4}-t}{2}\right)\Gamma\left(\frac{s+t-\Delta_{1}-\Delta_{4}}{2}\right)\Gamma\left(\frac{s+t-\Delta_{2}+\Delta_{3}}{2}\right),          \nonumber    \\
 s&=& -(k_{1}+k_{2})^{2}=\Delta_{1}+\Delta_{2}-2s_{12},   \;\;\;\;\;\;\;\;\;\;\;\;    t= -(k_{1}+k_{3})^{2}=\Delta_{1}+\Delta_{3}-2s_{13}.    \label{fourpoint2}
\end{eqnarray} 
For every conformal primary with twist $\tau$ contributing to the conformal block expansion of $\mathcal{A}(u,v)$ in the direct channel, $\mathcal{M}(s,t)$ has poles at $s=\tau+2m,$ $m=0,1,2,\cdots$ where $m=0$ corresponds to the primary and the leading twist descendants (and similarly for the other channels). \\
\\
The factors of gamma functions in ~\eqref{fourpoint2} also contribute poles, for example at $s=\Delta_{1}+\Delta_{2}+2m$. These poles correspond to operators of the form $O_{1}\partial^{l}\left(\partial^{2}\right)^{m}O_{2}$ that contribute to the conformal block expansion and have the said values of twist in a regime where anomalous dimensions are suppressed. In large $N$ gauge theories, these are the familiar double trace operators. The Mellin amplitude then accounts for the contributions from only single trace operators and is a meromorphic function.

\subsection{Tensor structures}
\label{subsec:tens}

In order to discuss a Mellin representation for fermionic conformal correlators, first we have to discuss the tensor structures that appear in these correlators and select a basis for each. We shall restrict the discussion to the case of 3d Minkowski spacetime for simplicity and also assume that all operators of the same spin have different conformal dimensions. Generalization to other signatures and spacetime is straightforward. We shall also assume that all the operators have different conformal dimensions. We shall be using the embedding formalism for spinors developed in \cite{Weinberg:2010fx, Iliesiu:2015qra}. There does not seem to be any canonical basis of tensor structures. It helps to choose basis elements of definite parity as we shall see later. One should also note that not every choice of basis is suitable for defining the Mellin amplitude such that the poles of the amplitude can be associated with operators contributing to the conformal block expansion of the correlator. This is because for certain choices of bases, as explained in Section 4.4 of \cite{Kravchuk:2016qvl}, there maybe spurious singularities in the conformal blocks. For example, in the context of the fermion four point function, the naive conformal blocks associated with the basis in Section 2.4 of \cite{Iliesiu:2015qra} have singularities at $z=\bar{z}$. A neat way to count the number of independent tensor structures and to find relations between tensor structures (when it is otherwise tedious to do so) is to shift to a conformal frame \cite{Kravchuk:2016qvl}. We have reviewed the general principle and the relevant results in Appendix \ref{countessa}.  We stick to the choice of bases made in \cite{Iliesiu:2015qra, Iliesiu:2015akf} making an independent choice of basis only for the fermion four point function.\\
\\
\textbf{Quick review of Embedding formalism}\\
\\
We are considering a theory, not of definite parity, in three spacetime dimensions with Minkowski signature $-++$. The double cover of $SO(2,1)$ is isomorphic to $Sp(2,\mathbb{R})$ and the smallest fundamental representation is that of a real two dimensional vector space which describes Majorana fermions and the fundamental generators preserve a $2\times 2$ symplectic tensor. We shall be following the conventions of \cite{Iliesiu:2015qra}. We shall review the gist of it here.  Conformal transformations which act non-linearly in 3d (signature $-++$) act linearly as Lorentz transformations in 5d (signature $-+++-$). Therefore we embed the 3d spacetime $x^{\mu}$ in 5d spacetime $X^{A}$ by identifying the 3d spacetime  with the projective null cone in 5d in the following manner,
\begin{eqnarray}
X=X^{+}\left(x^{\mu},1,x^{2}\right) .            \label{embedcoordinates}
\end{eqnarray}
5d spacetime coordinates are written in lightcone coordinates as $X=\left(X^{\mu},X^{+},X^{-}\right)$ with $X^{\pm}=X^{4}\pm X^{3}$. Gamma matrices ($\gamma_{\mu}$ in 3d and $\Gamma_{I}$ in 5d) are chosen to be real.\\
\\
For every spinor $\psi^{\alpha}$ (transforming in the fundamental representation), an auxilliary anti-fundamental spinor (primary of vanishing dimension) $s_{\alpha}$ is introduced, so that we can conveniently work with the scalar,
\begin{eqnarray}
\psi(x,s)=s_{\alpha}\psi^{\alpha}(x)     .             \label{aux3d}
\end{eqnarray} 
The spinorial 5d conformal group is isomorphic to $Sp(4,\mathbb{R})$ (double cover of $SO(3,2)$) and the fundamental generators now preserve a $4\times 4$ symplectic tensor. We embed $\psi^{\alpha}(x)$ into a 5d spinor on the lightcone $\Psi^{I}(X)$ (fundamental of $Sp(4,\mathbb{R})$), and again take an auxilliary anti-fundamental spinor $S_{I}$ to define,
\begin{eqnarray}
\Psi(X,S)=S_{I}\Psi^{I}\left(X\right),    
\qquad   S_I = \sqrt{X^+} \begin{pmatrix} s_\alpha \\ - x^{\alpha}_{\;\beta} s^{\beta} \end{pmatrix},        
\qquad  x^{\alpha}_{\;\beta}=x^{\mu}\left(\gamma_{\mu}\right)^{\alpha}_{\;\beta}.				\label{aux5d}
\end{eqnarray} 
Transformation properties of $\Psi^{I}(X)$ under rotations and boosts dictate the precise manner in which 3d spinors are embedded into 5d spinors in general and then the transversality condition $S_{I}X^{I}_{\;J}=0$ (where $X^{I}_{\;J}=X^{A}\left(\Gamma_{A}\right)^{I}_{\;J}$) fixes how $S_{I}$ can be expressed in terms of $s_{\alpha}$. Further, the requirement that $\Psi(X,S)$ is a Lorentz scalar in 5d iff $\psi(x,s)$ is a scalar primary in 3d with dimension $\Delta$ fixes $\Psi(X,S)$ and $\psi(x,s)$ to be related in the following manner,
\begin{eqnarray}
\Psi(X,S)=\frac{1}{\left(X^{+}\right)^{\Delta}}\psi(x,s).         \label{embed1}
\end{eqnarray}
$\Psi(X,S)$ has the homogeneity property,
\begin{eqnarray}
\Psi(aX,bS)=a^{-\Delta-\frac{1}{2}}b\; \Psi(X,S).           \label{homo}
\end{eqnarray}
The form of the correlators (alongwith the tensor structures) is then fixed by the requirements of 5d Lorentz invariance, homogeneity ~\eqref{homo} and transversality. For example the two point function can be taken to be,
\begin{eqnarray}
\Braket{\Psi\left(X_{1},S_{1}\right)\Psi\left(X_{2},S_{2}\right)}=i\frac{\Braket{S_{1}S_{2}}}{X_{12}^{\Delta+\frac{1}{2}}}.      \label{twopt5d}
\end{eqnarray}
$X_{ij}=-2X_{i}\cdot X_{j}$ and $\Braket{S_{1}X_{2}X_{3}\cdots X_{k-1}S_{k}}=(S_{1})_{I}(X_{2})^{I}_{\,J}\left(X_{3}\right)^{J}_{\,K}\cdots\left(X_{k-1}\right)^{K}_{\;L}\left(S_{k}\right)^{L}$. In 3d, the two point function looks like,
\begin{eqnarray}
\braket{\psi^{\alpha}(x_{1})\psi_{\beta}(x_{2})}= i\frac{\left(x_{12}\right)^{\alpha}_{\;\beta}}{\left(x_{12}^{2}\right)^{\Delta+\frac{1}{2}}}.    \label{twopt3d}            
\end{eqnarray}
In general, any real operator of spin $l$ can be represented as $\phi^{\alpha_{1}\alpha_{2}\cdots\alpha_{2l}}$ where the $\alpha_{i}$ are fundamental indices of $Spin(2,1)$. Here $\phi^{\alpha_{1}\alpha_{2}\cdots\alpha_{2l}}$ is symmetric\footnote{This is possible only in 3d} in all indices. As before, to work in an index free manner, we can introduce an auxilliary spinor $s_{\alpha}$ to form,
\begin{eqnarray}
\phi(x,s)=s_{\alpha_{1}}\cdots s_{\alpha_{2l}}\phi^{\alpha_{1}\alpha_{2}\cdots\alpha_{2l}}(x).        \label{hs}
\end{eqnarray} 
An analogous construction gives the associated 5d operator $\Phi(X,S)$.\\
\\
\textbf{Three point functions.}\\
\\
First let us state the tensor structures for the relevant three point functions which are those of two spin half fermions $\psi_1,\,\psi_2$ and a bosonic operator $O_{3,l}$ and of one spin half fermion $\psi_1$ one scalar $O_2$ and one fermionic operator of any spin $\psi_{3,l}$. The structures for the three point function of two fermions and a scalar $\Braket{\psi_{1}\psi_{2}O_{3}}$ can be taken to be,
\begin{eqnarray}
r_{di}^{+}=\frac{\Braket{S_{1}S_{2}}}{\sqrt{X_{12}}}\to \frac{\slashed{x}_{12}}{|x_{12}|},     \;\;\;\;\;\;\;\;\;\;\;\;  r_{di}^{-}=\frac{\Braket{S_{1}X_{3}S_{2}}}{\sqrt{X_{13}X_{32}}}\to \frac{\slashed{x}_{13}\slashed{x}_{32}}{|x_{13}||x_{23}|} .        \label{2spino1sc}
\end{eqnarray} 
For the three point function of two fermions and a spin $l$ bosonic operator $\Braket{\psi_{1}\psi_{2}O_{3,l}}$ ($l>0$), we have two parity even structures and two parity odd ones\footnote{For $l=0$, $r_{di,1}^{+}$ goes to $r_{di}^{+}$ and $r_{di,3}^{-}$ goes to $r_{di}^{-}$} which can taken to be,
\begin{eqnarray}
r_{di,1}^{+}&=&\frac{\Braket{S_{1}S_{2}}\Braket{S_{3}X_{1}X_{2}S_{3}}^{l}}{X_{12}^{\frac{l+1}{2}}X_{13}^{\frac{l}{2}}X_{23}^{\frac{l}{2}}},    \;\;\;\;\;\;\;\;\;\;  r_{di,2}^{+}=\frac{\Braket{S_{1}S_{3}}\Braket{S_{2}S_{3}}\Braket{S_{3}X_{1}X_{2}S_{3}}^{l-1}}{X_{12}^{\frac{l-1}{2}}X_{13}^{\frac{l}{2}}X_{23}^{\frac{l}{2}}} ,  \label{2spino1tens}       \\
r_{di,3}^{-}&=&\frac{\Braket{S_{3}X_{1}X_{2}S_{3}}^{l-1}}{X_{12}^{\frac{l}{2}}X_{13}^{\frac{l+1}{2}}X_{23}^{\frac{l+1}{2}}}\left[X_{23}\Braket{S_{1}S_{2}}\Braket{S_{2}X_{1}S_{3}}+X_{13}\Braket{S_{2}S_{3}}\Braket{S_{1}X_{2}S_{3}}\right] ,       \nonumber \\
r_{di,4}^{-}&=&\frac{\Braket{S_{3}X_{1}X_{2}S_{3}}^{l-1}}{X_{12}^{\frac{l}{2}}X_{13}^{\frac{l+1}{2}}X_{23}^{\frac{l+1}{2}}}\left[X_{23}\Braket{S_{1}S_{2}}\Braket{S_{2}X_{1}S_{3}}-X_{13}\Braket{S_{2}S_{3}}\Braket{S_{1}X_{2}S_{3}}\right].          \nonumber
\end{eqnarray}
The 3d expressions can be obtained from the rule $\Braket{S_{1}X_{2}X_{3}\cdots X_{k-1}S_{k}}$ $\rightarrow$ $\slashed{x}_{12}\slashed{x}_{23}\cdots\slashed{x}_{k-1,k}\slashed{x}_{k-1,k}$. We shall denote products of the form $\Braket{S_{i}X_{a}\cdots X_{b}S_{m}}\Braket{S_{k}X_{u}\cdots X_{v}S_{l}}$ in 3d \\
as $\left[ \slashed{x}_{ia}\cdots\slashed{x}_{bm} \right] \left[ \slashed{x}_{ku}\cdots\slashed{x}_{vl} \right]$. \\
\\
For the three point function of one spin half fermion one scalar and one fermionic operator of any spin $\Braket{\psi_{1}O_{2}\psi_{3,l}}$ ($l>\frac{1}{2}$), we can take the following tensor structures,
\begin{eqnarray}
r_{cr}^{+}&=&\frac{\Braket{S_{1}S_{3}}\Braket{S_{3}X_{1}X_{2}S_{3}}^{l-\frac{1}{2}}}{X_{12}^{\frac{l}{2}-\frac{1}{4}}X_{13}^{\frac{l}{2}+\frac{1}{4}}X_{23}^{\frac{l}{2}-\frac{1}{4}}},             \label{2ferm1sc}\\
r_{cr}^{-}&=&\frac{\Braket{S_{1}X_{2}S_{3}}\Braket{S_{3}X_{1}X_{2}S_{3}}^{l-\frac{1}{2}}}{X_{12}^{\frac{l}{2}+\frac{1}{4}}X_{13}^{\frac{l}{2}-\frac{1}{4}}X_{23}^{\frac{l}{2}+\frac{1}{4}}} .        \nonumber 
\end{eqnarray}
Note that we have chosen the same tensor structures for the three point functions as in \cite{Iliesiu:2015qra, Iliesiu:2015akf} only with different normalization.\\
\\
\textbf{Four point function of two scalars and two fermions.}\\
\\ 
Let us now consider correlators with two fermions and two scalars\footnote{We have briefly discussed the generalization to include more scalars in Appendix \ref{countessa}}. \\
\\
We choose the following tensor structures for the four point function of two spin half fermions and two scalars $\Braket{\psi_{1}\psi_{2}O_{3}O_{4}}$, as in \cite{Iliesiu:2015akf}\footnote{Two different choices of normalization for any tensor structure can result in a difference to the corresponding Mellin amplitude only if they are different by factors of invariants. When such an invariant is a product of cross-ratios, it would cause a simple shift in the poles of the Mellin amplitude.},
\begin{eqnarray}
t_{1}^{+}=\frac{\Braket{S_{1}S_{2}}}{\sqrt{X_{12}}}  ,  \;\;\;\;\;\;\; && \;\;\;\;\;\;\; t_{2}^{+}=\frac{\Braket{S_{1}X_{3}X_{4}S_{2}}}{\sqrt{X_{13}X_{34}X_{42}}}        ,    \label{tens4}  \\
t_{3}^{-}=\frac{\Braket{S_{1}X_{3}S_{2}}}{\sqrt{X_{13}X_{32}}},    \;\;\;\;\;\;\; && \;\;\;\;\;\;\; t_{4}^{-}=\frac{\Braket{S_{1}X_{4}S_{2}}}{\sqrt{X_{14}X_{42}}}.         \nonumber  
\end{eqnarray}
\textbf{Four point function of fermions}\\
\\
We present here the basis for the fermion four point function that we shall use. We shall justify our choice with more details in \ref{subsec:4ferm-pole}. In this case, the tensor structures are of the form $\Braket{S_{i}\cdots S_{j}}\Braket{S_{k}\cdots S_{l}}$. In general such tensor structures may be related by complicated identities. The idea of expressing the embedding space tensor structures in a chosen conformal frame is particularly useful for relating (or showing the mutual independence of) different tensor structures in this context.\\
\\
For four fermions there are sixteen independent tensor structures and we pick a basis with elements of definite parity. The parity even structures are taken to be\footnote{The choice is not unique.},
\begin{eqnarray}
p_{1}^{+}=\frac{\Braket{S_{1}S_{2}}\Braket{S_{3}S_{4}}}{\sqrt{X_{12}X_{34}}},           &&  \;\;\;\;\;\;\;\;\;\;\;\;\;\;\;\;   p_{2}^{+}=\frac{\Braket{S_{1}S_{2}}\Braket{S_{3}X_{1}X_{2}S_{4}}}{\sqrt{X_{12}^{2}X_{13}X_{24}}},          \label{mt6}          \\
p_{3}^{+}=\frac{\Braket{S_{1}X_{3}\Gamma^{A}S_{2}}\Braket{S_{3}X_{1}\Gamma_{A}S_{4}}}{\sqrt{X_{13}X_{32}X_{31}X_{14}}} ,          &&  \;\;\;\;\;\;\;\;\;\;\;\;\;\;\;\;   p_{4}^{+}=\frac{\Braket{S_{1}\Gamma^{A}\Gamma^{B}S_{2}}\Braket{S_{3}\Gamma_{A}\Gamma_{B}S_{4}}}{\sqrt{X_{12}X_{34}}},                 \nonumber \\ 
p_{5}^{+}=\frac{\Braket{S_{1}X_{3}S_{2}}\Braket{S_{3}X_{1}S_{4}}}{\sqrt{X_{13}^{2}X_{14}X_{23}}} ,          &&  \;\;\;\;\;\;\;\;\;\;\;\;\;\;\;\;   p_{6}^{+}=\frac{\Braket{S_{1}X_{3}S_{2}}\Braket{S_{3}X_{2}S_{4}}}{\sqrt{X_{23}^{2}X_{13}X_{24}}} ,               \nonumber \\
p_{7}^{+}=\frac{\Braket{S_{1}X_{4}S_{2}}\Braket{S_{3}X_{1}S_{4}}}{\sqrt{X_{13}X_{14}^{2}X_{24}}},           &&  \;\;\;\;\;\;\;\;\;\;\;\;\;\;\;\;   p_{8}^{+}=\frac{\Braket{S_{1}X_{4}S_{2}}\Braket{S_{3}X_{2}S_{4}}}{\sqrt{X_{14}X_{23}X_{24}^{2}}}   .               \nonumber    
\end{eqnarray} 
The parity odd part of the basis can be taken to be composed of the following structures,
\begin{eqnarray}
p_{9}^{-}=\frac{\Braket{S_{1}S_{2}}\Braket{S_{3}X_{1}S_{4}}}{\sqrt{X_{12}X_{13}X_{14}}}  ,         &&  \;\;\;\;\;\;\;\;\;\;\;\;\;\;\;\;   p_{10}^{-}=\frac{\Braket{S_{1}S_{2}}\Braket{S_{3}X_{2}S_{4}}}{\sqrt{X_{12}X_{23}X_{24}}} ,       \label{mt7}          \\
p_{11}^{-}=\frac{\Braket{S_{1}X_{3}S_{2}}\Braket{S_{3}S_{4}}}{\sqrt{X_{13}X_{23}X_{34}}}  ,         &&  \;\;\;\;\;\;\;\;\;\;\;\;\;\;\;\;   p_{12}^{-}=\frac{\Braket{S_{1}X_{4}S_{2}}\Braket{S_{3}S_{4}}}{\sqrt{X_{14}X_{24}X_{34}}}  ,        \nonumber   \\
p_{13}^{-}=\frac{\Braket{S_{1}\Gamma^{A}S_{2}}\Braket{S_{3}\Gamma_{A}X_{1}S_{4}}}{\sqrt{X_{12}X_{13}X_{14}}}    ,       &&  \;\;\;\;\;\;\;\;\;\;\;\;\;\;\;\;   p_{14}^{-}=\frac{\Braket{S_{1}\Gamma^{A}S_{2}}\Braket{S_{3}\Gamma_{A}X_{2}S_{4}}}{\sqrt{X_{12}X_{23}X_{24}}}   ,       \nonumber   \\
p_{15}^{-}=\frac{\Braket{S_{1}\Gamma^{A}X_{3}S_{2}}\Braket{S_{3}\Gamma_{A}S_{4}}}{\sqrt{X_{13}X_{23}X_{34}}}    ,       &&  \;\;\;\;\;\;\;\;\;\;\;\;\;\;\;\;   p_{16}^{-}=\frac{\Braket{S_{1}\Gamma^{A}X_{4}S_{2}}\Braket{S_{3}\Gamma_{A}S_{4}}}{\sqrt{X_{14}X_{42}X_{34}}}     .     \nonumber   
\end{eqnarray} 
\textbf{Four dimensions}\\
\\
Our discussion in three dimensions can be easily generalized. The nature of spinors changes with dimension and signature. The problem of counting tensor structures in four dimensions is again handled in the best manner by choosing a conformal frame \cite{Kravchuk:2016qvl}. 4d tensor structures in the embedding formalism and blocks have been discussed in \cite{SimmonsDuffin:2012uy, Elkhidir:2014woa}. The relevant setup is coherently presented in \cite{Cuomo:2017wme} and they also implement the setup in the freely available Mathematica package "CFTs4d". One can easily use this Mathematica package to obtain independent tensor structures (with expressions in both embedding space and the conformal frame) for upto four point functions for any kind of correlator. In three dimensions, all operators exchanged in the OPEs can be taken to be symmetric representations of the double cover of the Lorentz group. However, in higher dimensions, one has to also consider mixed symmetry representations (see \cite{Echeverri:2016dun}).

\subsection{Definition}
\label{subsec:defn}

After our discussion on tensor structures, we are equipped to define Mellin amplitudes for correlators of fermions and scalars. In general for a correlator of $2K$ fermions and $M$ scalars ($2K+M=n$), we can define the Mellin amplitude (in embedding space) with the following set of Mellin-Barnes integrals,
\be
&& \Braket{\Psi_{1}\left(X_{1},S_{1}\right)\cdots\Psi_{2k}\left(X_{2K},S_{2K}\right)\Phi_{2K+1}\left(X_{2K+1}\right)\cdots\Phi_{n}\left(X_{n}\right)}   \nonumber \\
=&&\sum_{k}\tilde{\mathbb{T}}_{k}\int[ds_{ij}]\prod_{1\leq i<j}^{n}X_{ij}^{-s_{ij}-a_{ij;k}}\Gamma\left(s_{ij}+a_{ij;k}+n_{ij;k}+\frac{1}{2}\sum_{m=1}^{K}\delta_{i,2m-1}\delta_{j,2m}\right)
\nonumber
\\
& & \times \mathcal{M}_{k}\left(\{s_{ij}\}\right)  \prod_{m=1}^{K}\frac{1}{\sqrt{X_{2m-1,2m}}},          \label{defu}
\ee
The set $\{\mathcal{M}_{k}\left(s_{ij}\right)\}$ is the Mellin amplitude. 
We demand the Mellin variables to satify the following constraints:
\be
\tau_{i}-\sum_{l\neq i}s_{li}=0,    \;\;\;\; \forall i.        \label{alcantara} 
\ee
\\
In the equation above, the tensor structures $\tilde{\mathbb{T}}_{i}$ do not have a denominator (not normalized) unlike those in ~\eqref{mt6} for example. The set $\{\tilde{\mathbb{T}}_{i}\}$ must form a basis of tensor structures for the given correlator and apart from being a Lorentz invariant in $d+2$ dimensions and satisfying the transversality condition, each $\tilde{\mathbb{T}}_{i}$ must satisfy the following homogeneity condition in $S_{i}$,
\be
\tilde{\mathbb{T}}_{k}\left(b_{1}S_{1},\cdots,b_{2K}S_{2K};X_{1},\cdots,X_{n}\right)=\left(\prod_{i=1}^{2K}b_{i}\right)\tilde{\mathbb{T}}_{k}\left(S_{1},\cdots,S_{2K};X_{1},\cdots,X_{n}\right).              \label{tenshomo}
\ee 
$a_{ij}$ are numbers which determine the normalization of the tensor structure and they are such that the dimension of the tensor structure (in physical space) is zero. Let us define,
\be
 \mathbb{T}_{k}= \tilde{\mathbb{T}}_{k}\prod_{1\leq i<j}^{n}X_{ij}^{-a_{ij;k}} .      \label{tensresc}
\ee
Concretely, the numbers $a_{ij}$ are fixed by the requirement that given $\lambda_{i}=\sqrt{\sigma_{i}}$, the following must hold,
\be
\mathbb{T}_{k}\left(\lambda_{1}S_{1},\cdots,\lambda_{2K}S_{2K};\sigma_{1}X_{1},\cdots,\sigma_{n}X_{n}\right)=\mathbb{T}_{k}\left(S_{1},\cdots,S_{2K};X_{1},\cdots,X_{n}\right) .           \label{tensca}
\ee 
$\tau_{i}$ is the twist of the operator at $X_{i}$. So $\tau_{i}=\Delta_{i}-\frac{1}{2}$ for $i\in\{1,\cdots,2K\}$ and $\tau_{j}=\Delta_{j}$ for $j\in\{2K+1,\ldots,n\}$.\\
\\
$n_{ij;k}$ are integers that we keep undetermined for now. The gamma functions in ~\eqref{defu} have been extracted in analogy with the case of scalars to simplify the asymptotics of the Mellin amplitude on the complex plane and the factorization formulae. In Chapter \ref{sec:wittendiagrams} we shall be computing Mellin amplitudes in the large $N$ limit of a strongly coupled CFT (through tree level Witten diagrams) and in Chapter \ref{sec:perturbative} we shall be computing Mellin amplitudes in a weakly interacting CFT. We shall choose $n_{ij;k}$ such that in either case the Mellin amplitude for the contact interaction are polynomials in the Mellin variables (constant for the contact Witten diagrams). This way, the Mellin amplitudes in the large $N$ limit of the strongly coupled CFT (dual to a quantum field theory in $AdS$) encodes only the bulk dynamics. In the perturbative regime, the singularities of the Mellin amplitude do not carry information on the trivial composite operators. \\   
\\
It can be checked that the correlator in ~\eqref{defu} is consistent with the homogeneity condition ~\eqref{homo}, given that ~\eqref{tenshomo} and ~\eqref{tensca} are satisfied. The conformality constraints imposed by ~\eqref{alcantara} in ~\eqref{defu} can be interpreted in terms of fictitious Mellin momenta $k_{i}$ with $k_{i}\cdot k_{j}=s_{ij}$ and an on-shell condition $k_{i}^{2}=-\tau_{i}$ as the overall conservation of Mellin momentum $\sum_i k_{i}=0$. This is a generalization of the corresponding scenario for scalar correlator as discussed in \ref{subsec:scalar}. This time, one can relate the Mellin variables to Mandelstam variables as
\begin{eqnarray}
S_{i_{1}\cdots i_{a}}=-\left(k_{i_{1}}+\cdots+k_{i_{a}}\right)^{2}=-2\sum_{l<k\leq a}s_{i_{l}i_{k}}+\sum_{j=1}^{a} \tau_{i_{j}} .         \label{mandelferm}
\end{eqnarray}
Since we have chosen to work in Minkowski spacetime, we shall always understand that $X_{ij}\rightarrow -(x_{i}-x_{j})^{2}+i\epsilon_{ij}$. The relative values of all the $\epsilon_{ij}$ is assumed to be consistent with the time ordering in the correlator. \\
\\
In this paper, we shall mainly be focussing on the four point function. We shall assume all operators of the same spin to have different conformal dimensions. Let us mention concretely, the definition for the two kinds of four point functions. Here we make a choice of $n_{ij;k}$. The Mellin amplitude for the four point function of two fermions and two scalars is defined by the following,
\be
\Braket{\Psi_{1}\Psi_{2}\Phi_{3}\Phi_{4}}=\int[ds_{ij}]\prod_{i<j}\left(X_{ij}\right)^{-s_{ij}}\frac{1}{\sqrt{X_{12}}} \left[\sum_{i=1}^{4}t_{i}\bar{M}_{i}\left(\{s_{ab}\}\right)\right] .       \label{directdef}
\ee
The Mellin variables satisfy the conformality constraints as mentioned in ~\eqref{alcantara}.\\
\\
The tensor structures $t_{i}$ for this correlator are chosen in ~\eqref{tens4}. In ~\eqref{directdef}, the superscript from ~\eqref{tens4} indicating the parity of the tensor structures $t_{i}$ has been suppressed. Following Mack \cite{Mack:2009mi}, we shall call $\bar{M}\equiv \{\bar{M}_{i}\}$ the reduced Mellin amplitude. In this case, we choose all the integers $n_{ij;k}$ to be zero and the Mellin amplitude $\{\mathcal{M}_{i}\}$ is related to $\{\bar{M}_{i}\}$ in the following way. 
\be
\mathcal{M}_{1}&=&\bar{M}_{1}\left[\Gamma\left(s_{12}+1\right)\Gamma\left(s_{13}\right)\Gamma\left(s_{14}\right)\Gamma\left(s_{23}\right)\Gamma\left(s_{24}\right)\Gamma\left(s_{34}\right)\right]^{-1},       \nonumber \\
\mathcal{M}_{2}&=&\bar{M}_{2}\left[\Gamma\left(s_{12}+\frac{1}{2}\right)\Gamma\left(s_{13}+\frac{1}{2}\right)\Gamma\left(s_{14}\right)\Gamma\left(s_{23}\right)\Gamma\left(s_{24}+\frac{1}{2}\right)\Gamma\left(s_{34}+\frac{1}{2}\right)\right]^{-1},       \nonumber \\
\mathcal{M}_{3}&=&\bar{M}_{3}\left[\Gamma\left(s_{12}+\frac{1}{2}\right)\Gamma\left(s_{13}+\frac{1}{2}\right)\Gamma\left(s_{14}\right)\Gamma\left(s_{23}+\frac{1}{2}\right)\Gamma\left(s_{24}\right)\Gamma\left(s_{34}\right)\right]^{-1}  ,   \nonumber\\
\mathcal{M}_{4}&=&\bar{M}_{4}\left[\Gamma\left(s_{12}+\frac{1}{2}\right)\Gamma\left(s_{13}\right)\Gamma\left(s_{14}+\frac{1}{2}\right)\Gamma\left(s_{23}\right)\Gamma\left(s_{24}+\frac{1}{2}\right)\Gamma\left(s_{34}\right)\right]^{-1} .       \label{concre1}  
\ee
Similarly, define the four point function of fermions in the following way,
\begin{eqnarray}
\Braket{\Psi_{1}\Psi_{2}\Psi_{3}\Psi_{4}}=\int [ds_{ij}]\prod_{i<j}\left(X_{ij}\right)^{-s_{ij}}\frac{1}{\sqrt{X_{12}X_{34}}} \left[\sum_{i=1}^{16}p_{i}\bar{M}_{i}\left(\{s_{ab}\}\right)\right] .       \label{4fermdef}
\end{eqnarray}
The tensor structures $p_{i}$ are as in  ~\eqref{mt6}, ~\eqref{mt7}. The choice of the integers $n_{ij;k}$ dictates the relation between the reduced Mellin amplitude $\{\bar{M}_{i}\}$ and the Mellin amplitude $\mathcal{M}_{i}$. We choose all the integers to be zero apart from the following,
\be
n_{12;2}=n_{13;3}=n_{13;5}=n_{23;6}=n_{14;7}=n_{24;8}=-1 .      \label{lahm}
\ee
We have presented the relations explicitly in Appendix \ref{mueller}.

\subsection{Pole structure: Fermion-scalar four point function}
\label{subsec:pole2f}

In this section, we will look at the pole structure of the mixed fermion scalar four point function in the direct and the crossed channels.

\subsubsection{Direct channel}
\label{subsec:2ferm-s}

The mixed fermion scalar four point function can be expressed in the following manner.
\begin{eqnarray}
\Braket{\psi_{1}\psi_{2}\phi_{3}\phi_{4}} &=& \left(\frac{X_{24}}{X_{14}}\right)^{\frac{\Delta_{1}-\Delta_{2}}{2}}\left(\frac{X_{14}}{X_{13}}\right)^{\frac{\Delta_{3}-\Delta_{4}}{2}} \frac{\mathcal{A}(u,v)}{\left(X_{12}\right)^{\frac{\Delta_{1}+\Delta_{2}}{2}}\left(X_{34}\right)^{\frac{\Delta_{3}+\Delta_{4}}{2}}},      \label{ferm4pt} \\
\mathcal{A}(u,v)&=&\int_{c_{s}-i\infty}^{c_{s}+i\infty} \frac{ds}{2\pi i} \int_{c_{t}-i\infty}^{c_{t}+i\infty} \frac{dt}{2\pi i}  \left[\sum_{i=1}^{4}t_{i}\bar{M}_{i}(s,t)\right]  u^{\frac{s}{2}}v^{-\frac{s+t-\tau_{1}-\tau_{4}}{2}},        \label{ferm4pt2}  \\
 s= -(k_{1}+k_{2})^{2}&=&\tau_{1}+\tau_{2}-2s_{12},   \;\;\;\;\;\;\;\;\;\;\;    t= -(k_{1}+k_{3})^{2}=\tau_{1}+\tau_{3}-2s_{13}.   \nonumber 
\end{eqnarray}  
We wish to compare ~\eqref{ferm4pt2} with the contribution to $\mathcal{A}(u,v)$ from a single operator exchanged in the direct channel. For this, one can do a ``dimensional analysis'' to check the power law behavior of $\mathcal{A}(u,v)$ in $u$ in the OPE limit $(u,v) \rightarrow (0,1)$ (with $\frac{1-v}{\sqrt{u}}$ held fixed). We have explicitly checked the leading behavior of the conformal blocks using the differential operators presented in \cite{Iliesiu:2015qra} that enable one to obtain these direct channel blocks from the corresponding blocks for scalar four point function. The contribution from one operator exchanged via the OPE has also been presented in general for any value of spin of the external operators in \cite{Kravchuk:2017dzd}. In this paper the Gelfand-Tsetlin basis for $Spin(d)$ representations has been used. We are using a basis defined by our choice of gamma matrices (as in \cite{Iliesiu:2015qra}) and are also using three point structures of definite parity (unlike in \cite{Kravchuk:2017dzd}). The operators contributing to the direct channel block expansion are those that appear in both the OPE of two scalars and that of two spin-half fermions, and hence are integer spin operators in symmetric traceless representations of the Lorentz group. \\
\\
Let $\mathcal{A}(u,v)=\sum_{i}t_{i}\mathcal{A}_{i}(u,v)$. The three point function of two fermions and an integer spin operator has in general four independent tensor structures~\eqref{2spino1tens} and hence four structure constants. Consequently each $\mathcal{A}_{i}(u,v)$ will in general receive contributions from four different conformal partial waves $g^{a}_{\Delta,l}$ (with covariant pre-factors stripped off). Let $g^{i,a}_{\Delta,l}$ be the contribution of $g^{a}_{\Delta,l}$ to $\mathcal{A}_{i}$. Here ``$a$'' lables the four tensor structures in the three point function $\Braket{\psi_{1}\psi_{2}\mathcal{O}_{l}}$. \\
\\
Let us recall from ~\eqref{tens4} that $t_{1},t_{2}$ are parity even and $t_{3},t_{4}$ are parity odd. Also from ~\eqref{2spino1tens}, $r_{1},r_{2}$ are parity even and $r_{3},r_{4}$ are parity odd. Considering this and the explicit form of the three point structures, we see that the only non-zero  $g^{i,a}_{\Delta,l}$ are $g^{1,1}_{\Delta,l}$, $g^{1,2}_{\Delta,l}$, $g^{2,2}_{\Delta,l}$, $g^{3,3}_{\Delta,l}$, $g^{3,4}_{\Delta,l}$, $g^{4,3}_{\Delta,l}$ and $g^{4,4}_{\Delta,l}$. For $l=0$, the only non-zero ones are $g^{1,1}_{\Delta,0}\equiv g^{1,+}_{\Delta,0}$, $g^{3,3}_{\Delta,0}\equiv g^{3,-}_{\Delta,0}$ and $g^{4,3}_{\Delta,0}\equiv g^{4,-}_{\Delta,0}$.  \\
\\
We summarise the limiting behavior of $g^{i,a}_{\Delta,l}$ in the OPE limit here. This is generically given by some combination of Gegenbauer polynomials. For $l\geq 1$,
\begin{eqnarray}
\mathcal{A}_{1}&\supset& \lambda_{\psi_{1}\psi_{2}\mathcal{O}_{l}}^{1}\lambda_{\mathcal{O}_{l}\phi_{3}\phi_{4}}g_{s,\Delta,l}^{1,1}+ \lambda_{\psi_{1}\psi_{2}\mathcal{O}_{l}}^{2}\lambda_{\mathcal{O}_{l}\phi_{3}\phi_{4}}g_{s,\Delta,l}^{1,2}     \label{pst1}  \\
      & \approx &  u^{\frac{\Delta}{2}}\sum_{k=0}^{\floor[\big]{\frac{l-2}{2}}}\left(\lambda_{\psi_{1}\psi_{2}\mathcal{O}_{l}}^{1}\lambda_{\mathcal{O}_{l}\phi_{3}\phi_{4}}\mathcal{K}_{1}^{1,k}+\lambda_{\psi_{1}\psi_{2}\mathcal{O}_{l}}^{2}\lambda_{\mathcal{O}_{l}\phi_{3}\phi_{4}}\mathcal{K}_{1}^{2,k}\right)\left(\frac{v-1}{2\sqrt{u}}\right)^{l-2k}      +  \cdots , \nonumber \\
\mathcal{A}_{2}&\supset& \lambda_{\psi_{1}\psi_{2}\mathcal{O}_{l}}^{2}\lambda_{\mathcal{O}_{l}\phi_{3}\phi_{4}}g_{s,\Delta,l}^{2,2}  \label{pst2} \\
         & \approx &  \lambda_{\psi_{1}\psi_{2}\mathcal{O}_{l}}^{2}\lambda_{\mathcal{O}_{l}\phi_{3}\phi_{4}} u^{\frac{\Delta}{2}}\sum_{k=0}^{\floor[\big]{\frac{l-1}{2}}}\mathcal{K}_{2}^{k}\left(\frac{v-1}{2\sqrt{u}}\right)^{l-1-2k}+\cdots ,                \nonumber   \\
\mathcal{A}_{3}&\supset& \lambda_{\psi_{1}\psi_{2}\mathcal{O}_{l}}^{3}\lambda_{\mathcal{O}_{l}\phi_{3}\phi_{4}}g_{s,\Delta,l}^{3,3}+ \lambda_{\psi_{1}\psi_{2}\mathcal{O}_{l}}^{4}\lambda_{\mathcal{O}_{l}\phi_{3}\phi_{4}}g_{s,\Delta,l}^{3,4}     \label{pst3} \\
        & \approx &  u^{\frac{\Delta}{2}}\sum_{k=0}^{\floor[\big]{\frac{l-1}{2}}}\lambda_{\psi_{1}\psi_{2}\mathcal{O}_{l}}^{3}\lambda_{\mathcal{O}_{l}\phi_{3}\phi_{4}}\mathcal{K}_{3}^{3,k}\left(\frac{v-1}{2\sqrt{u}}\right)^{l-2k}+ u^{\frac{\Delta-1}{2}}\sum_{k=0}^{\floor[\big]{\frac{l-1}{2}}}\lambda_{\psi_{1}\psi_{2}\mathcal{O}_{l}}^{4}\lambda_{\mathcal{O}_{l}\phi_{3}\phi_{4}}\mathcal{K}_{3}^{4,k}\left(\frac{v-1}{2\sqrt{u}}\right)^{l-1-2k} + \cdots,   \nonumber \\
\mathcal{A}_{4}&\supset& \lambda_{\psi_{1}\psi_{2}\mathcal{O}_{l}}^{3}\lambda_{\mathcal{O}_{l}\phi_{3}\phi_{4}}g_{s,\Delta,l}^{4,3}+ \lambda_{\psi_{1}\psi_{2}\mathcal{O}_{l}}^{4}\lambda_{\mathcal{O}_{l}\phi_{3}\phi_{4}}g_{s,\Delta,l}^{4,4}     \label{pst4}    \\
          &\approx & u^{\frac{\Delta}{2}}\sum_{k=0}^{\floor[\big]{\frac{l-1}{2}}}\lambda_{\psi_{1}\psi_{2}\mathcal{O}_{l}}^{3}\lambda_{\mathcal{O}_{l}\phi_{3}\phi_{4}}\mathcal{K}_{4}^{3,k}\left(\frac{v-1}{2\sqrt{u}}\right)^{l-2k}+ u^{\frac{\Delta-1}{2}}\sum_{k=0}^{\floor[\big]{\frac{l-1}{2}}}\lambda_{\psi_{1}\psi_{2}\mathcal{O}_{l}}^{4}\lambda_{\mathcal{O}_{l}\phi_{3}\phi_{4}}\mathcal{K}_{4}^{4,k}\left(\frac{v-1}{2\sqrt{u}}\right)^{l-1-2k} + \cdots  . \nonumber
\end{eqnarray}
$\lambda_{\psi_{1}\psi_{2}\mathcal{O}_{l}}^{a}$ are the structure constants of the three point function $\Braket{\psi_{1}\psi_{2}\mathcal{O}_{l}}$ associated to the tensor structure $r_{d,i}$ as in ~\eqref{2spino1tens}. $\mathcal{K}_{i}^{j,k}$ are constants.\\
\\
For a scalar exchange, matters simplify as $\lambda_{\psi_{1}\psi_{2}\mathcal{O}_{0}}^{1}\equiv \lambda_{\psi_{1}\psi_{2}\mathcal{O}}^{+}$, $\lambda_{\psi_{1}\psi_{2}\mathcal{O}_{0}}^{3}\equiv \lambda_{\psi_{1}\psi_{2}\mathcal{O}}^{-}$, $\lambda_{\psi_{1}\psi_{2}\mathcal{O}_{0}}^{2}\equiv 0$ and $\lambda_{\psi_{1}\psi_{2}\mathcal{O}_{0}}^{4}\equiv 0$. So we have,
\begin{eqnarray}
\mathcal{A}_{1} &\supset&  \lambda_{\psi_{1}\psi_{2}\mathcal{O}}^{+}\lambda_{\mathcal{O}\phi_{3}\phi_{4}}\mathcal{K}_{1}^{1,0}u^{\frac{\Delta}{2}}  +\cdots     , \label{ps1}  \\
\mathcal{A}_{3} &\supset&  \lambda_{\psi_{1}\psi_{2}\mathcal{O}}^{-}\lambda_{\mathcal{O}\phi_{3}\phi_{4}}\mathcal{K}_{3}^{3,0}u^{\frac{\Delta}{2}}  +\cdots , \label{ps2} \\
\mathcal{A}_{4} &\supset&  \lambda_{\psi_{1}\psi_{2}\mathcal{O}}^{-}\lambda_{\mathcal{O}\phi_{3}\phi_{4}}\mathcal{K}_{4}^{3,0}u^{\frac{\Delta}{2}}   + \cdots . \label{ps3} 
\end{eqnarray}
Comparing ~\eqref{pst1} -~\eqref{pst4} and ~\eqref{ps1} -~\eqref{ps3} with ~\eqref{ferm4pt2}, we can deduce the pole structure for $\bar{M}_{i}$. The Mellin amplitude $\{\mathcal{M}_{i}\}$ can be obtained from the reduced Mellin amplitude $\{\bar{M}_{i}\}$ as shown in ~\eqref{concre1} and thus we obtain the pole structure of the Mellin amplitude in this channel as summarised in Table \ref{table1}.
\begin{table}
\centering 
\begin{tabular}{|c|c|C{6.2cm}|C{5.2cm}|}
\hline
\textbf{Component of M.A.} & \textbf{Location of Poles} 
  & \textbf{Residues $\sim$}  \\ 
\hline\hline
\multirow{4}{*}{\vspace*{.5in}$\mathcal{M}_1$} & \multirow{4}{*}{\vspace*{.5in}$\tau +2k$}  
  &\multirow{4}{*}{\vspace*{1.2cm} $\lambda_{\psi_1\psi_2\mathcal{O}_\ell}^1\lambda_{\mathcal{O}_\ell\phi_3\phi_4}\;\;,\;\;\lambda_{\psi_1\psi_2\mathcal{O}_\ell}^2\lambda_{\mathcal{O}_\ell\phi_3\phi_4} $}  \\ [.4cm]
\hline\hline
\multirow{4}{*}{\vspace*{.5in}$\mathcal{M}_2$} & \multirow{4}{*}{\vspace*{.5in}$\tau +1+2k$}  
  &\multirow{4}{*}{\vspace*{1.2cm} $\lambda_{\psi_1\psi_2\mathcal{O}_\ell}^2\lambda_{\mathcal{O}_\ell\phi_3\phi_4}$}  \\ [.4cm]
\hline\hline
\multirow{4}{*}{\vspace*{.5in}$\mathcal{M}_3$} & \multirow{4}{*}{\vspace*{.5in}$\tau +2k$}  
  &\multirow{4}{*}{\vspace*{1.2cm} $\lambda_{\psi_1\psi_2\mathcal{O}_\ell}^3\lambda_{\mathcal{O}_\ell\phi_3\phi_4}\;\;,\;\;\lambda_{\psi_1\psi_2\mathcal{O}_\ell}^4\lambda_{\mathcal{O}_\ell\phi_3\phi_4}$}  \\ [.4cm]
\hline\hline
\multirow{4}{*}{\vspace*{.5in}$\mathcal{M}_4$} & \multirow{4}{*}{\vspace*{.5in}$\tau+2k$}  
  &\multirow{4}{*}{\vspace*{1.2cm} $\lambda_{\psi_1\psi_2\mathcal{O}_\ell}^3\lambda_{\mathcal{O}_\ell\phi_3\phi_4}\;\;,\;\;\lambda_{\psi_1\psi_2\mathcal{O}_\ell}^4\lambda_{\mathcal{O}_\ell\phi_3\phi_4}$}  \\ [.4cm]
\hline\hline
\end{tabular}
\caption{Fermion-scalar four point function: Direct channel poles }
\label{table1}
\end{table}
When the exchanged operator is a scalar $l=0$, we should take all structure constants apart from $\lambda_{\psi_{1}\psi_{2}\mathcal{O}_{l}}^{1}$, $\lambda_{\psi_{1}\psi_{2}\mathcal{O}_{l}}^{3}$, $\lambda_{\mathcal{O}_{l}\phi_{3}\phi_{4}}$ to be zero. \\
\\
$k=0$ corresponds to the exchange of the primary and the leading twist descendants while $k>0$ corresponds to the descendants with higher values of twist. Generically the singular terms in each component of the Mellin amplitude are of the following form,
\begin{eqnarray}
\frac{\lambda_{\psi_{1}\psi_{2}\mathcal{O}_{l}}^{a}\lambda_{\mathcal{O}_{l}\phi_{3}\phi_{4}}Q_{l,k}(t)}{s-\tau-2k}.      \label{polerep}
\end{eqnarray}
$Q_{l,k}(t)$ can be expected to be a polynomial in $t$ whose degree is determined by spin $l$ of the exchanged operator. Comparing with the case of the scalar four point function, we can expect the degree of $Q_{l,k}$ to be $l$ for $\bar{M}_{1}$, $\bar{M}_{3}$, $\bar{M}_{4}$ and $l-1$ for $\bar{M}_{2}$. We leave a rigorous derivation of this polynomial to future work.\\
\\ 
The Mellin amplitude of the three point function $\Braket{\psi_{1}\psi_{2}\mathcal{O}_{l}}$ has four components, and each one is a constant proportional to the corresponding structure constant $\lambda_{\psi_{1}\psi_{2}\mathcal{O}_{l}}^{a}$. The Mellin amplitude associated with  $\Braket{\mathcal{O}_{l}\phi_{3}\phi_{4}}$ is just a constant proportional to $\lambda_{\mathcal{O}_{l}\phi_{3}\phi_{4}}$. Therefore from ~\eqref{polerep}, it is clear that each component of the Mellin amplitude associated with the four point function $\Braket{\psi_{1}\psi_{2}\phi_{3}\phi_{4}}$ factorizes on the poles listed above onto components of Mellin amplitudes of the corresponding three point functions.

\subsubsection{Crossed channel}
\label{subsec:2ferm-t}

Now we consider the exchange of operators in the crossed channel, in particular the OPE channel $13-24$. The four point function can be expressed as follows\footnote{~\eqref{pfft1} is identical to equation 2.12 of \cite{Iliesiu:2015akf} with the following relabelling from ``theirs''$\rightarrow$``ours'' $1\rightarrow 1, 2\rightarrow 3, 3\rightarrow 4, 4\rightarrow 2$ and renaming $g^{I}\rightarrow \mathcal{A}_{i}$.},
\begin{eqnarray}
\Braket{\psi_{1}\psi_{2}\phi_{3}\phi_{4}}_{c}&=&\left(\frac{X_{34}}{X_{14}}\right)^{\frac{\Delta_{13}}{2}}\left(\frac{X_{14}}{X_{12}}\right)^{\frac{\Delta_{24}}{2}}\frac{\tilde{v}^{\frac{\Delta_{13}}{2}}}{X_{13}^{\frac{\Delta_{1}+\Delta_{3}}{2}}X_{24}^{\frac{\Delta_{2}+\Delta_{4}}{2}}}\sum_{i=1}^{4}t_{i}\tilde{A}_{i}\left(\tilde{u},\tilde{v}\right),          \label{pfft1}  \\
\tilde{u}=\frac{X_{13}X_{24}}{X_{12}X_{34}}   &&  \qquad  \tilde{v}=\frac{X_{14}X_{23}}{X_{12}X_{34}}, 
\qquad \Delta_{ij}=\Delta_i -\Delta_j,            \nonumber   \\
\tilde{A}_{i}\left(\tilde{u},\tilde{v}\right)&=& \int \frac{dt}{2\pi i}\int \frac{ds}{2\pi i} \bar{M}_{i}\left(s,t\right)\tilde{u}^{\frac{t+\frac{1}{2}}{2}}\tilde{v}^{\frac{s+t+\frac{1}{2}-\Delta_{1}-\Delta_{4}}{2}}.           \label{pfft2}  
\end{eqnarray}
The operators contributing to the block expansion in the crossed channel are fermionic operators. Once again, we shall compare ~\eqref{pfft2} with the leading behavior of the corresponding blocks in the OPE limit $x_{1}\rightarrow x_{3}$. These blocks are also a type of ``seed-blocks" in three dimensions, and have been computed in \cite{Iliesiu:2015akf, Karateev:2017jgd}. We have chosen the same tensor structures as they have for the relevant three point functions ~\eqref{2ferm1sc} and also the same tensor structures for the four point function.\\
\\
Three point functions of one spin-half fermion, one scalar and one generic fermion have one parity odd and one parity even tensor structure ~\eqref{2ferm1sc}. Hence each $\mathcal{\tilde{A}}_{i}$ will receive contributions from four different blocks $g^{i,\pm\pm}$. Let $g^{i,jk}_{\Delta,l}$ be the contribution to $\mathcal{\tilde{A}}_{i}$ from the block associated with the fusion of tensor structures $r_{c}^{j}$ and $r_{c}^{k}$ ~\eqref{2ferm1sc} of the three point functions. Therefore we can see that the following parity selection rule holds: the only non-zero $g^{i,jk}_{\Delta,l}$ are $g^{1,++}_{\Delta,l}$, $g^{1,--}_{\Delta,l}$, $g^{2,++}_{\Delta,l}$, $g^{2,--}_{\Delta,l}$, $g^{3,+-}_{\Delta,l}$, $g^{3,-+}_{\Delta,l}$, $g^{4,+-}_{\Delta,l}$ and $g^{4,-+}_{\Delta,l}$ \\
\\
We state the results on the pole structure here. Please refer to Appendix \ref{hazard} for the calculation. Let $\lambda_{\psi\phi\Psi_{l}}^{\pm}$ be the structure constant associated with the term with tensor structure $r_{c}^{\pm}$ (see ~\eqref{2ferm1sc}) in the three point function $\Braket{\psi\phi\Psi_{l}}$. Comparing ~\eqref{mero14} and ~\eqref{mero15}, we can conclude that the reduced Mellin amplitude and consequently the Mellin amplitude has the poles in $t$ as summarised in Table \ref{table2} for the exchange of fermionic operator $\Psi_{l}$ with twist $\tau$.
\\
\begin{table}
\centering 
\begin{tabular}{|c|c|C{6.2cm}|C{5.2cm}|}
\hline
\textbf{Component of M.A.} & \textbf{Location of Poles} 
  & \textbf{Residues $\sim$}  \\ 
\hline\hline
\multirow{4}{*}{\vspace*{.2in}$\mathcal{M}_1$} & \multirow{4}{*}{\vspace*{.5in}$t=\tau +2k$}  
  &\multirow{4}{*}{\vspace*{1.2cm} $\lambda_{\psi_1\phi_3\Psi_\ell}^+\lambda_{\Psi_\ell\phi_4\psi_2}^+$}  \\ [.4cm]
 \cline{2-3}
&\multirow{4}{*}{ \vspace*{.5in}$t=\tau+1 +2k$} & \vspace*{.1cm} $\lambda_{\psi_1\phi_3\Psi_\ell}^-\lambda_{\Psi_\ell\phi_4\psi_2}^-$  \\ [.3cm]
\hline\hline
\multirow{4}{*}{\vspace*{.2in}$\mathcal{M}_2$} & \multirow{4}{*}{\vspace*{.5in}$t=\tau +1+2k$}  
  &\multirow{4}{*}{\vspace*{1.2cm} $\lambda_{\psi_1\phi_3\Psi_\ell}^+\lambda_{\Psi_\ell\phi_4\psi_2}^+$}  \\ [.4cm]
 \cline{2-3}
&\multirow{4}{*}{ \vspace*{.5in}$t=\tau+2k$} & \vspace*{.1cm} $\lambda_{\psi_1\phi_3\Psi_\ell}^-\lambda_{\Psi_\ell\phi_4\psi_2}^-$  \\ [.3cm]
\hline\hline
\multirow{4}{*}{\vspace*{.2in}$\mathcal{M}_3$} & \multirow{4}{*}{\vspace*{.5in}$t=\tau +1+2k$}  
  &\multirow{4}{*}{\vspace*{1.2cm} $\lambda_{\psi_1\phi_3\Psi_\ell}^+\lambda_{\Psi_\ell\phi_4\psi_2}^-$}  \\ [.4cm]
 \cline{2-3}
&\multirow{4}{*}{ \vspace*{.5in}$t=\tau+2k$} & \vspace*{.1cm} $\lambda_{\psi_1\phi_3\Psi_\ell}^-\lambda_{\Psi_\ell\phi_4\psi_2}^+$  \\ [.3cm]
\hline\hline
\multirow{4}{*}{\vspace*{.2in}$\mathcal{M}_4$} & \multirow{4}{*}{\vspace*{.5in}$t=\tau+2k$}  
  &\multirow{4}{*}{\vspace*{1.2cm} $\lambda_{\psi_1\phi_3\Psi_\ell}^+\lambda_{\Psi_\ell\phi_4\psi_2}^-$}  \\ [.4cm]
 \cline{2-3}
&\multirow{4}{*}{ \vspace*{.5in}$t=\tau +1+2k$} & \vspace*{.1cm} $\lambda_{\psi_1\phi_3\Psi_\ell}^-\lambda_{\Psi_\ell\phi_4\psi_2}^+$  \\ [.3cm]
\hline\hline
\end{tabular}
\caption{Fermion-scalar four point function: Crossed channel poles.}
\label{table2}
\end{table}
\\
We see a novelty in the pole structure here. Each component of the Mellin amplitude has two series of poles for each primary exchanged. For $l>\frac{1}{2}$, the polynomials associated with the residues can be expected to be of degree $l-\frac{1}{2}$ when the corresponding pole is at $\tau+2k$ and of degree $l-\frac{3}{2}$ when the corresponding pole is at $\tau+2k+1$. When $l=\frac{1}{2}$, the polynomial should be a constant in all cases. It is clear that each component of the Mellin amplitude $\bar{M}_{i}$ factorizes at the poles onto components of the Mellin amplitudes of the corresponding three point functions as described above. \\
\\
There are also poles in the Mellin amplitude in the $u$-channel. The $u$-channel is related to the $s$- and $t$-channel by the relation $u=\sum_{i} \tau_{i}-s-t$. These correspond to operators exchanged in the OPE limit $x_{1}\rightarrow x_{4}$ i.e. operators appearing in the OPEs of both $\psi_{1}\phi_{4}$ and that of $\phi_{3}\psi_{4}$. The location of these poles can be worked out from the preceeding discussion. We have stated the results in Appendix \ref{chettri}.

\subsection{Pole structure: Four fermion correlator}
\label{subsec:4ferm-pole}

The Mellin amplitude ~\eqref{fourpoint2} associated with the correlator $\Braket{\phi_{1}\phi_{2}\phi_{3}\phi_{4}}$ has $s$-channel poles at $s=\tau+2k$ for each operator with twist $\tau$ contributing to the $s$-channel conformal block expansion of the correlator. The residue at the pole is $\lambda_{\phi_{1}\phi_{2}\mathcal{O}_{l}}\lambda_{\mathcal{O}_{l}\phi_{3}\phi_{4}}Q_{l,k}(t)$, $l$ being the spin of the exchanged operator. $Q_{l,k}$ are polynomials in $t$ of degree $l$. One way to explain this analyticity property is in terms of the expansion of the conformal block $G_{\Delta,l}$ around the OPE limit \cite{Dolan:2000ut},
\begin{eqnarray} 
G_{\Delta,l}=u^{\frac{\Delta-l}{2}}\sum_{k=0}^{\infty}u^{k}g_{k}(v)           \label{scblock}
\end{eqnarray}
where $g_{k}(v)$ has a power series expansion in $1-v$. \\
\\
For the correlator $\Braket{\psi_{1}\psi_{2}\psi_{3}\psi_{4}}$, the nature of the conformal blocks depends on the basis of tensor structures. As mentioned earlier in Section \ref{subsec:tens}, a generic basis of tensor structures may lead to the conformal blocks having spurious singularities. We will choose a basis such that each conformal block can be expanded around the OPE limit as follows,
\begin{eqnarray}
\sum_{i=1}^{I}u^{\frac{\tau-a_{i}}{2}}\sum_{k=0}^{\infty}u^{k}\tilde{g}_{k}(v)       \label{fermblock} 
\end{eqnarray} 
Here $I$ is some finite integer greater than zero, $a_{i}<\tau$ are integers and $\tilde{g}_{k}$ has a power series expansion in $1-v$. These would ensure that each component of the Mellin amplitude has finitely many series of poles corresponding to each exchanged primary and the residue at each pole is a product of the relevant structure constants and a polynomial whose degree is determined by the spin $l$. In particular, we do not want the conformal blocks to have spurious singularities. ~\eqref{mt6}, ~\eqref{mt7} is one possible choice of such a basis $\{p_{i}\}$. We state the results for the pole structure here.\\
\\
Corresponding to each integer spin $l$ primary $\mathcal{O}_{l}$ of twist $\tau$ contributing to the direct channel conformal block expansion of the correlator, the Mellin amplitude has poles and residues as summarised in Table \ref{table4}.\\
\\ 
When the exchanged operator is a scalar $l=0$, we should take all structure constants apart from $\lambda_{\psi_{1}\psi_{2}\mathcal{O}_{l}}^{1}$, $\lambda_{\psi_{1}\psi_{2}\mathcal{O}_{l}}^{3}$, $\lambda_{\mathcal{O}_{l}\psi_{3}\psi_{4}}^{1}$ and $\lambda_{\mathcal{O}_{l}\psi_{3}\psi_{4}}^{3}$ to be zero. The poles in the crossed channels can also be worked out. We state the results in Appendix \ref{ronaldinho}. 
\begin{table}[t]
\centering 
\begin{tabular}{|c|c|C{6.2cm}|C{5.2cm}|}
\hline
\textbf{Component of M.A.} & \textbf{Location of Poles} 
  & \textbf{Residues $\sim$}  \\ 
\hline\hline
\multirow{4}{*}{} & \multirow{4}{*}{\vspace*{.2in}$s=\tau +2k$}  
  &\vspace*{.1cm} $\lambda_{\psi_1\psi_2\mathcal{O}_\ell}^1\lambda_{\mathcal{O}_\ell\psi_3\psi_4}^1\;\;,\,\; \lambda_{\psi_1\psi_2\mathcal{O}_\ell}^1\lambda_{\mathcal{O}_\ell\psi_3\psi_4}^2 $  \\ [.2cm]
&  & \vspace*{.1cm} $\lambda_{\psi_1\psi_2\mathcal{O}_\ell}^2\lambda_{\mathcal{O}_\ell\psi_3\psi_4}^1  \;\,,\;\;\lambda_{\psi_1\psi_2\mathcal{O}_\ell}^2\lambda_{\mathcal{O}_\ell\psi_3\psi_4}^2$   \\[.2cm]
 \cline{2-3}$\mathcal{M}_1$
&\multirow{4}{*}{ \vspace*{.2in}$s=\tau+1 +2k$} & \vspace*{.1cm} $\lambda_{\psi_1\psi_2\mathcal{O}_\ell}^3\lambda_{\mathcal{O}_\ell\psi_3\psi_4}^3\;\;,\;\;\lambda_{\psi_1\psi_2\mathcal{O}_\ell}^3\lambda_{\mathcal{O}_\ell\psi_3\psi_4}^4$  \\ [.2cm]
&  & \vspace*{.1cm} $\lambda_{\psi_1\psi_2\mathcal{O}_\ell}^4\lambda_{\mathcal{O}_\ell\psi_3\psi_4}^3\;\;,\;\;\lambda_{\psi_1\psi_2\mathcal{O}_\ell}^4\lambda_{\mathcal{O}_\ell\psi_3\psi_4}^4$   \\ [.2cm]
\hline\hline
\multirow{4}{*}{ \vspace*{.2in}$\mathcal{M}_2 $}
&\multirow{4}{*}{ \vspace*{.2in}$s=\tau +1+2k$} & \vspace*{.1cm} $\lambda_{\psi_1\psi_2\mathcal{O}_\ell}^1\lambda_{\mathcal{O}_\ell\psi_3\psi_4}^2\;\;,\;\;\lambda_{\psi_1\psi_2\mathcal{O}_\ell}^2\lambda_{\mathcal{O}_\ell\psi_3\psi_4}^1$  \\ [.2cm]
&  & \vspace*{.1cm} $\lambda_{\psi_1\psi_2\mathcal{O}_\ell}^2\lambda_{\mathcal{O}_\ell\psi_3\psi_4}^2$   \\ [.2cm]
\hline\hline
\multirow{4}{*}{$\mathcal{M}_3\;\;,\;\;\mathcal{M}_5\;\;,\;\;\mathcal{M}_6\;\;,$} & \multirow{4}{*}{\vspace*{.2in}$s=\tau -1+2k$}  
  &\multirow{4}{*}{\vspace*{.5cm} $\lambda_{\psi_1\psi_2\mathcal{O}_\ell}^2\lambda_{\mathcal{O}_\ell\psi_3\psi_4}^1\;\;,\,\; \lambda_{\psi_1\psi_2\mathcal{O}_\ell}^2\lambda_{\mathcal{O}_\ell\psi_3\psi_4}^2 $}  \\ [.2cm]
&  &   \\[.2cm]
 \cline{2-3}$\mathcal{M}_7\;\;,\;\;\mathcal{M}_8$
&\multirow{4}{*}{ \vspace*{.2in}$s=\tau +2k$} & \vspace*{.1cm} $\lambda_{\psi_1\psi_2\mathcal{O}_\ell}^3\lambda_{\mathcal{O}_\ell\psi_3\psi_4}^3\;\;,\;\;\lambda_{\psi_1\psi_2\mathcal{O}_\ell}^3\lambda_{\mathcal{O}_\ell\psi_3\psi_4}^4$  \\ [.2cm]
&  & \vspace*{.1cm} $\lambda_{\psi_1\psi_2\mathcal{O}_\ell}^4\lambda_{\mathcal{O}_\ell\psi_3\psi_4}^3\;\;,\;\;\lambda_{\psi_1\psi_2\mathcal{O}_\ell}^4\lambda_{\mathcal{O}_\ell\psi_3\psi_4}^4$   \\ [.2cm]
\hline\hline
\multirow{4}{*}{} & \multirow{4}{*}{\vspace*{.2in}$s=\tau +2k$}  
  &\multirow{4}{*}{\vspace*{.5cm} $\lambda_{\psi_1\psi_2\mathcal{O}_\ell}^2\lambda_{\mathcal{O}_\ell\psi_3\psi_4}^1\;\;,\,\; \lambda_{\psi_1\psi_2\mathcal{O}_\ell}^2\lambda_{\mathcal{O}_\ell\psi_3\psi_4}^2 $}  \\ [.2cm]
&  &   \\[.2cm]
 \cline{2-3}$\mathcal{M}_4$
&\multirow{4}{*}{ \vspace*{.2in}$s=\tau+1 +2k$} & \vspace*{.1cm} $\lambda_{\psi_1\psi_2\mathcal{O}_\ell}^3\lambda_{\mathcal{O}_\ell\psi_3\psi_4}^3\;\;,\;\;\lambda_{\psi_1\psi_2\mathcal{O}_\ell}^3\lambda_{\mathcal{O}_\ell\psi_3\psi_4}^4$  \\ [.2cm]
&  & \vspace*{.1cm} $\lambda_{\psi_1\psi_2\mathcal{O}_\ell}^4\lambda_{\mathcal{O}_\ell\psi_3\psi_4}^3\;\;,\;\;\lambda_{\psi_1\psi_2\mathcal{O}_\ell}^4\lambda_{\mathcal{O}_\ell\psi_3\psi_4}^4$   \\ [.2cm]
\hline\hline
\multirow{4}{*}{$\mathcal{M}_9 \;\;,\;\;\mathcal{M}_{10}$}  & \multirow{4}{*}{\vspace*{.2in}$s=\tau +2k$}  
  &\vspace*{.1cm} $\lambda_{\psi_1\psi_2\mathcal{O}_\ell}^1\lambda_{\mathcal{O}_\ell\psi_3\psi_4}^3\;\;,\,\; \lambda_{\psi_1\psi_2\mathcal{O}_\ell}^1\lambda_{\mathcal{O}_\ell\psi_3\psi_4}^4 $  \\ [.2cm]
&  & \vspace*{.1cm} $\lambda_{\psi_1\psi_2\mathcal{O}_\ell}^2\lambda_{\mathcal{O}_\ell\psi_3\psi_4}^3  \;\,,\;\;\lambda_{\psi_1\psi_2\mathcal{O}_\ell}^2\lambda_{\mathcal{O}_\ell\psi_3\psi_4}^4$   \\[.2cm]
 \cline{2-3}
&\multirow{4}{*}{ \vspace*{.6in}$s=\tau+1 +2k$} & \vspace*{.1cm} $\lambda_{\psi_1\psi_2\mathcal{O}_\ell}^3\lambda_{\mathcal{O}_\ell\psi_3\psi_4}^2\;\;,\;\;\lambda_{\psi_1\psi_2\mathcal{O}_\ell}^4\lambda_{\mathcal{O}_\ell\psi_3\psi_4}^2$  \\ [.2cm]
\hline\hline
\multirow{4}{*}{$\mathcal{M}_{11} \;\;,\;\;\mathcal{M}_{12}$}  & \multirow{4}{*}{\vspace*{.2in}$s=\tau +2k$}  
  &\vspace*{.1cm} $\lambda_{\psi_1\psi_2\mathcal{O}_\ell}^3\lambda_{\mathcal{O}_\ell\psi_3\psi_4}^1\;\;,\,\; \lambda_{\psi_1\psi_2\mathcal{O}_\ell}^4\lambda_{\mathcal{O}_\ell\psi_3\psi_4}^1 $  \\ [.2cm]
&  & \vspace*{.1cm} $\lambda_{\psi_1\psi_2\mathcal{O}_\ell}^3\lambda_{\mathcal{O}_\ell\psi_3\psi_4}^2  \;\,,\;\;\lambda_{\psi_1\psi_2\mathcal{O}_\ell}^4\lambda_{\mathcal{O}_\ell\psi_3\psi_4}^2$   \\[.2cm]
 \cline{2-3}
&\multirow{4}{*}{ \vspace*{.6in}$s=\tau+1 +2k$} & \vspace*{.1cm} $\lambda_{\psi_1\psi_2\mathcal{O}_\ell}^2\lambda_{\mathcal{O}_\ell\psi_3\psi_4}^3\;\;,\;\;\lambda_{\psi_1\psi_2\mathcal{O}_\ell}^2\lambda_{\mathcal{O}_\ell\psi_3\psi_4}^4$  \\ [.2cm]
\hline\hline
\multirow{4}{*}{\vspace*{0.2in}$\mathcal{M}_{13} \;\;,\;\;\mathcal{M}_{14}$}  & \multirow{4}{*}{\vspace*{.6in}$s=\tau+1+2k$}  
  &\vspace*{.1cm} $\lambda_{\psi_1\psi_2\mathcal{O}_\ell}^3\lambda_{\mathcal{O}_\ell\psi_3\psi_4}^2\;\;,\,\; \lambda_{\psi_1\psi_2\mathcal{O}_\ell}^4\lambda_{\mathcal{O}_\ell\psi_3\psi_4}^2 $  \\ [.2cm]
 \cline{2-3}
&\multirow{4}{*}{ \vspace*{.6in}$s=\tau+2+2k$} & \vspace*{.1cm} $\lambda_{\psi_1\psi_2\mathcal{O}_\ell}^2\lambda_{\mathcal{O}_\ell\psi_3\psi_4}^3\;\;,\;\;\lambda_{\psi_1\psi_2\mathcal{O}_\ell}^2\lambda_{\mathcal{O}_\ell\psi_3\psi_4}^4$  \\ [.2cm]
\hline\hline
\multirow{4}{*}{\vspace*{0.2in}$\mathcal{M}_{15} \;\;,\;\;\mathcal{M}_{16}$}  & \multirow{4}{*}{\vspace*{.6in}$s=\tau+1+2k$}  
  &\vspace*{.1cm} $\lambda_{\psi_1\psi_2\mathcal{O}_\ell}^2\lambda_{\mathcal{O}_\ell\psi_3\psi_4}^3\;\;,\,\; \lambda_{\psi_1\psi_2\mathcal{O}_\ell}^2\lambda_{\mathcal{O}_\ell\psi_3\psi_4}^4 $  \\ [.2cm]
 \cline{2-3}
&\multirow{4}{*}{ \vspace*{.6in}$s=\tau+2+2k$} & \vspace*{.1cm} $\lambda_{\psi_1\psi_2\mathcal{O}_\ell}^3\lambda_{\mathcal{O}_\ell\psi_3\psi_4}^2\;\;,\;\;\lambda_{\psi_1\psi_2\mathcal{O}_\ell}^4\lambda_{\mathcal{O}_\ell\psi_3\psi_4}^2$  \\ [.2cm]
\hline\hline
\end{tabular}
\caption{Fermion four point function: Direct channel poles.}
\label{table4}
\end{table}

\FloatBarrier


\section{Witten diagrams}
\label{sec:wittendiagrams}

The AdS/CFT correspondence is a conjectured duality between String Theories in $d+1$ dimensional AdS spacetime and CFTs living on its $d$ dimensional boundary. When the bulk spacetime is weakly curved and the bulk theory is well approximated by the supergravity limit, we can use Witten diagrams to compute correlation functions in the dual strongly interacting CFT. These computations are quite cumbersome in position space. In the Mellin representation, they are simplified greatly \cite{Penedones:2010ue, Paulos:2011ie, Nandan:2011wc, Kharel:2013mka} and the corresponding Mellin amplitudes can be concretely related to scattering amplitudes in QFT in $d+1$ dimensions through the so-called ``flat-space limit'' \cite{Penedones:2010ue, Fitzpatrick:2011hu, Goncalves:2014rfa}. Here we shall present a few results for tree-level Witten diagrams with fermionic legs which serve to illustrate some of the general feature discussed in the previous chapter \ref{sec:generalities}. The calculations are simply reduced to calculations of scalar Witten diagrams \cite{Kawano:1999au, Ghezelbash:1998pf}, the results for which are available \cite{Penedones:2010ue, Paulos:2011ie}. Hence we do not need to set up these calculations in embedding space notation. We shall however present results in embedding space notation in order to relate to the tensor structures in Section \ref{subsec:tens}. We have provided a short review of Fermions in the AdS/CFT correspondence in Appendix \ref{app-fermi}. In the diagrams, solid lines with arrows denote fermion propagators and solid lines without arrows denote scalar propagators.

\subsection{Contact Witten diagram}

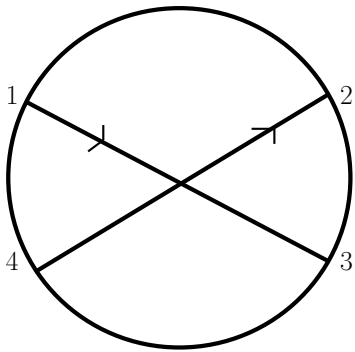
\begin{figure}
\begin{center}
\begin{tikzpicture}[scale=.5,transform shape]
\draw[ultra thick, black] (5.9,-2.2) -- (-2,2);
\draw[ultra thick, black] (-1.8,-2.5) -- (5.9,2.2);
\node at (-2.4,2.2) {\huge $1$};
\node at (-2.4,-2.2) {\huge $4$};
\node at (6.4,-2.2) {\huge $3$};
\node at (6.4,2.2) {\huge $2$};
\draw[thick, black] (0,1) -- (0,1.4);
\draw[thick, black] (-.4,0.7) -- (0,1);
\draw[thick, black] (4.5,1.3) -- (4.5,.9);
\draw[thick, black] (4.5,1.3) -- (3.9,1.3);
\draw [ultra thick] (2,0) circle (4.5cm);
\end{tikzpicture}
\end{center}
\caption{Four point contact Witten diagram with two fermions and two scalars.}
\label{diag:cont_witten}
\end{figure}

First we shall consider the contact Witten diagram involving two fermions and two scalars as shown in figure \ref{diag:cont_witten}. As described in Appendix \ref{app-fermi}, bulk-to-boundary spinor propagators are related to bulk-to-boundary scalar propagators in the following manner,
	\be
		\Sigma_{\Delta}\left(z, \vec{x} \right) &=& 
		\frac{\Gamma_{\mu}\left( z^{\mu} - x^{\mu} \right)}{\sqrt{z_0}} K_{\Delta + \frac{1}{2}}\left(z, \vec{x} \right) \mathcal{P}^-,
            	\non       
            		\\
		\bar{\Sigma}_{\Delta}\left(z, \vec{x} \right) &=& 
		\mathcal{P}^+ \frac{\Gamma_{\mu}\left( z^{\mu} - x^{\mu} \right)}{\sqrt{z_0}} K_{\Delta + \frac{1}{2}} \left(z, \vec{x} \right). 
			 \label{umititi}
	\ee
$\mathcal{P}^{\pm}=(1\pm\Gamma^{0})/2$, $\Gamma^{\mu}$ being gamma matrices of the bulk. Using ~\eqref{umititi} we see that the two bulk-to-boundary propagators for fermions $\bar{\Sigma}_{\Delta_1}$ and $\Sigma_{\Delta_2}$ can be reduced to a product of two scalar bulk-to-boundary propagators $K_{\Delta_1+\frac{1}{2}}$ and $K_{\Delta_2+\frac{1}{2}}$ with an additional tensor structure:
	\be
		\bar\Sigma_{\Delta_{1}}(z,\vec x_1) \Sigma_{\Delta_{2}}(z,\vec x_2)
		= \left(\vec{x}^{\mu}_{12}\Gamma_{\mu}\mathcal{P}^-\right)
		K_{\Delta_{1}+\frac{1}{2}}(z,\vec x_1)K_{\Delta_{2}+\frac{1}{2}}(z,\vec x_2).
		\label{sigmaprod}
	\ee
When contracted with polarization spinors localized on the boundary, $\vec{x}^{\mu}_{12}\Gamma_{\mu}\mathcal{P}^-$ is equivalent to $\vec{x}^{a}_{12}\gamma_{a}\equiv \slashed{x}_{12}$ (contracted with polarization spinors of the boundary) where $\gamma_{a}$ are gamma matrices in the boundary theory. Thus the position space expression for this diagram can be simplified to the evaluation of a scalar Witten diagram as,
	\be
		B^{\bar{\psi}_1 \psi_2}_{\phi_3 \phi_4} =  \Braket{S_{1}S_{2}}\int_{AdS} dZ \prod_{i=1}^{2}K_{\Delta_{i}+\f{1}{2}}(Z,X_{i})\prod_{i=3}^{4}K_{\Delta_{i}}(Z,X_i).						
				\label{intsyn}
	\ee
In the embedding space notation used in ~\eqref{intsyn}, $Z$ is a bulk point and $X_{i}$ are points on the boundary. The integral in \eqref{intsyn} is the expression of a contact Witten diagram of four scalars\footnote{For the scalar Green's functions in AdS, we will stick to the conventions in \cite{Penedones:2010ue, Paulos:2011ie}.}. This can be expressed in Mellin space \cite{Penedones:2010ue} to obtain%
	\footnote{The normalization for the delta function reads $\mdelta{x} = 2\pi i \; \delta(x)$ and the measure is given by $\left( d s_{il} \right) = \frac{d s_{il}}{2\pi i}$.}
	\begin{eqnarray}
		B^{\bar{\psi}_1 \psi_2}_{\phi_3 \phi_4} & = &
		\Braket{S_{1}S_{2}}
		\Mellint{1}{4} X_{il}^{-s_{il}}\Gamma\left(s_{il}\right)		
		\mathbb{M}_{2,2}
		\prod\limits_{i=1}^{4}\mdelta{\Delta_{i}+\f{1}{2}(\delta_{i1}+\delta_{i2})-\sum\limits_{j\not=i}s_{ij}}
						\non
				\\
		&=& \frac{\Braket{S_{1}S_{2}}}{\sqrt{X_{12}}}
		\Mellint{1}{4} X_{il}^{-s_{il}-\frac{1}{2}\delta_{1i}\delta_{2l}}\Gamma\left(s_{il}+\delta_{1i}\delta_{2l}\right)		
		\mathbb{M}_{2,2}		
		\prod\limits_{i=1}^{4}\mdelta{\tau_{i}-\sum\limits_{j\not=i}s_{ij}},
						\label{konte}
	\end{eqnarray}
The only non-zero component of the Mellin amplitude of the contact interaction is $\mathcal{M}_{1}=\mathbb{M}_{2,2}$. In general, the corresponding result for $2n$ fermions and $m$ scalars is given by,
	\be
		\mathcal{M}_{1}&=&\mathbb{M}_{2n,m}=\frac{\pi^h}{2}\Gamma\left(\frac{1}{2}\sum\limits_{i=1}^{2n+m}\Delta_{i}+ \frac{n}{2}-h\right)\prod_{i=1}^{2n}
		\left[\frac{C_{\Delta_{i}+\f{1}{2}}}{\Gamma(\Delta_{i}+\f{1}{2})}
		\right]\prod_{i=2n+1}^{2n+m}
		\left[\frac{C_{\Delta_{i}}}{\Gamma(\Delta_{i})}
		\right],
		      \label{kontemm}
		    \\
		C_{\Delta}&=&\frac{\Gamma(\Delta)}{2\pi^{h}\Gamma(\Delta-h+1)},		\qquad 
		h \equiv \f{d}{2} .             \nonumber     
	\ee
This is a constant independent of the Mellin variables as in the case of scalar contact interaction. From ~\eqref{kontemm}, we also know that the three point function of two fermions and a scalar is parity even.

\subsection{Scalar Exchange Witten Diagram with two external fermions }

	\begin{figure}
		\begin{center}
		\hspace*{-.6in}
			\begin{tikzpicture}[scale=.5,transform shape]
\draw[ultra thick, black] (0,0) -- (-2,2);
\draw[ultra thick, black] (0,0) -- (-2,-2);
\draw[ultra thick, black] (0,0) -- (4,0);
\draw[ultra thick, black] (4,0) -- (6,2);
\draw[ultra thick, black] (4,0) -- (6,-2);
\node at (-2.4,2.2) {\huge $1$};
\node at (-2.4,-2.2) {\huge $2$};
\node at (6.4,-2.2) {\huge $3$};
\node at (6.4,2.2) {\huge $4$};
\draw[thick, black] (-1,1) -- (-1,1.4);
\draw[thick, black] (-1,1) -- (-1.4,1);
\draw[thick, black] (-1,-1) -- (-1,-0.6);
\draw[thick, black] (-1,-1) -- (-0.6,-1);
\draw[ultra thick] (2,0) ellipse (4.5cm and 4.5cm);
			\end{tikzpicture}
\hspace*{1.2in}
			\begin{tikzpicture}[scale=.5,transform shape]
\draw[ultra thick, black] (0,0) -- (-2,2);
\draw[ultra thick, black] (0,0) -- (-2,-2);
\draw[ultra thick, black] (0,0) -- (4,0);
\draw[ultra thick, black] (4,0) -- (6,2);
\draw[ultra thick, black] (4,0) -- (6,-2);
\node at (-2.4,2.3) {\huge $1$};
\node at (-2.4,-2.2) {\huge $2$};
\node at (6.4,-2.2) {\huge $3$};
\node at (6.4,2.2) {\huge $4$};
\draw[thick, black] (-1,1) -- (-1,1.4);
\draw[thick, black] (-1,1) -- (-1.4,1);
\draw[thick, black] (-1,-1) -- (-1,-0.6);
\draw[thick, black] (-1,-1) -- (-0.6,-1);
\draw[thick, black] (5,1) -- (5,1.4);
\draw[thick, black] (5,1) -- (5.4,1);
\draw[thick, black] (5,-1) -- (5,-0.6);
\draw[thick, black] (5,-1) -- (4.6,-1);
\draw[ultra thick] (2,0) ellipse (4.5cm and 4.5cm);
			\end{tikzpicture}
		\end{center}
	\caption{Scalar exchange Witten diagrams with two and four external fermions.}
			\label{diag:scal_witt}
	\end{figure}
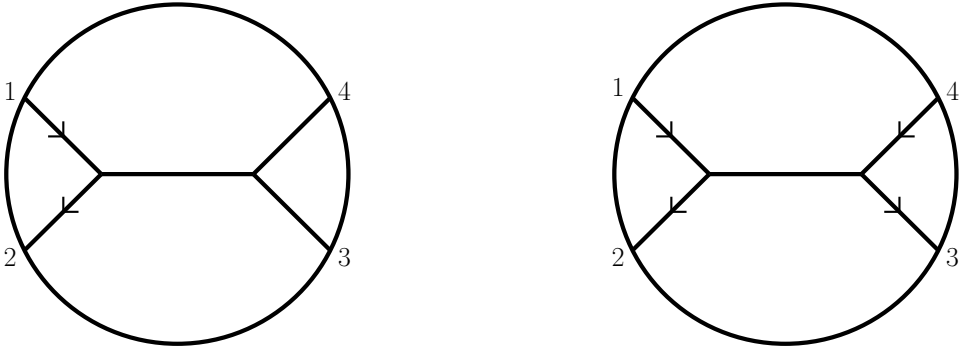

Next we consider the four point scalar exchange Witten diagram with two external fermions as shown in figure \ref{diag:scal_witt}. This expression in position space is
	\be
		\int d^{d+1}z_1 \sqrt{g(z_1)}\int d^{d+1}z_2 \sqrt{g(z_2)}\;\bar\Sigma_{\Delta_{1}}(z_1,\vec x_1) 
		\Sigma_{\Delta_{2}}(z_1,\vec x_2)G_{\Delta}(z_1,z_2)\prod_{i=3}^{4}K_{\Delta_{i}}(z_2,\vec x_i).
				\non
	\ee
Using \eqref{sigmaprod} and switching to embedding space notation, we get,
	\be
		A^{\bar{\psi}_1 \psi_2}_{\phi_3 \phi_4}=
		\Braket{S_{1}S_{2}} \int_{AdS} dZ_1 \int_{AdS} dZ_2
		\prod_{i=1}^{2}K_{\Delta_{i}+\f{1}{2}}(Z_1, X_i)G_{\Delta}(Z_1,Z_2)\prod_{i=3}^{4}K_{\Delta_{i}}(Z_2,X_i).
				\non
	\ee
The integral in the above expression was studied in \cite{Penedones:2010ue}. Using this result\footnote{Please note that in \cite{Penedones:2010ue}, Mellin variables are denoted by $\delta_{ij}$ and Mandelstam variables as $s_{i_{1}\cdots i_{k}}$.}, we obtain
	\begin{eqnarray}
				\non
		A^{\bar{\psi}_1 \psi_2}_{\phi_3 \phi_4} =
		\frac{\Braket{S_{1}S_{2}}}{\sqrt{X_{12}}}
		\Mellint{1}{4} X_{il}^{-s_{il}-\frac{1}{2}\delta_{1i} \delta_{2l}}\Gamma\left(s_{il}+\delta_{1i} \delta_{2l}\right)		
		\mathbb{N}^{\bar{\psi}_1 \psi_2}_{\phi_3 \phi_4}\left(s_{il}\right)
		\prod\limits_{i=1}^{4}\mdelta{\tau_{i}-\sum\limits_{j\not=i}s_{ij}},  \\   \label{mourinho}
	\end{eqnarray}
where in terms of the Mandelstam variable $s=\tau_{1}+\tau_{2}-2s_{12}$, we have
	\be
	\mathcal{M}_{1}=\mathbb{N}^{\bar{\psi}_1 \psi_2}_{\phi_3 \phi_4}(s_{il})&=&
		\f{\mathbb{M}_{2,2}}{\Gamma\left(\f{\sum_i\Delta_i}{2} + \frac{1}{2} - h\right)
		\Gamma\left(\frac{\Delta_1+\Delta_2+1-s}{2}\right)
		\Gamma\left(\frac{\Delta_3+\Delta_4-s}{2} \right)}
				\non
					\\
		& & \times \int_{-i\infty}^{i\infty}\frac{dc}{2\pi i}\f{l(c)l(-c)}{(\Delta-h)^2-c^2},
		\label{Msij}
	\ee
	\be
		l(c) & = &
		\f{\Gamma\left(\f{h+c-s}{2}\right)\Gamma\left(\f{\Delta_1+\Delta_2-h+c}{2}+\f{1}{2}\right)
		\Gamma\left(\f{\Delta_3+\Delta_4-h+c}{2}\right)}{2\Gamma(c)},
				\non
	\ee
where $\Delta$ is the conformal dimension of the exchanged operator.\\
\\
The poles in ~\eqref{Msij} occur when the contour of the integral is pinched between two colliding poles of the integrand. These poles are at\footnote{There are other such poles from the integral but these are cancelled by zeroes in the pre-factor.} $s=\Delta+2m$ which is exactly as predicted for $\mathcal{M}_{1}$ in Section \ref{subsec:2ferm-s}. As shown in \cite{Penedones:2010ue, Paulos:2011ie}, the Mellin amplitude can be in fact written as a series over these poles and the residues follow from a simple shift in the corresponding residue there.

\subsection{Scalar Exchange Witten Diagram with four external fermions}

The diagram in which four external fermions interact via an exchange of a scalar operator is shown in figure \ref{diag:scal_witt}. It can also be manipulated in a similar way as the previous examples. The expression for this is then given by 
	\be
		A^{\bar{\psi}_1 \psi_2}_{\bar{\psi}_4 \psi_3} = -
		\Braket{S_{1}S_{2}}\Braket{S_{3}S_{4}} \hspace{-0.15cm}
		\int_{AdS} \hspace{-0.3cm} dZ_1 
		\int_{AdS} \hspace{-0.3cm} dZ_2 
		\prod_{i=1}^{2}K_{\Delta_{i}+\f{1}{2}} \hspace{-0.1cm}  (Z_1,X_i)
		 G_{\Delta}(Z_1,Z_2)\prod_{i=3}^{4}K_{\Delta_{i}+\f{1}{2}} \hspace{-0.1cm} (Z_2,X_i).
				\label{wenger}
	\ee
The Mellin amplitude can be calculated just like in the previous example and the only non-zero component is $\mathcal{M}_{1}$.
	\be
	\mathcal{M}_{1}=\mathbb{N}^{\bar{\psi}_1 \psi_2}_{\bar{\psi}_4 \psi_3}(s_{il})&=&
		\f{-\mathbb{M}_{4,0}}{\Gamma\left(\f{\sum_i\Delta_i}{2}+1-h\right)
		\Gamma\left(\frac{\Delta_1+\Delta_2-s}{2}+\f{1}{2}\right)
		\Gamma\left(\frac{\Delta_3+\Delta_4-s}{2}+\f{1}{2}\right)}
				\non
					\\
		& & \times \int_{-i\infty}^{i\infty}\frac{dc}{2\pi i}\f{l(c)l(-c)}{(\Delta-h)^2-c^2},     \label{fergusson}
	\ee
	\be
		l(c) & = &
		\f{\Gamma\left(\f{h+c-s}{2}\right)\Gamma\left(\f{\Delta_1+\Delta_2-h+c}{2}+\f{1}{2}\right)
		\Gamma\left(\f{\Delta_3+\Delta_4-h+c}{2} +\f{1}{2}\right)}{2\Gamma(c)}.
				\non 
	\ee
The poles of $\mathcal{M}_{1}$ are at $s=\Delta+2m$. In Section \ref{subsec:4ferm-pole} predict another series of poles at $s=\Delta+1+2m$. One can explain the absence of this second series simply by looking at the relevant three point functions. From ~\eqref{kontemm}, we know that the three point function here is of positive definite parity and hence the second series of poles is absent.

\subsection{Spinor Exchange Witten Diagrams }

\begin{figure}
\begin{center}
\begin{tikzpicture}[scale=.5,transform shape]
\draw[ultra thick, black] (0,0) -- (-2,2);
\draw[ultra thick, black] (0,0) -- (-2,-2);
\draw[ultra thick, black] (0,0) -- (4,0);
\draw[ultra thick, black] (4,0) -- (6,2);
\draw[ultra thick, black] (4,0) -- (6,-2);
\node at (-2.4,2.2) {\huge $1$};
\node at (-2.4,-2.2) {\huge $3$};
\node at (6.4,-2.2) {\huge $4$};
\node at (6.4,2.2) {\huge $2$};
\draw[thick, black] (-1,1) -- (-1,1.4);
\draw[thick, black] (-1,1) -- (-1.4,1);
\draw[thick, black] (5,1) -- (5,.6);
\draw[thick, black] (4.6,1) -- (5,1);
\draw[thick, black] (2,.3) -- (2.4,0);
\draw[thick, black] (2,-.3) -- (2.4,0);
\draw[ultra thick] (2,0) ellipse (4.5cm and 4.5cm);
\end{tikzpicture}
\end{center}
\caption{ Spinor exchange Witten diagram}
\label{diag:spin_witt}
\end{figure}
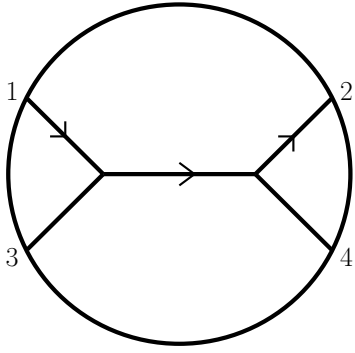 


Next, we present the Mellin amplitude for the spinor exchange diagram \ref{diag:spin_witt}.%
	\footnote{Note the non-standard labeling of the external legs.}
 \cite{Kawano:1999au} has shown that the calculation of the spinor exchange diagram can effectively be reduced to the calculation of a scalar exchange diagram \cite{Penedones:2010ue}. This calculation is presented in details in Appendix \ref{sec:SpExAdS}. Here we quote the final result,
	\be
					\nonumber
		A^{\bar{\psi}_1 \phi_3}_{\psi_2 \phi_4}
		& = & \frac{\Braket{S_{1}S_{2}}}{\sqrt{X_{12}}}
		\Mellint{1}{4} X_{il}^{-s_{il}-\frac{1}{2}\delta_{1i}\delta_{2l}}\Gamma\left(s_{il}+\delta_{1i}\delta_{2l}\right) 
		\left( \Delta_{1}+\Delta_{3}+\Delta + 1 -d -2s_{13} \right)
				 \\
				 	\nonumber
		& & \times \mathbb{N}^{\bar{\psi}_1 \phi_3}_{\psi_2 \phi_4} \left(s_{il}\right) 
		\prod\limits_{i=1}^{4}\mdelta{\tau_{i}-\sum\limits_{j\not=i}s_{ij}}
		 		\\
 					\nonumber
		&& + 2 \frac{\Braket{S_{1}X_{3}X_{4}S_{2}}}{\sqrt{X_{13}X_{34}X_{42}}} 
 		\Mellint{1}{4} X_{il}^{-s_{il}-\frac{1}{2}\delta_{1i}\delta_{2l}}
 		\Gamma\left(s_{il} + \frac{1}{2}\left(\delta_{i1}\delta_{l2} + \delta_{i1}\delta_{l3} + \delta_{i3}\delta_{l4} + \delta_{i2}\delta_{l4} \right)\right) 
 				\\
 		& & \times \bar{\mathbb{N}}^{\bar{\psi}_1 \phi_3}_{\psi_2 \phi_4} \left(s_{il}\right) 
 		\prod\limits_{i=1}^{4}\mdelta{\tau_{i}-\sum\limits_{j\not=i}s_{ij}},          \label{klinnsmann}
	\ee
The Mellin amplitude has two non-zero components $\mathcal{M}_{1}$ and $\mathcal{M}_{2}$. In terms of the Mandelstam variable $t=\tau_{1}+\tau_{3}-2s_{13}$, $\mathcal{M}_{1}$ is given by,

	\be
		\mathcal{M}_{1}&=& \left( t+\tau + 2 -d \right)\mathbb{N}^{\bar{\psi}_1 \phi_3}_{\psi_2 \phi_4} \left(s_{il}\right)  \non \\
		&=& \f{\left( t+\tau + 2 -d \right)\mathbb{M}_{2,2}}{\Gamma\left(\f{\sum_i\Delta_i}{2} + \frac{1}{2} - h \right)
		\Gamma\left(\frac{\tau_1 +\tau_3-t}{2} \right) \Gamma\left(\frac{\tau_2 +\tau_4-t}{2}\right)}
		 \int_{-i\infty}^{i\infty}\frac{dc}{2\pi i}\f{l(c)l(-c)}{(\tau+1-h)^2-c^2},      \label{jurgen1}  \\
	l(c) & = &
		\f{\Gamma\left(\f{h+c-t-1}{2}\right)\Gamma\left(\f{\tau_1+\tau_3-h+c+1}{2}\right)
		\Gamma\left(\f{\tau_2+\tau_4-h+c+1}{2} \right)}{2\Gamma(c)}.
				\non 
	\ee
Thus $\mathcal{M}_{1}$ has poles at $t=\tau+2m$, $\tau$ being the twist of the exchanged spinor. Considering that the relevant three point function is parity even, these poles match with our predictions in Section \ref{subsec:2ferm-t}. $\mathcal{M}_{2}$ is given by,
	\be
		\mathcal{M}_{2}&=& 2 \; \bar{\mathbb{N}}^{\bar{\psi}_1 \phi_3}_{\psi_2 \phi_4} \left(s_{il}\right)  \non \\
		&=& \f{2 \; \mathbb{M}_{2,2}}{\Gamma\left(\f{\sum_i\Delta_i}{2} + \frac{1}{2} - h \right)
		\Gamma\left(\frac{\tau_1 +\tau_3-t+1}{2} \right) \Gamma\left(\frac{\tau_2 +\tau_4-t+1}{2}\right)}
		 \int_{-i\infty}^{i\infty}\frac{dc}{2\pi i}\f{l(c)l(-c)}{(\tau+1-h)^2-c^2},      \label{jurgen2}  \\
	l(c) & = &
		\f{\Gamma\left(\f{h+c-t}{2}\right)\Gamma\left(\f{\tau_1+\tau_3-h+c+1}{2}\right)
		\Gamma\left(\f{\tau_2+\tau_4-h+c+1}{2} \right)}{2\Gamma(c)}.
				\non 
	\ee
$\mathcal{M}_{2}$ has poles at $t=\tau+1+2m$ which is the series predicted in Section \ref{subsec:2ferm-t} for $\mathcal{M}_{2}$ when the corresponding three point functions are parity even.

\section{Conformal Feynman integrals}
\label{sec:perturbative}

Like Witten diagrams, conformal Feynman integrals take very simple forms in Mellin space \cite{Paulos:2012nu, Nandan:2013ip, Nizami:2016jgt}. In \cite{Nizami:2016jgt} Mellin space Feynman rules for tree level interactions in the weak coupling regime were derived for scalar operators. The diagrammatic rules in Mellin space showed that assuming an interaction without derivatives, the Mellin amplitude associated with a tree level diagram is given by a product of beta functions, each of which is associated with an internal propagator. Each vertex yields the trivial contribution $1$. The beta function propagator is a function of the Mandelstam variables composed of the fictitious Mellin momenta and have the right kind of poles as expected from the Mellin amplitude.\\
\\
In this section we will extend these calculations to Mellin amplitudes associated with tree level interactions with two or four external fermions. We shall assume a Yukawa-like interaction without derivatives. These calculations can be simply done in physical space without the need for embedding space notation. However, we shall present the final result in embedding space notation in order the comparison with the tensor structures in Section \ref{subsec:tens} becomes more transparent. We can also assume that  our tree level calculations are done in Euclidean signature so that we do not have to worry about the $i\epsilon$ for the convergence of our integrals, and the final result can be Wick rotated with the correct $i\epsilon$ prescription (implicitly) to  Minkowski signature.\\
\\	
The Mellin amplitudes for the conformal Feynman integrals with one or more internal propagators are computed using a recursive method\footnote{This method was developed by Arnab Rudra for scalar conformal integrals while working on \cite{Nizami:2016jgt}} that we describe in details in Appendix \ref{sec:RecMethScal}. We shall stick to four point calculations in this section to keep the notation simple even though these calculations can easily be extended to include diagrams with any number of scalar legs with more than one scalar or fermion propagator. \\
\\
In our Feynman diagrams, solid lines with arrows will denote fermion propagators and solid lines without arrows will denote scalar propagators.

\subsection{Fermion-scalar four point function: Contact diagram }

\begin{figure}[h]
		\begin{center}
			\begin{tikzpicture}[scale=.8,transform shape]
		\draw[ultra thick, black] (0,0) -- (0,2);
		\draw[ultra thick, black] (0,0) -- (0,-2);
		\draw[ultra thick, black] (0,0) -- (2,0);
		\draw[ultra thick, black] (0,0) -- (-2,0);
		\node at (0,2.5) {$4$};
		\node at (0,-2.5) {$2$};
		\node at (2.5,0) {$3$};
		\node at (-2.5,0) {$1$};
		\draw[thick, black] (-1.2,0.2) -- (-1,0);
		\draw[thick, black] (-1.2,-0.2) -- (-1,0);
		\draw[thick, black] (-0.2,-0.8) -- (0,-1);
		\draw[thick, black] (0.2,-0.8) -- (0,-1);
			\end{tikzpicture}
		\hspace{1cm}
			\begin{tikzpicture}[scale=.8,transform shape]
		\draw[ultra thick, black] (0,0) -- (0,2);
		\draw[ultra thick, black] (0,0) -- (0,-2);
		\draw[ultra thick, black] (0,0) -- (2,0);
		\draw[ultra thick, black] (0,0) -- (-2,0);
		\node at (0,2.5) {$4$};
		\node at (0,-2.5) {$2$};
		\node at (2.5,0) {$3$};
		\node at (-2.5,0) {$1$};
		\draw[thick, black] (-1.2,0.2) -- (-1,0);
		\draw[thick, black] (-1.2,-0.2) -- (-1,0);
		\draw[thick, black] (-0.2,-0.8) -- (0,-1);
		\draw[thick, black] (0.2,-0.8) -- (0,-1);
		\draw[thick, black] (-0.2,1.2) -- (0,1);
		\draw[thick, black] (0.2,1.2) -- (0,1);
		\draw[thick, black] (0.8,0.2) -- (1,0);
		\draw[thick, black] (0.8,-0.2) -- (1,0);
			\end{tikzpicture}
				\end{center}
		\caption{Contact diagrams with two and four fermions.}
		\label{diag:2f2s}
	\end{figure}
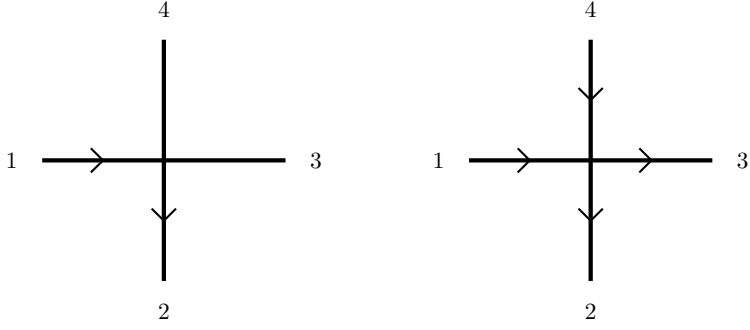

To apply the recursive method for diagrams with fermionic legs, the results for the Mellin amplitude associated to the corresponding contact interaction diagrams have to be known. In this section, we shall consider the contact interaction with two fermions represented by diagram \ref{diag:2f2s}.
This calculation was presented by Symanzik \cite{Symanzik:1972wj} and we shall state the result here. \\
\\
The conformal integral for the contact interaction of two fermions and two scalars is given by%
	\footnote{The measure reads $\mathcal{D}u = \frac{1}{2} \frac{d^d u}{\pi^{d/2}}$ and the delta functions are appropriately normalized $\mdelta{x} = 2\pi i \; \delta(x)$. Further the integration measure of the Mellin variables is given by $\left( d s_{il} \right) = \frac{ds_{il}}{2\pi i}$. }
	\be
				\nonumber
		C^{\bar{\psi}_1 \psi_2}_{\phi_3 \phi_4} 
		& = & \Cooint{u} \Singferprop{x}{1}{u}{}{\Delta_1} \Singferprop{u}{}{x}{2}{\Delta_2} \Singprop{x_1}{-u}{\Delta_3} \Singprop{x_4}{-u}{\Delta_4}
			\ee
The sum of all the scaling dimensions should equal to the spacetime dimension $d$. In embedding space notation, and in conformity with the definition ~\eqref{directdef}, the Mellin representation of this conformal integral is given by,
\be
\sum_{j=3}^{4}\frac{\Braket{S_{1}X_{j}S_{2}}}{\sqrt{X_{1j}X_{j2}}}\Mellint{1}{4}X_{il}^{-s_{il}-\frac{1}{2}\delta_{i1}\delta_{l2}}\Gamma\left(s_{il}+\frac{1}{2}\left(\delta_{i1}+\delta_{i2}\right)\delta_{jl}+\frac{1}{2}\delta_{i1}\delta_{l2}\right) \prod_{i}\hat{\delta}\left(\tau_{i}-\sum_{j\neq i}s_{ij}\right)     \nonumber \\        \label{contactu}
\ee
\textbf{Result:} The Mellin amplitude has two non-zero components, $\mathcal{M}_{3}$ and $\mathcal{M}_{4}$ both of which are proportional to $1$.\\
\\
It is easy to generalize this result by adding more (or less) scalar legs. In particular, this tells us that the three point function of two spin one-half fermions and one scalar is parity odd in this case.

\subsection{Fermion four point function: Contact interaction}
\label{4con}

The Mellin amplitude associated to the contact diagram \ref{diag:2f2s} with four fermionic legs is necessary for computing Mellin amplitudes associated with conformal integrals with four fermionic legs. The corresponding conformal integral is
	\be
				\nonumber
		C^{\bar{\psi}_1 \psi_2}_{\bar{\psi}_4 \psi_3} & = & 
		\Cooint{u} \prod_{i=1}^4 \Ga{\Delta_i +\frac{1}{2}} \spinpr{\Singferpro{x}{1}{u}{}{\Delta_1} \Singferpro{u}{}{x}{2}{\Delta_2}}
		\spinpr{\Singferpro{x}{4}{u}{}{\Delta_4} \Singferpro{u}{}{x}{3}{\Delta_3}}
			\label{cont4}
	\ee
This calculation was also presented in \cite{Symanzik:1972wj}. The Mellin representation of this integral, in conformity with the definition ~\eqref{4fermdef} is given in embedding space notation%
\footnote{\be
							\Braket{S_{1}\Gamma^{A}S_{2}}\Braket{S_{3}\Gamma_{A}S_{4}}  \xrightarrow[\text{physical 							space}]\; \left[\slashed{x}_{1}\gamma^{\mu}+\gamma^{\mu}\slashed{x}_{2}\right]
							\left[\slashed{x}_{3}\gamma_{\mu}+\gamma_{\mu}\slashed{x}_{4}\right]-2\left[\slashed{x}_{1}							\slashed{x}_{2}\right][\mathbb{I}]-2[\mathbb{I}]\left[\slashed{x}_{3}\slashed{x}_{4}\right] 
									\non
							\ee} by the following,
\be
&& \Mellint{1}{4}X_{il}^{-s_{il}-\frac{1}{2}\delta_{i1}\delta_{l2}-\frac{1}{2}\delta_{i3}\delta_{l4}}\left[\frac{1}{2}\frac{\Braket{S_{1}\Gamma^{A}S_{2}}\Braket{S_{3}\Gamma_{A}S_{4}}}{\sqrt{X_{12}X_{34}}}\Gamma\left(s_{il}+\delta_{i1}\delta_{l2}+\delta_{i3}\delta_{l4}\right) + \right.     \nonumber    \\
&& \left. \sum_{j=3}^{4}\sum_{k=1}^{2}\frac{\Braket{S_{1}X_{j}S_{2}}\Braket{S_{3}X_{k}S_{4}}}{\sqrt{X_{1j}X_{j2}X_{3k}X_{k4}}}\Gamma\left(s_{il}+\frac{1}{2}\left(\delta_{i1}+\delta_{i2}\right)\delta_{lj}+\frac{1}{2}\left(\delta_{3l}+\delta_{4l}\right)\delta_{ik}+\frac{1}{2}\delta_{i1}\delta_{l2}+\frac{1}{2}\delta_{i3}\delta_{l4}\right)\right]    \nonumber  \\   
&& \prod_{i}\hat{\delta}\left(\tau_{i}-\sum_{j\neq i}s_{ij}\right)          \label{cont4u}
\ee
\textbf{Result:} In our chosen basis of tensor structures ~\eqref{mt6}, the non-zero components of the Mellin amplitude are the following,
\be
&& \mathcal{M}_{1}=\frac{1}{4} \;\;\;\;\;\;\;\; \mathcal{M}_{3}=1 \;\;\;\;\;\;\;\; \mathcal{M}_{4}=\frac{1}{4} \;\;\;\;\;\;\;\; \mathcal{M}_{5}=s_{13}-1 \non \\
&& \mathcal{M}_{6}=s_{23} \;\;\;\;\;\;\;\; \mathcal{M}_{7}=s_{14} \;\;\;\;\;\;\;\; \mathcal{M}_{8}=s_{24}      \label{suarez}
\ee
We expanded the first tensor structure in ~\eqref{cont4u} in our basis ~\eqref{mt6} as follows, 
\be
\frac{\Braket{S_{1}\Gamma^{A}S_{2}}\Braket{S_{3}\Gamma_{A}S_{4}}}{\sqrt{X_{12}X_{34}}}=\frac{1}{2}p_{1}+2\sqrt{\frac{v}{u}}p_{3}+\frac{1}{2}p_{4}-2\sqrt{\frac{v}{u}}p_{5}             \label{rexpress}
\ee

\subsection{Fermion-scalar four point function: Scalar exchange. }

\begin{figure}[h]
		\begin{center}
			\begin{tikzpicture}[scale=.8,transform shape]
		\draw[ultra thick, black] (0,0) -- (-2,2);
		\draw[ultra thick, black] (0,0) -- (-2,-2);
		\draw[ultra thick, black] (0,0) -- (4,0);
		\draw[ultra thick, black] (4,0) -- (6,2);
		\draw[ultra thick, black] (4,0) -- (6,-2);
		\node at (-2.2,2.2) {$1$};
		\node at (-2.2,-2.2) {$2$};
		\node at (6.2,-2.2) {$3$};
		\node at (6.2,2.2) {$4$};
		\draw[thick, black] (-1,1) -- (-1,1.2);
		\draw[thick, black] (-1,1) -- (-1.2,1);
		\draw[thick, black] (-1,-1) -- (-1,-0.8);
		\draw[thick, black] (-1,-1) -- (-0.8,-1);
			\end{tikzpicture}
		\hspace{1cm}
			\begin{tikzpicture}[scale=.8,transform shape]
		\draw[ultra thick, black] (0,0) -- (3,0);
		\draw[ultra thick, black] (0,0) -- (3,3);
		\draw[ultra thick, black] (3,0) -- (0,3);
		\draw[ultra thick, black] (0,0) -- (-1,-1.5);
		\draw[ultra thick, black] (3,0) -- (4,-1.5);
		\node at (-0.2,3.2) {$1$};
		\node at (-1.2,-1.7) {$2$};
		\node at (4.2,-1.7) {$3$};
		\node at (3.2,3.2) {$4$};
		\draw[thick, black] (2,1) -- (2,1.2);
		\draw[thick, black] (2,1) -- (1.8,1);
		\draw[thick, black] (1.5,0) -- (1.7,0.2);
		\draw[thick, black] (1.5,0) -- (1.7,-0.2);
		\draw[thick, black] (-0.7,-1) -- (-0.7,-0.75);
		\draw[thick, black] (-0.7,-1) -- (-0.4,-0.95);
			\end{tikzpicture}
		\end{center}
			\caption{Fermion scalar four point diagrams with scalar and fermionic exchange.}
		\label{diag:2fschan}
	\end{figure}
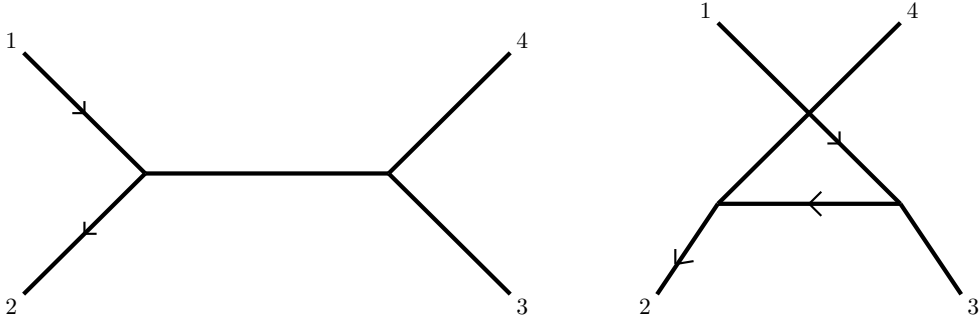

Now we wish to calculate the Mellin amplitude corresponding to the scalar propagator in a diagram \ref{diag:2fschan} with two external fermions and two scalars.
The conformal integral represented by the diagram \ref{diag:2fschan} is given by,
\be
I^{\bar{\psi}_1 \psi_2}_{\phi_3 \phi_4}
	& = & 
	\Cooint{u_1} \Cooint{u_2} \Singferprop{x}{1}{u}{1}{\Delta_1} \Singferprop{u}{1}{x}{2}{\Delta_2} 
			\nonumber  \\
	& & \times
	\prod_{i=3}^4 \Singprop{x_i}{-u_2}{\Delta_i} \frac{1}{\abs{u_1-u_2}^{2\gamma}}          \label{confPSc}
\ee
Conformality of the integral requires $\Delta_{1}+\Delta_{2}=\Delta_{3}+\Delta_{4}=d-\gamma$. The computation of the associated Mellin amplitude using the recursive method follows exactly as described for the scalar case in \ref{sec:RecMethScal}. We state the result here,
\be
	\sum_{j=3}^{4}\frac{\Braket{S_{1}X_{j}S_{2}}}{\sqrt{X_{1j}X_{j2}}} 
	&& \Mellint{1}{4}X_{il}^{-s_{il}-\frac{1}{2}\delta_{i1}\delta_{l2}}
	\Gamma\left(s_{il}+\frac{1}{2}\left(\delta_{1i}+\delta_{2i}\right)\delta_{lj}+\frac{1}{2}\delta_{i1}\delta_{l2}\right)      
		\nonumber 
			\\
      && \frac{1}{2\Gamma\left(\gamma\right)}
      B\left(\frac{\gamma-s}{2},\frac{d}{2}-\gamma\right)\prod_{i}\hat{\delta}\left(\tau_{i}-\sum_{j\neq i}s_{ij}\right)          
      	\label{scpropu}
\ee
\textbf{Result:} The non-zero components of the Mellin amplitude are,
\be
\mathcal{M}_{3}=\mathcal{M}_{4}=\frac{1}{2\Gamma\left(\gamma\right)}B\left(\frac{\gamma-s}{2},\frac{d}{2}-\gamma\right)       \label{dybala}
\ee
The poles are located at $-(p_{1}+p_{2})^{2}=s=\gamma+2m$ which matches our predictions in Section \ref{subsec:2ferm-s}. This result can be easily generalized to one where we have more scalars at either of the two interaction vertices.

\subsection{Fermion-scalar four point function: Fermion exchange.} 

We wish to calculate the Mellin amplitude associated to the diagram \ref{diag:2fschan} with a propagating spin-half fermion. The relevant conformal integral is,
\be
\nonumber
	I^{\bar{\psi}_1 \phi_3}_{\psi_2 \phi_4}
	& = & \Cooint{u_1} \Cooint{u_2} \Singferprop{x}{1}{u}{1}{\Delta_1} \Singferpro{u}{1}{u}{2}{\gamma} \Singferprop{u}{2}{x}{2}{\Delta_2}
			\\
				& & \times \Singprop{x_3}{-u_1}{\Delta_3} \Singprop{x_4}{-u_2}{\Delta_4}         
						\label{crossiprop}
\ee 
We state the result for the associated Mellin amplitude. 
	\be
		&& \Mellint{1}{4}X_{il}^{-s_{il}-\frac{1}{2}\delta_{1i}\delta_{2l}}\prod_{i}\hat{\delta}\left(\tau_{i}-\sum_{j\neq i}s_{ij}\right)   
			\nonumber 
			\\
		&& \left[\frac{\Braket{S_{1}X_{3}X_{4}S_{2}}}{\sqrt{X_{13}X_{34}X_{42}}}
		\frac{1}{2\Gamma(\tau+1)}B\left(\frac{\tau-t}{2},\frac{d}{2}-\tau\right)
		\prod_{1\leq i<l}
		\Gamma\left(s_{il}+\frac{1}{2}\left(\delta_{i1} \delta_{2l} + \delta_{i1}\delta_{l3}+\delta_{i3}\delta_{l4}+\delta_{i2}\delta_{l4}\right)\right)   \right.  
			\nonumber 
			\\
		&& \left. -\frac{\Braket{S_{1}S_{2}}}{\sqrt{X_{12}}}
		\frac{s_{13}}{2\Gamma(\tau+1)}
		B\left(\frac{\tau+1-t}{2},\frac{d}{2}-\tau\right)\prod_{1\leq i<l}
		\Gamma\left(s_{il}+\delta_{i1}\delta_{l2}\right)\right]          
					\label{croiproMe}
\ee
\textbf{Result:} The two non-zero components of the Mellin amplitude are,
\be
\mathcal{M}_{1}&=& -\frac{\tau_{1}+\tau_{3}-t}{4\Gamma(\tau+1)}B\left(\frac{\tau+1-t}{2},\frac{d}{2}-\tau\right)          \label{mbappe}   \\
\mathcal{M}_{2}&=& \frac{1}{2 \Gamma(\tau+1)}B\left(\frac{\tau-t}{2},\frac{d}{2}-\tau\right)                       \label{griezmann}
\ee
$\mathcal{M}_{1}$ has poles at $- \left(p_1 + p_3\right)^2=t=\tau+1+2m$ while $\mathcal{M}_{2}$ has poles at $t=\tau+2m$, where $\tau=\gamma-\frac{1}{2}$ is the twist of the propagating operator. From Section \ref{subsec:2ferm-t}, we see that these are the ``parity odd'' series of poles for $\mathcal{M}_{1}$ and $\mathcal{M}_{2}$ respectively. And indeed this is what we expect when the three point function of two spinors and a scalar is parity odd.

\subsection{Fermion four point function: Scalar propagator}

We wish to calculate the Mellin amplitude for a scalar exchange Feynman diagram ($s$-channel) with four external fermions. The position space integral is,
	\be
				\nonumber
		&&\Cooint{u_1} \Cooint{u_2} \spinpr{\Singferpro{x}{1}{u}{1}{\Delta_1} \Singferpro{u}{1}{x}{2}{\Delta_2}} 					\Singpro{u_1} {-u_2}{\gamma} \non\\
		&&\times \spinpr{\Singferpro{x}{4}{u}{2}{\Delta_4} \Singferpro{u}{2}{x}{3}{\Delta_3}} 					\prod_{i=1}^4 \Ga{\Delta_i + \frac{1}{2}}            \label{debruyne}
	\ee 
The Mellin representation of this integral is given by,
	\be
				\nonumber
		I^{\bar{\psi}_1 \psi_2}_{\bar{\psi}_4 \psi_3} & =& \Mellint{1}{4} X_{il}^{-s_{il}-\frac{1}{2}\delta_{i1}\delta_{j2}-\frac{1}{2}	\delta_{i3}\delta_{j4}}\prod_{i}\hat{\delta}\left(\tau_{i}-\sum_{j\neq i}s_{ij}\right)          \non\\ 
		&& \left[\frac{1}{2}\frac{\Braket{S_{1}\Gamma^{A}S_{2}}\Braket{S_{3}\Gamma_{A}S_{4}}}{\sqrt{X_{12}X_{34}}}\frac{1}{2\Gamma(\gamma)}B\left(\frac{\gamma - s +1}{2},\frac{d}{2}-\gamma\right)\prod_{i<l}\Gamma\left(s_{il} + \delta_{1i}\delta_{2j} + \delta_{3i}\delta_{4j} \right)      \right.          \non\\
                && + \sum_{j=3}^{4}\sum_{k=1}^{2}\frac{\Braket{S_{1}X_{j}S_{2}}\Braket{S_{3}X_{k}S_{4}}}{\sqrt{X_{1j}X_{j2}X_{3k}X_{k4}}}\frac{1}{2\Gamma(\gamma)}B\left(\frac{\gamma - s}{2},\frac{d}{2}-\gamma\right)    \non\\
		&& \left. \prod_{i<l}\Gamma\left(s_{il} + \frac{1}{2}\left(\delta_{1i}+\delta_{2i}\right)\delta_{jl}+ \frac{1}{2}\delta_{ik}\left(\delta_{3l}+\delta_{4l}\right)+ \frac{1}{2}\left(\delta_{1i}\delta_{2j} + \delta_{3i}\delta_{4j}\right) \right) \right]       \label{aguero}
	\ee 
As in Section \ref{4con}, we have to expand the integral above in our chosen basis of tensor structures ~\eqref{mt6} using ~\eqref{rexpress}. \\
\\
\textbf{Result:} The Mellin amplitude has the following non-zero components.
\be
\mathcal{M}_{1}&=&\mathcal{M}_{4}=\frac{1}{8\Gamma(\gamma)}B\left(\frac{\gamma+1-s}{2},\frac{d}{2}-\gamma\right)           \non\\
\mathcal{M}_{3}&=& \frac{1}{2\Gamma(\gamma)}B\left(\frac{\gamma - s}{2},\frac{d}{2}-\gamma\right)         \non\\
\mathcal{M}_{5}&=& \frac{s_{13}-1}{2\Gamma(\gamma)}B\left(\frac{\gamma - s}{2},\frac{d}{2}-\gamma\right)   \non\\
\mathcal{M}_{6}&=& \frac{s_{23}}{2\Gamma(\gamma)}B\left(\frac{\gamma - s}{2},\frac{d}{2}-\gamma\right)   \non\\
\mathcal{M}_{7}&=& \frac{s_{14}}{2\Gamma(\gamma)}B\left(\frac{\gamma - s}{2},\frac{d}{2}-\gamma\right)   \non\\
\mathcal{M}_{8}&=& \frac{s_{24}}{2\Gamma(\gamma)}B\left(\frac{\gamma - s}{2},\frac{d}{2}-\gamma\right)   \label{asensio}
\ee
The poles of all the non-zero $\mathcal{M}_{i}$ above are exactly as predicted in Section \ref{subsec:4ferm-pole}. Since the three point functions here are of odd parity, we only see the ``parity odd'' series of poles.

\section{Discussion}
\label{sec:disco}

In this paper, we have introduced Mellin amplitudes associated with correlators of spin half fermions and scalars. Such Mellin amplitudes have multiple components, each component being associated with an element of a chosen basis of tensor structures.  We have explained that not all choices of bases are suitable for defining a Mellin amplitude with the desired analyticity properties because in certain bases, the associated conformal blocks have spurious singularities. We have examined the pole structure of the Mellin amplitudes defined in suitable bases which also makes clear how the components of the Mellin amplitude factorize onto components of three point Mellin amplitudes (which are just the structure constants). \\
\\
Given a scalar Mellin amplitude, one can read off the twist of the primary exchanged from the leading pole of a given series. One can then check the degree of the polynomial in the residue at this pole which gives the spin of the operator. Thus one knows the dimension of the primary. With knowledge about the explicit form of the residue polynomials one can also read off the OPE coefficients from the residues. For the fermionic Mellin amplitude there is an additional feature as generically there are two distinct poles for each primary. If the theory at hand has three point functions of definite parity, then one of the series of poles will be absent. If the three point functions are not of definite parity, then the leading and sub-leading poles in a series give information on two different OPE coefficients. We must also emphasize that the pole structure of the Mellin amplitude is intimately connected to the chosen basis of tensor structures. One can see multiple series of poles in a given component of the Mellin amplitude because the corresponding component of the correlator receives contributions from multiple conformal partial waves. In particular, it may be possible to choose bases for the four point function and the three point functions such that at least in one given channel, the tensor structures and the conformal blocks align perfectly such that each component of the Mellin amplitude has a single series of poles in this given channel. We leave a careful analysis of this direction to future work.\\
\\
After these general considerations, we have computed a few tree-level Witten diagrams with fermionic legs in Mellin space. These computations are easily reduced to the computation of tree-level Witten diagrams with all scalar legs. The corresponding Mellin amplitudes obtained nicely illustrate the general principles described, in particular one of the two series of poles. Finally we have considered position space conformal integrals corresponding to tree level Feynman diagrams and computed the associated Mellin amplitudes. These Mellin amplitudes demonstrate the existence of the other of the two series of poles as compared to the Witten diagrams. \\
\\
Let us now discuss future directions. In this project, we have not, for example, undertaken a detailed study of the residues at the poles, in particular the polynomials appearing there. It is necessary to do a more comprehensive analysis of the factorization properties of these Mellin amplitudes. Thereafter it would be natural to employ the Mellin bootstrap \cite{Gopakumar:2016cpb} on fermionic CFTs for example the Gross-Neveu model in three dimensions or the Gross-Neveu-Yukawa theory in $4-\epsilon$ dimensions. We hope to report on this in the future. It will be interesting to undertake an independent study of these Witten diagrams and write down the associated Feynman rules \cite{Paulos:2011ie} and also see if one can study the pole structure of the loop level Witten diagrams. An interesting topic to be addressed is the ``flat-space limit'' \cite{Fitzpatrick:2011hu, Goncalves:2014rfa}. \\
\\
Last but not the least, we must emphasize that our definition of the Mellin amplitude associated with correlators of fermionic operators cannot be claimed to be a canonical choice. We would be interested to know if there exists such a canonical definition and if it can be generalized in a natural way to incorporate operators of any spin in general dimension.

\acknowledgments

We are thankful to K. Jaswin and A. Rudra for collaboration during the initial stages of this work. We would like to thank R. Gopakumar for useful discussions; D. Karateev, P. Kravchuk and D. Simmons-Duffin for answering our questions relating to tensor structures and conformal blocks for spinning conformal correlators and T. Kawano for answering questions relating to spinors in AdS. JF would like to thank J. Plefka for useful discussions. SS is thankful to A. De, B. Eden, S. Ghosh, A. A. Nizami, J. Penedones, D. Ponomarev, V. Schomerus, E. Skvortsov, M. Staudacher and especially to A. Bissi and S. Jain for fruitful discussions. MV would like to acknowledge HRI string group members for useful discussions. The work of MV was supported by the Infosys scholarship for senior students.

\appendix

\section{Appendix: Mellin Amplitudes for Fermionic correlators}
\label{sec:app-general}

\subsection{Counting tensor structures}
\label{countessa}

Correlators expressed in embedding space variables are manifestly covariant with conformal transformations and are easy to work with. However the downside is that there is a great deal of redundancy in all the possible tensor structures one can write. Sometimes it is easy to see relations between different tensor structures through gamma matrix commutation relations or simple Fierz identities, but in general this is a tedious matter. A neat way to count independent tensor structures and figure out the web of relations relating the different tensor structures in embedding space is to go to a suitable conformal frame by Lorentz transformations as depicted elegantly in \cite{Kravchuk:2016qvl}. In this paper, they prove that independent tensor structures in a $n$-point function are in one-to-one correspondence with the singlets (scalars for parity even tensor structures and pseudo-scalars for parity odd tensor structures) of the little group that leaves the configuration of points (at which operators in the correlator are inserted) in this conformal frame invariant. These singlets can be represented by 
\begin{eqnarray}
\text{Res}^{O(d)}_{O(d+2-m)} \bigotimes_{i=1}^{n}\rho_{i}      \;\;\;\;\;\;\;\;\;\;\;\;  m=\text{Min}\{n,d+2\}           \label{count}
\end{eqnarray} 
$\text{Res}^{G}_{H}$ denotes the restriction of a representation of $G$ to a representation of $H\subseteq G$. $\rho_{i}$ is the representation of the Lorentz group in which the operator at the $i$th position in the correlator transforms. If parity is not a symmetry of the theory, then we should replace $O(\cdot)$ with $SO(\cdot)$. To consider half integer spin representations one has to use the double cover of $SO(\cdot)$ which is $Spin(\cdot)$ and for parity symmetric theory one has to make a choice of the $Pin(\cdot)$ group. If $n\geq d+2$, one can form a parity odd invariant and consequently restrict to using only parity even tensor structures. The following is a parity odd tensor structure suggested in \cite{Kravchuk:2016qvl},
\begin{eqnarray}
w=\frac{\epsilon_{\mu_{1}\cdots\mu_{d+2}}X_{1}^{\mu_{1}}\cdots X_{d+2}^{\mu_{d+2}}}{\sqrt{X_{12}X_{23}\cdots X_{d+1,d+2}X_{d+2,1}}}       \label{oddinv}
\end{eqnarray}
If there are identical operators in the correlator, permutation symmetries result in further reductions in the number of independent tensor structures as explained concretely in \cite{Kravchuk:2016qvl}, but we shall stick to assuming operators with different dimensions.\\
\\
A conformal frame for $n$ points is any fixed configuration of points to which one can always map any $n$ points using conformal transformations. The most familiar example of this is probably the conformal frame where four points are mapped to $0$, $1$ (along any axis $x$), $\infty$ and $(z,\bar{z})$ (on a chosen plane containing the axis $x$). In general, relations between embedding space tensor structures can be obtained by choosing a conformal frame and expressing them in terms of the conformal frame tensor structures which are free of redundancies, and then simple linear algebra gives relations between the different embedding space tensor structures. \\
\\
The counting of tensor structures for 3d fermions has already been done in Section 4.2 of \cite{Kravchuk:2016qvl}. We quote the relevant results here. The number of independent tensor structures (of definite parity, indicated by the signs in the superscript) for the 3-pt function of operators with spins $l_{i}$ is given by,
\begin{eqnarray}
N_{3d}^{\pm}&=&\frac{N_{3d}\left(l_{1},l_{2},l_{3}\right)\pm \kappa}{2}               \label{count3pt}    \\
N_{3d}&=& (2l_{1}+1)(2l_{2}+1)-p(p+1)       \;\;\;\;\;\;\;\;\;\;\;\;\;\; p=\text{Max}\left(l_{1}+l_{2}-l_{3},0\right) \;\;\;\; l_{1}\leq l_{2}\leq l_{3}             \nonumber 
\end{eqnarray} 
The number of independent $n$-point tensor structures for $n\geq 4$ \cite{Kravchuk:2016qvl} is given by,
\begin{eqnarray}
N_{3d}\left(l_{1},l_{2},l_{3},l_{4}\right)=\prod_{i=1}^{n}(2l_{i}+1)           \label{count4}
\end{eqnarray}
If there is at least one half-integer spin operator, we can take an equal number of parity odd structures and parity even structures. Thus one can choose two parity even and two parity odd tensor structures for $\Braket{\psi_{1}\psi_{2}O_{3}O_{4}}$. \\
\\ 
For $n\geq 5$ too, the number of independent tensor structures is just 4 and as mentioned before, it suffices to work with only parity even tensor structures. For example for $n=5$ (with the fermions inserted at $x_{1}$ and $x_{2}$), we can take these to be,
\begin{eqnarray} 
\frac{\Braket{S_{1}S_{2}}}{\sqrt{X_{12}}}  \;\;\;\;\;\;\;\;\;\; \frac{\Braket{S_{1}X_{i}X_{j}S_{2}}}{\sqrt{X_{1i}X_{ij}X_{k2}}}    \;\;\;\; i<j\in \{3,4,5\}   \label{tens5}
\end{eqnarray}
Parity odd tensor structures can be expressed in terms of parity even ones using the invariant $w$ in ~\eqref{oddinv}. For example, let's take the parity odd tensor structure $\Braket{S_{1}XS_{2}}$. We can multiply this with odd invariant $w$ and then use $\Gamma^{A_{1}}\sim\epsilon^{A_{1}\cdots A_{5}}\Gamma_{A_{2}}\cdots\Gamma_{A_{5}}$. Then we can apply the following identity.
\begin{eqnarray}
\epsilon^{A_{1}\cdots A_{k}}\epsilon^{B_{1}\cdots B_{k}}
                             =\begin{vmatrix} 
                                            \eta^{A_{1}B_{1}} & \eta^{A_{2}B_{1}} & \cdots & \eta^{A_{k}B_{1}} \\
                                            \eta^{A_{1}B_{2}} & \eta^{A_{2}B_{2}} & \cdots & \eta^{A_{k}B_{2}} \\
                                            \vdots                &  \vdots               & \ddots & \vdots                \\
                                            \eta^{A_{1}B_{k}} & \eta^{A_{2}B_{k}} & \cdots & \eta^{A_{k}B_{k}} \\   
                                           \end{vmatrix}         \label{oddtoeven}          
\end{eqnarray}. 
Also note that for a correlator of two fermions and $n-2$ scalars, we do not need tensor structures of the form $\Braket{S_{1}X_{i_{1}}\cdots X_{i_{k}}S_{2}}$ with $k>\floor{\frac{d+2}{2}}$. We would also need to relate the parity odd invariant $w$ to cross-ratios as the standard way to define Mellin amplitudes involves cross-ratios. We can easily express $w^{2}$ in terms of cross-ratios using ~\eqref{oddtoeven}.

\subsection{Reduced Mellin amplitude and Mellin amplitude for the four fermion correlator}
\label{mueller}

In ~\eqref{lahm}, we made a choice of $n_{ij;k}$ as in ~\eqref{defu} for the four point function of fermions. This choice decides how the reduced Mellin amplitude is related to the Mellin amplitude in ~\eqref{4fermdef}. We present the relations explicitly here.
\be
\mathcal{M}_{1}&=&\bar{M}_{1}\left[\Gamma\left(s_{12}+1\right)\Gamma\left(s_{13}\right)\Gamma\left(s_{14}\right)\Gamma\left(s_{23}\right)\Gamma\left(s_{24}\right)\Gamma\left(s_{34}+1\right)\right]^{-1}       \label{robben}  \\
\mathcal{M}_{2}&=&\bar{M}_{2}\left[\Gamma\left(s_{12}+\frac{1}{2}\right)\Gamma\left(s_{13}+\frac{1}{2}\right)\Gamma\left(s_{14}\right)\Gamma\left(s_{23}\right)\Gamma\left(s_{24}+\frac{1}{2}\right)\Gamma\left(s_{34}+\frac{1}{2}\right)\right]^{-1}       \nonumber \\
\mathcal{M}_{3}&=&\bar{M}_{3}\left[\Gamma\left(s_{12}+\frac{1}{2}\right)\Gamma\left(s_{13}\right)\Gamma\left(s_{14}+\frac{1}{2}\right)\Gamma\left(s_{23}+\frac{1}{2}\right)\Gamma\left(s_{24}\right)\Gamma\left(s_{34}+\frac{1}{2}\right)\right]^{-1}     \nonumber\\
\mathcal{M}_{4}&=&\bar{M}_{4}\left[\Gamma\left(s_{12}+1\right)\Gamma\left(s_{13}\right)\Gamma\left(s_{14}\right)\Gamma\left(s_{23}\right)\Gamma\left(s_{24}\right)\Gamma\left(s_{34}+1\right)\right]^{-1}      \non
\ee
\be
\mathcal{M}_{5}&=&\bar{M}_{5}\left[\Gamma\left(s_{12}+\frac{1}{2}\right)\Gamma\left(s_{13}\right)\Gamma\left(s_{14}+\frac{1}{2}\right)\Gamma\left(s_{23}+\frac{1}{2}\right)\Gamma\left(s_{24}\right)\Gamma\left(s_{34}+\frac{1}{2}\right)\right]^{-1}   \label{vanpersie}\\
\mathcal{M}_{6}&=&\bar{M}_{6}\left[\Gamma\left(s_{12}+\frac{1}{2}\right)\Gamma\left(s_{13}+\frac{1}{2}\right)\Gamma\left(s_{14}\right)\Gamma\left(s_{23}\right)\Gamma\left(s_{24}+\frac{1}{2}\right)\Gamma\left(s_{34}+\frac{1}{2}\right)\right]^{-1}  \non \\
\mathcal{M}_{7}&=&\bar{M}_{7}\left[\Gamma\left(s_{12}+\frac{1}{2}\right)\Gamma\left(s_{13}+\frac{1}{2}\right)\Gamma\left(s_{14}\right)\Gamma\left(s_{23}\right)\Gamma\left(s_{24}+\frac{1}{2}\right)\Gamma\left(s_{34}+\frac{1}{2}\right)\right]^{-1}   \non\\
\mathcal{M}_{8}&=&\bar{M}_{8}\left[\Gamma\left(s_{12}+\frac{1}{2}\right)\Gamma\left(s_{13}\right)\Gamma\left(s_{14}+\frac{1}{2}\right)\Gamma\left(s_{23}+\frac{1}{2}\right)\Gamma\left(s_{24}\right)\Gamma\left(s_{34}+\frac{1}{2}\right)\right]^{-1}   \non
\ee
\be
\mathcal{M}_{9}&=&\bar{M}_{9}\left[\Gamma\left(s_{12}+1\right)\Gamma\left(s_{13}+\frac{1}{2}\right)\Gamma\left(s_{14}+\frac{1}{2}\right)\Gamma\left(s_{23}\right)\Gamma\left(s_{24}\right)\Gamma\left(s_{34}+\frac{1}{2}\right)\right]^{-1}     \label{boateng}\\
\mathcal{M}_{10}&=&\bar{M}_{10}\left[\Gamma\left(s_{12}+1\right)\Gamma\left(s_{13}\right)\Gamma\left(s_{14}\right)\Gamma\left(s_{23}+\frac{1}{2}\right)\Gamma\left(s_{24}+\frac{1}{2}\right)\Gamma\left(s_{34}+\frac{1}{2}\right)\right]^{-1}       \non\\
\mathcal{M}_{11}&=&\bar{M}_{11}\left[\Gamma\left(s_{12}+\frac{1}{2}\right)\Gamma\left(s_{13}+\frac{1}{2}\right)\Gamma\left(s_{14}\right)\Gamma\left(s_{23}+\frac{1}{2}\right)\Gamma\left(s_{24}\right)\Gamma\left(s_{34}+1\right)\right]^{-1}       \non\\
\mathcal{M}_{12}&=&\bar{M}_{12}\left[\Gamma\left(s_{12}+\frac{1}{2}\right)\Gamma\left(s_{13}\right)\Gamma\left(s_{14}+\frac{1}{2}\right)\Gamma\left(s_{23}\right)\Gamma\left(s_{24}+\frac{1}{2}\right)\Gamma\left(s_{34}+1\right)\right]^{-1}     \non
\ee
\be
\mathcal{M}_{13}&=&\bar{M}_{13}\left[\Gamma\left(s_{12}+1\right)\Gamma\left(s_{13}+\frac{1}{2}\right)\Gamma\left(s_{14}+\frac{1}{2}\right)\Gamma\left(s_{23}\right)\Gamma\left(s_{24}\right)\Gamma\left(s_{34}+\frac{1}{2}\right)\right]^{-1}     \label{hummels}\\
\mathcal{M}_{14}&=&\bar{M}_{14}\left[\Gamma\left(s_{12}+1\right)\Gamma\left(s_{13}\right)\Gamma\left(s_{14}\right)\Gamma\left(s_{23}+\frac{1}{2}\right)\Gamma\left(s_{24}+\frac{1}{2}\right)\Gamma\left(s_{34}+\frac{1}{2}\right)\right]^{-1}       \non\\
\mathcal{M}_{15}&=&\bar{M}_{15}\left[\Gamma\left(s_{12}+\frac{1}{2}\right)\Gamma\left(s_{13}+\frac{1}{2}\right)\Gamma\left(s_{14}\right)\Gamma\left(s_{23}+\frac{1}{2}\right)\Gamma\left(s_{24}\right)\Gamma\left(s_{34}+1\right)\right]^{-1}       \non\\
\mathcal{M}_{16}&=&\bar{M}_{16}\left[\Gamma\left(s_{12}+\frac{1}{2}\right)\Gamma\left(s_{13}\right)\Gamma\left(s_{14}+\frac{1}{2}\right)\Gamma\left(s_{23}\right)\Gamma\left(s_{24}+\frac{1}{2}\right)\Gamma\left(s_{34}+1\right)\right]^{-1}  \non
\ee

\subsection{Mixed fermion scalar conformal blocks}
\label{hazard}

The leading behavior of the mixed fermion scalar conformal blocks can be found in \cite{Iliesiu:2015akf}. These blocks are expressed in invariants $r,\theta$ introduced in \cite{Pappadopulo:2012jk}.
\begin{eqnarray}
\tilde{u}=\frac{16r^{2}}{(1+r^{2}-2r\eta)^{2}}   \;\;\;\;\;\;\;\;\;\;\;\;\;\;\;\;    \tilde{v}=\frac{(1+r^{2}+2r\eta)^{2}}{(1+r^{2}-2r\eta)^{2}}          \label{mero7}
\end{eqnarray} 
The OPE limit in these coordinates is now given by $r\rightarrow 0$ with $\eta$ held constant. One can check that for small $r$, $\tilde{u}\approx r^{2}$ and $\eta\approx -\frac{1-\tilde{v}}{2\sqrt{\tilde{u}}}$.\\
\\
$g^{i,jk}_{\Delta,l}$ is the contribution to $\mathcal{\tilde{A}}_{i}$ from the block associated with the fusion of tensor structures $r_{c}^{j}$ and $r_{c}^{k}$ ~\eqref{2ferm1sc} of the three point functions. Only non-zero ones are $g^{i,jk}_{\Delta,l}$ are $g^{1,++}_{\Delta,l}$, $g^{1,--}_{\Delta,l}$, $g^{2,++}_{\Delta,l}$, $g^{2,--}_{\Delta,l}$, $g^{3,+-}_{\Delta,l}$, $g^{3,-+}_{\Delta,l}$, $g^{4,+-}_{\Delta,l}$ and $g^{4,-+}_{\Delta,l}$. The leading behavior of these blocks as $r\rightarrow 0$ is given by (from \cite{Iliesiu:2015akf}), 
\begin{eqnarray}
g^{(1,++)}_{\Delta,l}(r,\eta)&=&-r^{\Delta}\left(P_{l-\frac{1}{2}}^{(0,1)}(\eta)+P_{l-\frac{1}{2}}^{(1,0)}(\eta)\right)+O\left(r^{\Delta+1}\right)    \label{mero8}\\
g^{(1,--)}_{\Delta,l}(r,\eta)&=&-r^{\Delta}\left(P_{l-\frac{1}{2}}^{(0,1)}(\eta)-P_{l-\frac{1}{2}}^{(1,0)}(\eta)\right)+O\left(r^{\Delta+1}\right)    \nonumber \\
g^{(2,++)}_{\Delta,l}(r,\eta)&=&r^{\Delta}\left(P_{l-\frac{1}{2}}^{(0,1)}(\eta)-P_{l-\frac{1}{2}}^{(1,0)}(\eta)\right)+O\left(r^{\Delta+1}\right)   \nonumber \\
g^{(2,--)}_{\Delta,l}(r,\eta)&=&r^{\Delta}\left(P_{l-\frac{1}{2}}^{(0,1)}(\eta)+P_{l-\frac{1}{2}}^{(1,0)}(\eta)\right)+O\left(r^{\Delta+1}\right)    \nonumber \\
g^{(3,+-)}_{\Delta,l}(r,\eta)&=&r^{\Delta}\left(P_{l-\frac{1}{2}}^{(0,1)}(\eta)-P_{l-\frac{1}{2}}^{(1,0)}(\eta)\right)+O\left(r^{\Delta+1}\right)     \nonumber \\
g^{(3,-+)}_{\Delta,l}(r,\eta)&=&r^{\Delta}\left(P_{l-\frac{1}{2}}^{(0,1)}(\eta)+P_{l-\frac{1}{2}}^{(1,0)}(\eta)\right)+O\left(r^{\Delta+1}\right)      \nonumber \\
g^{(4,+-)}_{\Delta,l}(r,\eta)&=&r^{\Delta}\left(P_{l-\frac{1}{2}}^{(0,1)}(\eta)+P_{l-\frac{1}{2}}^{(1,0)}(\eta)\right)+O\left(r^{\Delta+1}\right)    \nonumber \\
g^{(4,-+)}_{\Delta,l}(r,\eta)&=&r^{\Delta}\left(P_{l-\frac{1}{2}}^{(0,1)}(\eta)-P_{l-\frac{1}{2}}^{(1,0)}(\eta)\right)+O\left(r^{\Delta+1}\right)     \nonumber
\end{eqnarray}
$P_{n}^{(\alpha,\beta)}(z)$ are Jacobi polynomials. We note the symmetry property of Jacobi polynomials,
\begin{eqnarray}
P_{n}^{(\alpha,\beta)}(-z)=(-1)^{n}P_{n}^{(\beta,\alpha)}         \label{symprop}
\end{eqnarray}
~\eqref{symprop} implies that  $P_{n}^{(\alpha,\beta)}(z)+P_{n}^{(\beta,\alpha)}(z)$ has only even powers of $z$ for even $n$ and only odd powers of $z$ for odd $n$; and $P_{n}^{(\alpha,\beta)}(z)-P_{n}^{(\beta,\alpha)}(z)$ has only odd powers of $z$ for even $n$ and even powers of $z$ for odd $n$. Considering this, and the series expansion of the Jacobi polynomials, we can express the leading behavior of the blocks (for $l>\frac{1}{2}$ and with $\ell=l-\frac{1}{2}$) in the following manner ,
\begin{eqnarray}
g^{(1,++)}_{\Delta,l}(\tilde{u},\tilde{v})&\approx &-\tilde{u}^{\frac{\Delta}{2}}\sum_{k=0}^{\floor[\big]{\frac{\ell}{2}}}H_{\ell,k}^{+(0,1)}\left(\frac{1-\tilde{v}}{2\sqrt{\tilde{u}}}\right)^{\ell-2k} +\cdots  \label{mero14}\\
g^{(1,--)}_{\Delta,l}(\tilde{u},\tilde{v})&\approx &-\tilde{u}^{\frac{\Delta}{2}}\sum_{k=0}^{\ceil[\big]{\frac{\ell}{2}}-1}H_{\ell,k}^{-(0,1)}\left(\frac{1-\tilde{v}}{2\sqrt{\tilde{u}}}\right)^{\ell-2k-1} +\cdots  \nonumber \\
g^{(2,++)}_{\Delta,l}(\tilde{u},\tilde{v})&\approx &\tilde{u}^{\frac{\Delta}{2}}\sum_{k=0}^{\ceil[\big]{\frac{\ell}{2}}-1}H_{\ell,k}^{-(0,1)}\left(\frac{1-\tilde{v}}{2\sqrt{\tilde{u}}}\right)^{\ell-2k-1}  +\cdots\nonumber \\
g^{(2,--)}_{\Delta,l}(\tilde{u},\tilde{v})&\approx &\tilde{u}^{\frac{\Delta}{2}}\sum_{k=0}^{\floor[\big]{\frac{\ell}{2}}}H_{\ell,k}^{+(0,1)}\left(\frac{1-\tilde{v}}{2\sqrt{\tilde{u}}}\right)^{\ell-2k}   +\cdots\nonumber \\
g^{(3,+-)}_{\Delta,l}(\tilde{u},\tilde{v})&\approx &\tilde{u}^{\frac{\Delta}{2}}\sum_{k=0}^{\ceil[\big]{\frac{\ell}{2}}-1}H_{\ell,k}^{-(0,1)}\left(\frac{1-\tilde{v}}{2\sqrt{\tilde{u}}}\right)^{\ell-2k-1}    +\cdots \nonumber \\
g^{(3,-+)}_{\Delta,l}(\tilde{u},\tilde{v})&\approx &\tilde{u}^{\frac{\Delta}{2}}\sum_{k=0}^{\floor[\big]{\frac{\ell}{2}}}H_{\ell,k}^{+(0,1)}\left(\frac{1-\tilde{v}}{2\sqrt{\tilde{u}}}\right)^{\ell-2k}    +\cdots \nonumber \\
g^{(4,+-)}_{\Delta,l}(\tilde{u},\tilde{v})&\approx &\tilde{u}^{\frac{\Delta}{2}}\sum_{k=0}^{\floor[\big]{\frac{\ell}{2}}}H_{\ell,k}^{+(0,1)}\left(\frac{1-\tilde{v}}{2\sqrt{\tilde{u}}}\right)^{\ell-2k}  +\cdots \nonumber \\
g^{(4,-+)}_{\Delta,l}(\tilde{u},\tilde{v})&\approx &\tilde{u}^{\frac{\Delta}{2}}\sum_{k=0}^{\ceil[\big]{\frac{\ell}{2}}-1}H_{\ell,k}^{-(0,1)}\left(\frac{1-\tilde{v}}{2\sqrt{\tilde{u}}}\right)^{\ell-2k-1}     +\cdots \nonumber
\end{eqnarray}
$H_{n,k}^{\pm(\alpha,\beta)}$ are coefficients in the series expansions of $P_{n}^{(\alpha,\beta)}(z)\pm P_{n}^{(\beta,\alpha)}(z)$. For $l=\frac{1}{2}$, we have,
\begin{eqnarray}
g^{(1,++)}_{\Delta,\frac{1}{2}}(\tilde{u},\tilde{v})&\approx & -2\tilde{u}^{\frac{\Delta}{2}}  +\cdots  \;\;\;\;\;\;\;\;\;\;\;\;      g^{(1,--)}_{\Delta,\frac{1}{2}}(\tilde{u},\tilde{v})\approx  -2\tilde{u}^{\frac{\Delta+1}{2}}  +\cdots          \label{mero15} \\
g^{(2,++)}_{\Delta,\frac{1}{2}}(\tilde{u},\tilde{v})&\approx & 2\tilde{u}^{\frac{\Delta+1}{2}} +\cdots \;\;\;\;\;\;\;\;\;\;\;\;\;\;   g^{(2,--)}_{\Delta,\frac{1}{2}}(\tilde{u},\tilde{v})\approx  2\tilde{u}^{\frac{\Delta}{2}}   +\cdots  \nonumber  \\
g^{(3,+-)}_{\Delta,\frac{1}{2}}(\tilde{u},\tilde{v})&\approx & 2\tilde{u}^{\frac{\Delta+1}{2}} +\cdots \;\;\;\;\;\;\;\;\;\;\;\;\;\;   g^{(3,-+)}_{\Delta,\frac{1}{2}}(\tilde{u},\tilde{v})\approx  2\tilde{u}^{\frac{\Delta}{2}}  +\cdots           \nonumber \\
g^{(4,+-)}_{\Delta,\frac{1}{2}}(\tilde{u},\tilde{v})&\approx & 2\tilde{u}^{\frac{\Delta}{2}} +\cdots \;\;\;\;\;\;\;\;\;\;\;\;\;\;  g^{(4,-+)}_{\Delta,\frac{1}{2}}(\tilde{u},\tilde{v})\approx  2\tilde{u}^{\frac{\Delta+1}{2}}  +\cdots   \nonumber 
\end{eqnarray}

\subsection{$u$-channel poles in the fermion-scalar four point Mellin amplitude}
\label{chettri}

The $u$-channel poles in this correlator are summarised in Table \ref{table3}.
\begin{table}
\centering 
\begin{tabular}{|c|c|C{6.2cm}|C{5.2cm}|}
\hline
\textbf{Component of M.A.} & \textbf{Location of Poles} 
  & \textbf{Residues $\sim$}  \\ 
\hline\hline
\multirow{4}{*}{\vspace*{.2in}$\mathcal{M}_1$} & \multirow{4}{*}{\vspace*{.5in}$s+t=\sum_i\tau_i-\tau -2k$}  
  &\multirow{4}{*}{\vspace*{1.2cm} $\lambda_{\psi_1\phi_4\Psi_\ell}^+\lambda_{\Psi_\ell\phi_3\psi_2}^+$}  \\ [.4cm]
 \cline{2-3}
&\multirow{4}{*}{ \vspace*{.5in}$s+t=\sum_i\tau_i-\tau+1 -2k$} & \vspace*{.1cm} $\lambda_{\psi_1\phi_4\Psi_\ell}^-\lambda_{\Psi_\ell\phi_3\psi_2}^-$  \\ [.3cm]
\hline\hline
\multirow{4}{*}{\vspace*{.2in}$\mathcal{M}_2$} & \multirow{4}{*}{\vspace*{.5in}$s+t=\sum_i\tau_i-\tau -2k$}  
  &\multirow{4}{*}{\vspace*{1.2cm} $\lambda_{\psi_1\phi_4\Psi_\ell}^+\lambda_{\Psi_\ell\phi_3\psi_2}^+$}  \\ [.4cm]
 \cline{2-3}
&\multirow{4}{*}{ \vspace*{.5in}$s+t=\sum_i\tau_i-\tau+1 -2k$} & \vspace*{.1cm} $\lambda_{\psi_1\phi_4\Psi_\ell}^-\lambda_{\Psi_\ell\phi_3\psi_2}^-$  \\ [.3cm]
\hline\hline
\multirow{4}{*}{\vspace*{.2in}$\mathcal{M}_3$} & \multirow{4}{*}{\vspace*{.5in}$s+t=\sum_i\tau_i-\tau -2k$}  
  &\multirow{4}{*}{\vspace*{1.2cm} $\lambda_{\psi_1\phi_4\Psi_\ell}^+\lambda_{\Psi_\ell\phi_3\psi_2}^-$}  \\ [.4cm]
 \cline{2-3}
&\multirow{4}{*}{ \vspace*{.5in}$s+t=\sum_i\tau_i-\tau-1 -2k$} & \vspace*{.1cm} $\lambda_{\psi_1\phi_4\Psi_\ell}^-\lambda_{\Psi_\ell\phi_3\psi_2}^+$  \\ [.3cm]
\hline\hline
\multirow{4}{*}{\vspace*{.2in}$\mathcal{M}_4$} & \multirow{4}{*}{\vspace*{.5in}$s+t=\sum_i\tau_i-\tau-1 -2k$}  
  &\multirow{4}{*}{\vspace*{1.2cm} $\lambda_{\psi_1\phi_4\Psi_\ell}^+\lambda_{\Psi_\ell\phi_3\psi_2}^-$}  \\ [.4cm]
 \cline{2-3}
&\multirow{4}{*}{ \vspace*{.5in}$s+t=\sum_i\tau_i-\tau -2k$} & \vspace*{.1cm} $\lambda_{\psi_1\phi_4\Psi_\ell}^-\lambda_{\Psi_\ell\phi_3\psi_2}^+$  \\ [.3cm]
\hline\hline
\end{tabular}
\caption{Fermion-scalar four point function: $u$-channel poles. }
\label{table3}
\end{table}

\subsection{Crossed channel poles in the four fermion Mellin amplitude}
\label{ronaldinho}

Corresponding to each integer spin $l$ primary $\mathcal{O}_{l}$ of twist $\tau$ contributing to the $\psi_{1}\psi_{3}$ and $\psi_{2}\psi_{4}$ OPE, the Mellin amplitude has poles and residues as summarised in ~\ref{table5}.
\begin{table}
\centering 
\begin{tabular}{|c|c|C{6.2cm}|C{5.2cm}|}
\hline
\textbf{Component of M.A.} & \textbf{Location of Poles} 
  & \textbf{Residues $\sim$}  \\ 
\hline\hline
\multirow{4}{*}{} & \multirow{4}{*}{\vspace*{.2in}$t=\tau -1+2k$}  
  &\vspace*{.1cm} $\lambda_{\psi_1\psi_3\mathcal{O}_\ell}^1\lambda_{\mathcal{O}_\ell\psi_2\psi_4}^1\;\;,\,\; \lambda_{\psi_1\psi_3\mathcal{O}_\ell}^1\lambda_{\mathcal{O}_\ell\psi_2\psi_4}^2 $  \\ [.2cm]
&  & \vspace*{.1cm} $\lambda_{\psi_1\psi_3\mathcal{O}_\ell}^2\lambda_{\mathcal{O}_\ell\psi_2\psi_4}^1  \;\,,\;\;\lambda_{\psi_1\psi_3\mathcal{O}_\ell}^2\lambda_{\mathcal{O}_\ell\psi_2\psi_4}^2$   \\[.2cm]
 \cline{2-3}$\mathcal{M}_1\;,\;\mathcal{M}_3\;,\;\mathcal{M}_4\;,\;\mathcal{M}_5$
&\multirow{4}{*}{ \vspace*{.2in}$t=\tau +2k$} & \vspace*{.1cm} $\lambda_{\psi_1\psi_3\mathcal{O}_\ell}^3\lambda_{\mathcal{O}_\ell\psi_2\psi_4}^3\;\;,\;\;\lambda_{\psi_1\psi_3\mathcal{O}_\ell}^3\lambda_{\mathcal{O}_\ell\psi_2\psi_4}^4$  \\ [.2cm]
&  & \vspace*{.1cm} $\lambda_{\psi_1\psi_3\mathcal{O}_\ell}^4\lambda_{\mathcal{O}_\ell\psi_2\psi_4}^3\;\;,\;\;\lambda_{\psi_1\psi_3\mathcal{O}_\ell}^4\lambda_{\mathcal{O}_\ell\psi_2\psi_4}^4$   \\ [.2cm]
\hline\hline
\multirow{4}{*}{} & \multirow{4}{*}{\vspace*{.2in}$t=\tau +2k$}  
  &\multirow{4}{*}{\vspace*{.5cm} $\lambda_{\psi_1\psi_3\mathcal{O}_\ell}^2\lambda_{\mathcal{O}_\ell\psi_2\psi_4}^1\;\;,\,\; \lambda_{\psi_1\psi_3\mathcal{O}_\ell}^2\lambda_{\mathcal{O}_\ell\psi_2\psi_4}^2 $}  \\ [.2cm]
&  &    \\[.2cm]
 \cline{2-3}$\mathcal{M}_2\;\;,\;\;\mathcal{M}_6\;\;,\;\;\mathcal{M}_7$
&\multirow{4}{*}{ \vspace*{.2in}$t=\tau+1 +2k$} & \vspace*{.1cm} $\lambda_{\psi_1\psi_3\mathcal{O}_\ell}^3\lambda_{\mathcal{O}_\ell\psi_2\psi_4}^3\;\;,\;\;\lambda_{\psi_1\psi_3\mathcal{O}_\ell}^3\lambda_{\mathcal{O}_\ell\psi_2\psi_4}^4$  \\ [.2cm]
&  & \vspace*{.1cm} $\lambda_{\psi_1\psi_3\mathcal{O}_\ell}^4\lambda_{\mathcal{O}_\ell\psi_2\psi_4}^3\;\;,\;\;\lambda_{\psi_1\psi_3\mathcal{O}_\ell}^4\lambda_{\mathcal{O}_\ell\psi_2\psi_4}^4$   \\ [.2cm]
\hline\hline
\multirow{4}{*}{} & \multirow{4}{*}{\vspace*{.2in}$t=\tau +1+2k$}  
  &\multirow{4}{*}{\vspace*{.5cm} $\lambda_{\psi_1\psi_3\mathcal{O}_\ell}^2\lambda_{\mathcal{O}_\ell\psi_2\psi_4}^1\;\;,\,\; \lambda_{\psi_1\psi_3\mathcal{O}_\ell}^2\lambda_{\mathcal{O}_\ell\psi_2\psi_4}^2 $}  \\ [.2cm]
&  &   \\[.2cm]
 \cline{2-3}$\mathcal{M}_8$
&\multirow{4}{*}{ \vspace*{.2in}$t=\tau+2 +2k$} & \vspace*{.1cm} $\lambda_{\psi_1\psi_3\mathcal{O}_\ell}^3\lambda_{\mathcal{O}_\ell\psi_2\psi_4}^3\;\;,\;\;\lambda_{\psi_1\psi_3\mathcal{O}_\ell}^3\lambda_{\mathcal{O}_\ell\psi_2\psi_4}^4$  \\ [.2cm]
&  & \vspace*{.1cm} $\lambda_{\psi_1\psi_3\mathcal{O}_\ell}^4\lambda_{\mathcal{O}_\ell\psi_2\psi_4}^3\;\;,\;\;\lambda_{\psi_1\psi_3\mathcal{O}_\ell}^4\lambda_{\mathcal{O}_\ell\psi_2\psi_4}^4$   \\ [.2cm]
\hline\hline
\multirow{4}{*}{$\mathcal{M}_9 \;,\;\mathcal{M}_{11}\;,\;\mathcal{M}_{13}\;,\;\mathcal{M}_{15}$}  & \multirow{4}{*}{\vspace*{.2in}$t=\tau +2k$}  
  &\vspace*{.1cm} $\lambda_{\psi_1\psi_3\mathcal{O}_\ell}^1\lambda_{\mathcal{O}_\ell\psi_2\psi_4}^3\;\;,\,\; \lambda_{\psi_1\psi_3\mathcal{O}_\ell}^1\lambda_{\mathcal{O}_\ell\psi_2\psi_4}^4 $  \\ [.2cm]
&  & \vspace*{.1cm} $\lambda_{\psi_1\psi_3\mathcal{O}_\ell}^2\lambda_{\mathcal{O}_\ell\psi_2\psi_4}^3  \;\,,\;\;\lambda_{\psi_1\psi_3\mathcal{O}_\ell}^2\lambda_{\mathcal{O}_\ell\psi_2\psi_4}^4$   \\[.2cm]
 \cline{2-3}
&\multirow{4}{*}{ \vspace*{.6in}$t=\tau+1 +2k$} & \vspace*{.1cm} $\lambda_{\psi_1\psi_3\mathcal{O}_\ell}^3\lambda_{\mathcal{O}_\ell\psi_2\psi_4}^2\;\;,\;\;\lambda_{\psi_1\psi_3\mathcal{O}_\ell}^4\lambda_{\mathcal{O}_\ell\psi_2\psi_4}^2$  \\ [.2cm]
\hline\hline
\multirow{4}{*}{$\mathcal{M}_{10} \;,\;\mathcal{M}_{12}\;,\;\mathcal{M}_{14}\;,\;\mathcal{M}_{16}$}  & \multirow{4}{*}{\vspace*{.2in}$t=\tau +2k$}  
  &\vspace*{.1cm} $\lambda_{\psi_1\psi_3\mathcal{O}_\ell}^3\lambda_{\mathcal{O}_\ell\psi_2\psi_4}^1\;\;,\,\; \lambda_{\psi_1\psi_3\mathcal{O}_\ell}^4\lambda_{\mathcal{O}_\ell\psi_2\psi_4}^1 $  \\ [.2cm]
&  & \vspace*{.1cm} $\lambda_{\psi_1\psi_3\mathcal{O}_\ell}^3\lambda_{\mathcal{O}_\ell\psi_2\psi_4}^2  \;\,,\;\;\lambda_{\psi_1\psi_3\mathcal{O}_\ell}^4\lambda_{\mathcal{O}_\ell\psi_2\psi_4}^2$   \\[.2cm]
 \cline{2-3}
&\multirow{4}{*}{ \vspace*{.6in}$t=\tau+1 +2k$} & \vspace*{.1cm} $\lambda_{\psi_1\psi_3\mathcal{O}_\ell}^2\lambda_{\mathcal{O}_\ell\psi_2\psi_4}^3\;\;,\;\;\lambda_{\psi_1\psi_3\mathcal{O}_\ell}^2\lambda_{\mathcal{O}_\ell\psi_2\psi_4}^4$  \\ [.2cm]
\hline\hline
\end{tabular}
\caption{Fermion four point function: $t$-channel poles. }
\label{table5}
\end{table}
The $u$-channel poles are summarised in Table ~\ref{table6}.\\
\\
When the exchanged operator is a scalar $l=0$, we should take all structure constants apart from $\lambda^{1}$, $\lambda^{3}$ to be zero. 
\begin{table}
\centering 
\begin{tabular}{|c|c|C{6.2cm}|C{5.2cm}|}
\hline
\textbf{Component of M.A.} & \textbf{Location of Poles} 
  & \textbf{Residues $\sim$}  \\ 
\hline\hline
\multirow{4}{*}{} & \multirow{4}{*}{\vspace*{.2in}$s+t=\sum_i \tau_i -\tau +1-2k$}  
  &\vspace*{.1cm} $\lambda_{\psi_1\psi_4\mathcal{O}_\ell}^1\lambda_{\mathcal{O}_\ell\psi_3\psi_2}^1\;\;,\,\; \lambda_{\psi_1\psi_4\mathcal{O}_\ell}^1\lambda_{\mathcal{O}_\ell\psi_3\psi_2}^2 $  \\ [.2cm]
&  & \vspace*{.1cm} $\lambda_{\psi_1\psi_4\mathcal{O}_\ell}^2\lambda_{\mathcal{O}_\ell\psi_3\psi_2}^1  \;\,,\;\;\lambda_{\psi_1\psi_4\mathcal{O}_\ell}^2\lambda_{\mathcal{O}_\ell\psi_3\psi_2}^2$   \\[.2cm]
 \cline{2-3}$\mathcal{M}_1$
&\multirow{4}{*}{ \vspace*{.2in}$s+t=\sum_i \tau_i -\tau -2k$} & \vspace*{.1cm} $\lambda_{\psi_1\psi_4\mathcal{O}_\ell}^3\lambda_{\mathcal{O}_\ell\psi_3\psi_2}^3\;\;,\;\;\lambda_{\psi_1\psi_3\mathcal{O}_\ell}^3\lambda_{\mathcal{O}_\ell\psi_3\psi_2}^4$  \\ [.2cm]
&  & \vspace*{.1cm} $\lambda_{\psi_1\psi_4\mathcal{O}_\ell}^4\lambda_{\mathcal{O}_\ell\psi_3\psi_2}^3\;\;,\;\;\lambda_{\psi_1\psi_4\mathcal{O}_\ell}^4\lambda_{\mathcal{O}_\ell\psi_3\psi_2}^4$   \\ [.2cm]
\hline\hline
\multirow{4}{*}{} & \multirow{4}{*}{\vspace*{.2in}$s+t=\sum_i \tau_i -\tau +1-2k$}  
  &\multirow{4}{*}{\vspace*{.5cm} $\lambda_{\psi_1\psi_4\mathcal{O}_\ell}^2\lambda_{\mathcal{O}_\ell\psi_3\psi_2}^1\;\;,\,\; \lambda_{\psi_1\psi_4\mathcal{O}_\ell}^2\lambda_{\mathcal{O}_\ell\psi_3\psi_2}^2 $}  \\ [.2cm]
&  &    \\[.2cm]
 \cline{2-3}$\mathcal{M}_2\;\;,\;\;\mathcal{M}_4 $
&\multirow{4}{*}{ \vspace*{.2in}$s+t=\sum_i \tau_i -\tau -2k$} & \vspace*{.1cm} $\lambda_{\psi_1\psi_4\mathcal{O}_\ell}^3\lambda_{\mathcal{O}_\ell\psi_3\psi_2}^3\;\;,\;\;\lambda_{\psi_1\psi_3\mathcal{O}_\ell}^3\lambda_{\mathcal{O}_\ell\psi_3\psi_2}^4$  \\ [.2cm]
&  & \vspace*{.1cm} $\lambda_{\psi_1\psi_4\mathcal{O}_\ell}^4\lambda_{\mathcal{O}_\ell\psi_3\psi_2}^3\;\;,\;\;\lambda_{\psi_1\psi_4\mathcal{O}_\ell}^4\lambda_{\mathcal{O}_\ell\psi_3\psi_2}^4$   \\ [.2cm]
\hline\hline
\multirow{4}{*}{} & \multirow{4}{*}{\vspace*{.2in}$s+t=\sum_i \tau_i -\tau -2k$}  
  &\vspace*{.1cm} $\lambda_{\psi_1\psi_4\mathcal{O}_\ell}^1\lambda_{\mathcal{O}_\ell\psi_3\psi_2}^1\;\;,\,\; \lambda_{\psi_1\psi_4\mathcal{O}_\ell}^1\lambda_{\mathcal{O}_\ell\psi_3\psi_2}^2 $  \\ [.2cm]
&  & \vspace*{.1cm} $\lambda_{\psi_1\psi_4\mathcal{O}_\ell}^2\lambda_{\mathcal{O}_\ell\psi_3\psi_2}^1  \;\,,\;\;\lambda_{\psi_1\psi_4\mathcal{O}_\ell}^2\lambda_{\mathcal{O}_\ell\psi_3\psi_2}^2$   \\[.2cm]
 \cline{2-3}$\mathcal{M}_3\;\;,\;\;\mathcal{M}_5$
&\multirow{4}{*}{ \vspace*{.2in}$s+t=\sum_i \tau_i -\tau-1 -2k$} & \vspace*{.1cm} $\lambda_{\psi_1\psi_4\mathcal{O}_\ell}^3\lambda_{\mathcal{O}_\ell\psi_3\psi_2}^3\;\;,\;\;\lambda_{\psi_1\psi_3\mathcal{O}_\ell}^3\lambda_{\mathcal{O}_\ell\psi_3\psi_2}^4$  \\ [.2cm]
&  & \vspace*{.1cm} $\lambda_{\psi_1\psi_4\mathcal{O}_\ell}^4\lambda_{\mathcal{O}_\ell\psi_3\psi_2}^3\;\;,\;\;\lambda_{\psi_1\psi_4\mathcal{O}_\ell}^4\lambda_{\mathcal{O}_\ell\psi_3\psi_2}^4$   \\ [.2cm]
\hline\hline
\multirow{4}{*}{} & \multirow{4}{*}{\vspace*{.2in}$s+t=\sum_i \tau_i -\tau -1-2k$}  
  &\multirow{4}{*}{\vspace*{.5cm} $\lambda_{\psi_1\psi_4\mathcal{O}_\ell}^2\lambda_{\mathcal{O}_\ell\psi_3\psi_2}^1\;\;,\,\; \lambda_{\psi_1\psi_4\mathcal{O}_\ell}^2\lambda_{\mathcal{O}_\ell\psi_3\psi_2}^2 $}  \\ [.2cm]
&  &   \\[.2cm]
 \cline{2-3}$\mathcal{M}_6\;\;,\;\;\mathcal{M}_7$
&\multirow{4}{*}{ \vspace*{.2in}$s+t=\sum_i \tau_i -\tau-2 -2k$} & \vspace*{.1cm} $\lambda_{\psi_1\psi_4\mathcal{O}_\ell}^3\lambda_{\mathcal{O}_\ell\psi_3\psi_2}^3\;\;,\;\;\lambda_{\psi_1\psi_3\mathcal{O}_\ell}^3\lambda_{\mathcal{O}_\ell\psi_3\psi_2}^4$  \\ [.2cm]
&  & \vspace*{.1cm} $\lambda_{\psi_1\psi_4\mathcal{O}_\ell}^4\lambda_{\mathcal{O}_\ell\psi_3\psi_2}^3\;\;,\;\;\lambda_{\psi_1\psi_4\mathcal{O}_\ell}^4\lambda_{\mathcal{O}_\ell\psi_3\psi_2}^4$   \\ [.2cm]
\hline\hline
\multirow{4}{*}{} & \multirow{4}{*}{\vspace*{.2in}$s+t=\sum_i \tau_i -\tau -2k$}  
  &\multirow{4}{*}{\vspace*{.5cm} $\lambda_{\psi_1\psi_4\mathcal{O}_\ell}^2\lambda_{\mathcal{O}_\ell\psi_3\psi_2}^1\;\;,\,\; \lambda_{\psi_1\psi_4\mathcal{O}_\ell}^2\lambda_{\mathcal{O}_\ell\psi_3\psi_2}^2 $}  \\ [.2cm]
&  &   \\[.2cm]
 \cline{2-3}$\mathcal{M}_8$
&\multirow{4}{*}{ \vspace*{.2in}$s+t=\sum_i \tau_i -\tau-1 -2k$} & \vspace*{.1cm} $\lambda_{\psi_1\psi_4\mathcal{O}_\ell}^3\lambda_{\mathcal{O}_\ell\psi_3\psi_2}^3\;\;,\;\;\lambda_{\psi_1\psi_3\mathcal{O}_\ell}^3\lambda_{\mathcal{O}_\ell\psi_3\psi_2}^4$  \\ [.2cm]
&  & \vspace*{.1cm} $\lambda_{\psi_1\psi_4\mathcal{O}_\ell}^4\lambda_{\mathcal{O}_\ell\psi_3\psi_2}^3\;\;,\;\;\lambda_{\psi_1\psi_4\mathcal{O}_\ell}^4\lambda_{\mathcal{O}_\ell\psi_3\psi_2}^4$   \\ [.2cm]
\hline\hline
\multirow{4}{*}{$\mathcal{M}_9 \;,\;\mathcal{M}_{12}\;,\;\mathcal{M}_{13}\;,\;\mathcal{M}_{16}$}  & \multirow{4}{*}{\vspace*{.2in}$s+t=\sum_{i}\tau_{i}-\tau-2k$}  
  &\vspace*{.1cm} $\lambda_{\psi_1\psi_4\mathcal{O}_\ell}^1\lambda_{\mathcal{O}_\ell\psi_3\psi_2}^3\;\;,\,\; \lambda_{\psi_1\psi_4\mathcal{O}_\ell}^1\lambda_{\mathcal{O}_\ell\psi_3\psi_2}^4 $  \\ [.2cm]
&  & \vspace*{.1cm} $\lambda_{\psi_1\psi_4\mathcal{O}_\ell}^2\lambda_{\mathcal{O}_\ell\psi_3\psi_2}^3  \;\,,\;\;\lambda_{\psi_1\psi_4\mathcal{O}_\ell}^2\lambda_{\mathcal{O}_\ell\psi_3\psi_2}^4$   \\[.2cm]
 \cline{2-3}
&\multirow{4}{*}{ \vspace*{.6in}$s+t=\sum_{i}\tau_{i}-\tau-1-2k$} & \vspace*{.1cm} $\lambda_{\psi_1\psi_4\mathcal{O}_\ell}^3\lambda_{\mathcal{O}_\ell\psi_3\psi_2}^2\;\;,\;\;\lambda_{\psi_1\psi_4\mathcal{O}_\ell}^4\lambda_{\mathcal{O}_\ell\psi_3\psi_2}^2$  \\ [.2cm]
\hline\hline
\multirow{4}{*}{$\mathcal{M}_{10} \;,\;\mathcal{M}_{11}\;,\;\mathcal{M}_{14}\;,\;\mathcal{M}_{15}$}  & \multirow{4}{*}{\vspace*{.2in}$s+t=\sum_{i}\tau_{i}-\tau-2k$}  
  &\vspace*{.1cm} $\lambda_{\psi_1\psi_4\mathcal{O}_\ell}^3\lambda_{\mathcal{O}_\ell\psi_3\psi_2}^1\;\;,\,\; \lambda_{\psi_1\psi_4\mathcal{O}_\ell}^4\lambda_{\mathcal{O}_\ell\psi_3\psi_2}^1 $  \\ [.2cm]
&  & \vspace*{.1cm} $\lambda_{\psi_1\psi_4\mathcal{O}_\ell}^3\lambda_{\mathcal{O}_\ell\psi_3\psi_2}^2  \;\,,\;\;\lambda_{\psi_1\psi_4\mathcal{O}_\ell}^4\lambda_{\mathcal{O}_\ell\psi_3\psi_2}^2$   \\[.2cm]
 \cline{2-3}
&\multirow{4}{*}{ \vspace*{.6in}$s+t=\sum_{i}\tau_{i}-\tau-1-2k$} & \vspace*{.1cm} $\lambda_{\psi_1\psi_4\mathcal{O}_\ell}^2\lambda_{\mathcal{O}_\ell\psi_3\psi_2}^3\;\;,\;\;\lambda_{\psi_1\psi_4\mathcal{O}_\ell}^2\lambda_{\mathcal{O}_\ell\psi_3\psi_2}^4$  \\ [.2cm]
\hline\hline
\end{tabular}
\caption{Fermion four point function: $u$-channel poles.}
\label{table6}
\end{table}

\FloatBarrier

\section{Appendix: Witten diagrams}

\label{sec:app-witt}

\subsection{Fermions in AdS}
\label{app-fermi}

In the AdS/CFT correspondence, an operator $\mathcal{O}$ of the CFT is sourced by an appropriately defined boundary value $\phi_0$ of the dual field $\phi$ in the QFT in AdS. In the planar limit of the strongly interacting CFT, the AdS partition function can be approximated by the saddle point method, as the action evaluated at the classical bulk field $\phi_{\text{cl}}$ that obeys the equation of motion (e.o.m.), and the CFT correlation functions are just given by tree level Witten diagrams. To evaluate $S[\phi_{\text{cl}}]$, the bulk field $\phi_{\text{cl}}$ can be written as a perturbative expansion in terms of the boundary fields. The $n$-point planar level correlation function can be obtained by taking $n$-times functional derivatives of the boundary field \cite{Ghezelbash:1998pf}
		\begin{eqnarray}
				\label{eq:AdSCFTtreecorr}
			 \left< \mathcal{O}(x_1) \ldots \mathcal{O}(x_n) \right>  \approx
			\left.
			e^{ S[\phi_{\text{cl}}]} \frac{\delta}{\delta \phi_0(x_1)} \ldots \frac{\delta}{\delta \phi_0 (x_n)} e^{- S[\phi_{\text{cl}}]}
			\right|_{\phi_0=0}.
		\end{eqnarray}		
In the following discussion, we work in the Poincar\'e patch, which is given by
		\be
					\nonumber
			ds^2 = \frac{1}{z_0^2}\left( dz_0^2 + d\vec{z}^2 \right)
			= \frac{1}{z_0^2} dz^{\mu} dz_{\mu}.
		\ee
However, as noted in \cite{Henningson:1998cd, Muck:1998rr, Henneaux:1998ch} to analyze the AdS/CFT correspondence around the classical solutions for spinor fields requires a careful analysis of the boundary terms of the Dirac action $S_D$. To obtain consistent classical solutions for the spinor fields  $\psi$ or $\bar{\psi}$ 
\footnote{Now we have dropped the subscript cl for the classical fields.}
from the Dirac action a surface term $S_{\text{F}}$ has to be added, which obeys the symmetries of AdS geometry amongst other things. The requirement of adding a boundary term is necessary such that the action is stationary on the classical path. This procedure is common for theories defined on spaces with boundaries.\\
\\
To be concrete, let us assume that the mass $m\geq 0$ of the spinor is non-negative. Studying the classical solutions of $\psi(z) = \psi^+(z) + \psi^-(z)$ close at the boundary gives solutions of the form  $\psi^-(z) = z_{0}^{\frac{d}{2}-m}\psi_{0}^{-}(\vec{z})+O(z_{0}^{\frac{d}{2}-m})$ and $\psi(z)^+ = z_{0}^{\frac{d}{2}+m}\psi_{0}^{+}(\vec{z})+O(z_{0}^{\frac{d}{2}+m})$, where $\psi^+(z)$ and $\psi^-(z)$ are eigenfunctions of $\Gamma^0$: $\Gamma^0 \psi^{\pm}(z) = \pm \psi^{\pm}(z)$. This shows that for positive mass, $\psi^-_0$ is the leading contribution if one approaches the boundary. Furthermore, demanding regularity of the solutions in the bulk upto $z_{0}\rightarrow \infty$ we obtain relations between $\psi_{0}^{-}$ and $\bar{\psi_{0}^{+}}$ and similarly between $\bar{\psi}_{0}^{+}$ and $\bar{\psi}_{0}^{-}$. This establishes that the boundary data is given only in terms of $\psi_{0}^{-}$ and $\psi_{0}^{+}$. This means that when the boundary is odd dimensional, the boundary value of a bulk spinor is exactly a Dirac spinor of the boundary CFT, and when the boundary is even dimensional, the boundary value is a Weyl spinor of the boundary theory.\\
\\
For concreteness, let us consider Yukawa theory in AdS described by the action
	\begin{eqnarray}
						\label{eq:AdSYukawa}
		S[\psi, \bar{\psi}, \phi] & = & S_{\text{D}} + S_{\text{GK}} + S_{\text{int}} + S_{\text{F}}
					\\
						\nonumber
		& = & \int_M \hspace{-0.2cm} d^{d+1} z\; \sqrt{g} \left[ \bar{\psi} \left( \slashed{D} - m \right) \psi 
		+ \frac{1}{2} \left(  \left( \nabla_{\mu} \phi\right)^2 + M^2 \phi^2 \right) + \lambda \phi \bar{\psi} \psi \right] 
					\\
						\nonumber
		& & + \int_{\partial M_{\epsilon}} \hspace{-0.3cm} d^d \vec{x}\; \sqrt{h_{\epsilon}} \bar{\psi} \psi,
	\end{eqnarray}	
$h_{\epsilon;ij}$ is the induced metric on the surface $\partial M_{\epsilon}$. $\partial M_{\epsilon}$ is the regularized boundary of the AdS space $M$, which approches the boundary for $z_0 = \epsilon \to 0$ \cite{Kawano:1999au, Muck:1998rr, Henneaux:1998ch, Henningson:1998cd}. Solving the bulk fields in terms of the boundary fields now leads to a recursion relation for the fields
		\begin{eqnarray}
						\nonumber
			\phi(z) & = & \phi_{\epsilon}^{(0)} (z) - \lambda \int d^{d+1} w \sqrt{g(w)} G_{\epsilon}(z,w) \bar{\psi}(w) \psi(w)
					\\
						\nonumber
			\psi(z) & = & \psi_{\epsilon}^{(0)} (z) - \lambda \int d^{d+1}w \sqrt{g(w)} S_{\epsilon}(z,w) \phi(w) \psi(w)
					\\
						\label{eq:AdSRecRel}
			\bar{\psi}(z) & = & \bar{\psi}_{\epsilon}^{(0)}(z) - \lambda \int d^{d+1}w \sqrt{g(w)} \bar{\psi}(w) \phi(w) S_{\epsilon}(z,w).
		\end{eqnarray}
Here $\phi_{\epsilon}^{(0)} $, $\psi_{\epsilon}^{(0)}$ and $\bar{\psi}_{\epsilon}^{(0)}$ denote the regularized solutions to the e.o.m in free theory. Further, $G_{\epsilon}(z,w)$ and $S_{\epsilon}(z,w)$ are the regularized scalar and spinorial bulk-to-bulk operators \cite{Muck:1998rr,Freedman:1998tz}. Eventually one takes the limit $\epsilon \to 0$ and now the regularized free theory solutions can be expressed in terms of the boundary values:%
\footnote{The conformal dimension of the scalar field satisfies $\Delta_s\left(\Delta_s-D\right) = M^2$ \cite{Witten:1998qj} and for the spinor fields $\Delta = m + \frac{d}{2}$ \cite{Henningson:1998cd,Muck:1998rr}.}
		\begin{eqnarray}
							\label{eq:WittRules}
	\phi^{(0)}&=&	\lim_{\epsilon \to 0} \phi_{\epsilon}^{(0)} (z)  =  \int d^d \vec{x} \; K_{\Delta_s} (z, \vec{x}) \phi_0(\vec{x})
					\\
							\nonumber
	\psi^{(0)}&=&	\lim_{\epsilon \to 0} \psi_{\epsilon}^0 (z)  =  - \int d^d \vec{x} \; \Sigma_{\Delta} \left(z, \vec{x} \right) \psi_0^- \left(\vec{x}\right) \qquad \text{with} \quad  
			\Sigma_{\Delta}\left(z, \vec{x} \right) = \frac{\Gamma_{\mu}\left( z^{\mu} - x^{\mu} \right)}{\sqrt{z_0}} K_{\Delta + \frac{1}{2}}\left(z, \vec{x} \right) \mathcal{P}^-
					\\
							\nonumber
	\bar{\psi}^{(0)}&=&	\lim_{\epsilon \to 0} \bar{\psi}_{\epsilon}^0(z)  =  \int d^d \vec{x} \; \bar{\psi}_0^+ \left(\vec{x}\right) \bar{\Sigma}_{\Delta} \left(z, \vec{x} \right) \qquad \text{with} \quad  
			\bar{\Sigma}_{\Delta}\left(z, \vec{x} \right) = \mathcal{P}^+ \frac{\Gamma_{\mu}\left( z^{\mu} - x^{\mu} \right)}{\sqrt{z_0}} K_{\Delta + \frac{1}{2}} \left(z, \vec{x} \right)
		\end{eqnarray}
Here $K_{\Delta} (z, \vec{x})$ and $\Sigma_{\Delta}\left(z, \vec{x} \right)$ are the scalar and fermionic bulk-to-boundary propagator, respectively (see \cite{Kawano:1999au}). $\Gamma^{\mu}$ are gamma matrices of the bulk.\\
\\
Using the recursion relation \eqref{eq:AdSRecRel} (and taking $\epsilon\rightarrow 0$ at the end), the action can be written in a perturbation series in terms of the boundary fields $\phi_0$, $\psi_0^-$ and $\bar{\psi}_0^+$. Now, taking the functional derivative with respect to the corresponding boundary fields according to \eqref{eq:AdSCFTtreecorr} shall give the corresponding correlator (in the planar limit) in the boundary CFT.

\subsection{Spinor exchange in AdS}
	\label{sec:SpExAdS}
	
In this section we calculate the spinor exchange diagram. Note that in this calculation the two scalars are switched, i.e. we calculate $A^{\bar{\psi}_1 \phi_4}_{\psi_2 \phi_3}$.
\\
\\
Plugging the perturbative solution \eqref{eq:AdSRecRel} into a generalized action \eqref{eq:AdSYukawa} where all fields might have a different mass we obtain
	\begin{eqnarray}
					\nonumber
		S_{\bar{\psi}\phi  \psi \phi } = - 2 \lambda^2 G \prod_{i=1}^4 \int_{-\infty}^{\infty} d^d \vec{x}_i\;
		\bar{\psi}_0^+\left(\vec{x}_1 \right)  \phi_0\left( \vec{x}_4 \right)
		A\left(\vec{x}_1,\vec{x}_2,\vec{x}_3,\vec{x}_4 \right)
		\psi^-_0 \left(\vec{x}_2\right) \phi_0\left(\vec{x}_3 \right).
	\end{eqnarray}
The actual diagram $A\left(\vec{x}_1,\vec{x}_2,\vec{x}_3,\vec{x}_4 \right)= A^{\bar{\psi}_1 \phi_4}_{\psi_2 \phi_3}$ we have to calculate is given by
	\be 
					\nonumber
		A^{\bar{\psi}_1 \phi_4}_{\psi_2 \phi_3} & = &
		- \int d^{d+1} z \sqrt{g(z)}  d^{d+1} w \sqrt{g(w)}
		K_{\Delta_4} ( z, \vec{x}_4) \bar{\Sigma}_{\Delta_1}(z, \vec{x}_1) S(z,w) \Sigma_{\Delta_2}(w, \vec{x}_2) K_{\Delta_3}(w,\vec{x}_3).
	\ee
Following the discussion \cite{Kawano:1999au} we can effectively reduce the calculation of this diagram to the evaluation of a four scalar diagram with scalar exchange. The first step is to use the conformal symmetry on the boundary to translate all coordinates by $\vec{x}_2$ such that the new coordinates on the boundary are given by $\vec{y}_i= \vec{x}_i - \vec{x}_2$ for $i\neq 2$. Afterwards these coordinates are inverted $\vec{y}_i^{\prime} = \vec{y}_i/\abs{\vec{y}_i}^2$. Since the bulk measure is invariant under inversion and due to the definite transformation behaviour of the propagators, the amplitude can be rewritten as
	\be 
				\label{eq:spinex2}
		A^{\bar{\psi}_1 \phi_4}_{\psi_2 \phi_3} & = & \frac{\slashed{\vec{y}}_1}{\abs{\vec{y}_1}^{2 \Delta_1 +1} \abs{\vec{y}_3}^{2 \Delta_3} \abs{\vec{y}_4}^{2 \Delta_4}} 
		\left[-\slashed{\vec{y}}_{14}^{\prime} \slashed{\partial}_4^{\prime} + \left( \Delta_1 +\frac{1}{2} + \Delta_4 + \Delta_+ -d\right) \right]
		I\left( \vec{y}_1^{\prime}, \vec{y}_3^{\prime}, \vec{y}_4^{\prime}  \right)
	\ee
with $\Delta_+ = d/2 + m + 1/2$ and $m$ is the mass of the exchanged fermion. In \cite{Kawano:1999au} the explicit expression for $I$ is given by
	\be
				\nonumber
		I \left( \vec{y_1}^{\prime}, \vec{y}_3^{\prime}, \vec{y}_4^{\prime}\right) = 
		\int \hspace{-0.15cm} d^{d+1} z \sqrt{g(z)}\,  d^{d+1} w \sqrt{g(w)}
		K_{\Delta_4} ( z, \vec{y}_4^{\prime}) K_{\Delta_1 +\frac{1}{2}}(z, \vec{y}_1^{\prime}) G_{\Delta_+}(z,w) K_{\Delta_2 + \frac{1}{2}}\left(w^{\prime},0\right) K_{\Delta_3}(w,\vec{y}_3^{\prime})
	\ee	
Now we note that the AdS measure is invariant under inversion and that scalar bulk-to-boundary propagator transform covariantly under inversion: $K_{\Delta}(z^{\prime}, \vec{x}^{\prime}) = \abs{\vec{x}}^{2\Delta} K_{\Delta}\left(z,\vec{x} \right)$. Furthermore, the scalar bulk-to-bulk propagator only depends on the chordial distance $u = \frac{(z-w)^2}{z_0^2 w_0^2}$ and is therefore invariant under inversion $G_{\Delta}(z^{\prime}, w^{\prime})= G_{\Delta}(z,w)$. These properties allow us to rewrite $I$ as a scalar exchange diagram with four external scalars:
	\be
				\nonumber
		I 
		& = &\abs{\vec{y}_1}^{2 \Delta_1 + 1} \abs{\vec{y}_3}^{2 \Delta_3} \abs{\vec{y}_4}^{2 \Delta_4}	
		 \int d^{d+1} z \sqrt{g(z)}  d^{d+1} w\sqrt{g(w)}
		K_{\Delta_4} ( z, \vec{y}_4) K_{\Delta_1 +\frac{1}{2}}(z, \vec{y}_1) 
			\\
				\nonumber		
		& & \times G_{\Delta_+}(z,w)
		K_{\Delta_2 +\frac{1}{2}} \left(w,0 \right) K_{\Delta_3}(w,\vec{y}_3)
		 		 		 \\
		 	\nonumber
		 & = & \abs{\vec{y}_1}^{2 \Delta_1 + 1} \abs{\vec{y}_3}^{2 \Delta_3} \abs{\vec{y}_4}^{2 \Delta_4}	
		 \int d^{d+1} z \sqrt{g(z)}  d^{d+1} w\sqrt{g(w)}
		K_{\Delta_4} ( z, \vec{x}_4) K_{\Delta_1 +\frac{1}{2}}(z, \vec{x}_1) 
					\\
				\nonumber	
		& & \times G_{\Delta_+}(z,w)
		K_{\Delta_2 +\frac{1}{2}} \left(w, \vec{x}_2 \right) K_{\Delta_3}(w,\vec{x}_3).
	\ee
In the last	step we have translated the bulk coordinates $\vec{z} \to \vec{z} - \vec{x}_2$ and $\vec{w} \to \vec{w} - \vec{x}_2$. For this expression the Mellin amplitude is known \cite{Penedones:2010ue}. Further, we see that $I$ depends explicitly only on the unprimed coordinates. Thus we define a new quantity $\tilde{I} \left( \vec{y_1}, \vec{y}_2, \vec{y}_4\right) := I \left( \vec{y_1}^{\prime}, \vec{y}_2^{\prime}, \vec{y}_4^{\prime}\right)$ such that we obtain
		\be 
 					\nonumber
 		\tilde{I}
  		 & = & \abs{\vec{y}_1}^{2 \Delta_1 + 1} \abs{\vec{y}_3}^{2 \Delta_3} \abs{\vec{y}_4}^{2 \Delta_4}
 		\Mellint{1}{4} \Gamma\left(s_{il} \right) \mathcal{M}(s_{il}) 
 		\frac{1}{\abs{\vec{y}_{13}}^{2 s_{13}} \abs{\vec{y}_{14}}^{2 s_{14}} \abs{\vec{y}_{34}}^{2 s_{34}}} 
 					\\
 		& & \times
 		\frac{1}{\abs{\vec{y}_{1}}^{2 s_{12}} \abs{\vec{y}_{3}}^{2 s_{23}} \abs{\vec{y}}_{4}^{2 s_{24}}} 
 		\prod_{i=1}^4 \hat{\delta}\left( \Delta_i + \frac{1}{2}\left(\delta_{1i} + \delta_{2i} \right)- \sum_{k=1, k \neq i}^4 s_{ik} \right).
 	\ee	 	
\\
\\
To evaluate \eqref{eq:spinex2} we note that tensor structure is generated by the derivative and $\slashed{\vec{y}}_{14}$. After inverting these
		\begin{eqnarray*}
		\slashed{\vec{y}}_{14}^{\prime} & = & \frac{\slashed{\vec{y}}_{1}}{\abs{\vec{y}_{1}}^2} -\frac{\slashed{\vec{y}}_{4}}{\abs{\vec{y}_{4}}^2} \qquad \text{and} 
				\\
		\slashed{\partial}_{\vec{y}_4^{\prime}} 
		& = &\gamma^{\mu} \frac{\partial}{\partial \vec{y}_{4,\mu}^{\prime}} \frac{\vec{y}_4^{ \prime \, \nu}}{\abs{\vec{y}_4^{\prime}}^2}
		\frac{\partial}{\partial \vec{y}_4^{\nu}}
		 =  \vec{y}_{4\, \mu}^{\prime}  \frac{\partial}{\partial \vec{y}_4^{\nu}}
		= \abs{\vec{y}_4 }^2 \slashed{\partial}_{\vec{y}_4} - 2 \slashed{\vec{y}}_4\vec{y}_4 \cdot \frac{\partial}{\partial \vec{y}_4}
	\end{eqnarray*}
we obtain the following three types of terms 
	\be
					\nonumber
		\slashed{\vec{y}}_1 \left(\frac{\slashed{\vec{y}}_{1}}{\abs{\vec{y}_{1}}^2} -\frac{\slashed{\vec{y}}_{4}}{\abs{\vec{y}_{4}}^2} \right)
		\left(
		\abs{\vec{y}_4 }^2 \gamma^{\mu} - 2 \slashed{\vec{y}}_4 \vec{y}_4^{\mu}
		\right)
		\frac{\vec{y}_{4 \, \mu}}{\abs{\vec{y}_4}^2}
		& = & \slashed{\vec{y}}_{14} 
		\quad \text{with coefficient} \quad 2\Delta_4 -2 s_{24},
				\\
					\nonumber
		\slashed{\vec{y}}_1 \left(\frac{\slashed{\vec{y}}_{1}}{\abs{\vec{y}_{1}}^2} -\frac{\slashed{\vec{y}}_{4}}{\abs{\vec{y}_{4}}^2} \right)
		\left(
		\abs{\vec{y}_4 }^2 \gamma^{\mu} - 2 \slashed{\vec{y}}_4 \vec{y}_4^{\mu}
		\right)
		\frac{-\vec{y}_{14 \, \mu}}{\abs{\vec{y}_{14}}^2}
		& = & \slashed{\vec{y}}_4 
		\quad \text{with coefficient} \quad -2 s_{14},
				\\
					\nonumber			
		\slashed{\vec{y}}_1 \left(\frac{\slashed{\vec{y}}_{1}}{\abs{\vec{y}_{1}}^2} -\frac{\slashed{\vec{y}}_{4}}{\abs{\vec{y}_{4}}^2} \right)
		\left(
		\abs{\vec{y}_4 }^2 \gamma^{\mu} - 2 \slashed{\vec{y}}_4 \vec{y}_4^{\mu}
		\right)
		\frac{\vec{y}_{43 \, \mu}}{\abs{\vec{y}_{34}}^2}
		& = & \frac{ \slashed{\vec{y}}_{14} \slashed{\vec{y}}_{43} \slashed{\vec{y}}_{4}}{\abs{\vec{y}_{34}}^2}
		\quad \text{with coefficient} \quad -2 s_{34}.
	\ee
\\
\\	
This gives
	\be
					\nonumber
		A^{\bar{\psi}_1 \phi_4}_{\psi_2 \phi_3}
		& = &  
		\Mellint{1}{4} \Singprop{\vec{x}_{il}}{}{s_{il}} \mathcal{M}(s_{il}) 
 		\prod_{i=1}^4 \hat{\delta}\left( \Delta_i + \frac{1}{2}\left( \delta_{1i} + \delta_{2i} \right) - \sum_{k=1, k \neq i}^4 s_{ik} \right)	
 				\\
 					\nonumber
 		& & \times \left( \slashed{\vec{x}}_{12} \left(\Delta_1 +\frac{1}{2} + \Delta_4 + \Delta_+ -d \right)
 		+ 2 \slashed{\vec{x}}_{14} \left(s_{24} - \Delta_4 \right)
 		+ 2 \slashed{\vec{x}}_{42} s_{14} 
 		+ 2  \frac{ \slashed{\vec{x}}_{14} \slashed{\vec{x}}_{43} \slashed{\vec{x}}_{42}}{\abs{\vec{x}_{43}}^2} s_{34} \right)
				\\
					\nonumber
		& = & 
		\Mellint{1}{4} \Singprop{\vec{x}_{il}}{}{s_{il}} \mathcal{M}(s_{il}) 
 		\prod_{i=1}^4 \hat{\delta}\left( \Delta_i + \frac{1}{2}\left( \delta_{1i} + \delta_{2i} \right) - \sum_{k=1, k \neq i}^4 s_{ik} \right)	
 				\\
 		& & \times \left( \slashed{\vec{x}}_{12} \left(\Delta_1 +\frac{1}{2} + \Delta_4 + \Delta_+ -d  -2 s_{14} \right)
 		+ 2  \frac{ \slashed{\vec{x}}_{14} \slashed{\vec{x}}_{43} \slashed{\vec{x}}_{32}}{\abs{\vec{x}_{43}}^2} s_{34} \right)
	\ee
after using $s_{14} + s_{24} + s_{34} = \Delta_4$. Rearranging the delta constraints in canonical form and using $\Delta_1 + \Delta_4 - 2 s_{14} + \Delta_+ - d + \frac{1}{2} = t + \tau - d +2$%
	\footnote{Note that here the 't-channel' is given by (14)-(23).} 
yields finally
	\be
					\nonumber
		A^{\bar{\psi}_1 \phi_4}_{\psi_2 \phi_3}
		& = &  
				\left(
		\ten{\vec{x}}{12} \left( t + \tau - d + 2 \right)
		\Mellint{1}{4} \Singprop{\vec{x}_{il}}{}{s_{il} + \delta_{i1} \delta_{2l}} \mathcal{M}(s_{il} + \delta_{i1} \delta_{2l}) 
				\right.
 				\\
 					\nonumber
 		& & 
 		+ 2 \frac{ \slashed{\vec{x}}_{14} \slashed{\vec{x}}_{43} \slashed{\vec{x}}_{32}}{\abs{\vec{x}_{14}} \abs{\vec{x}_{43}}\abs{\vec{x}_{32}}}
 		\Mellint{1}{4} \frac{1}{\abs{\vec{x}_{12}}}
 		\Multiprop{\vec{x}_{il}}{}{s_{il}}{+ \frac{1}{2}\left( \delta_{i1} \delta_{2l} + \delta_{i1} \delta_{4l} +\delta_{2i} \delta_{3l} + \delta_{i3} \delta_{4l} \right)} 
				\\
		& & \times  
				\left.
		\mathcal{M}\left(s_{il} + \frac{1}{2}\left( \delta_{i1} \delta_{2l} + \delta_{i1} \delta_{4l} +\delta_{2i} \delta_{3l} + \delta_{i3} \delta_{4l} \right)\right)
				\right)
		\Melldeltwist{1}{4}.
	\ee
For the definition of $\mathcal{M}(s_{il})$ see \ref{jurgen1} and \ref{jurgen2}.

\section{Appendix: Feynman diagrams}
\label{sec:app-feyn}

\subsection{A Recursive method }
		\label{sec:RecMethScal}

\begin{figure}[h]
\begin{center}
\begin{tikzpicture}[scale=.8,transform shape]
\draw[ultra thick, black] (0,0) -- (-2,2);
\draw[ultra thick, black] (0,0) -- (-2,-2);
\draw[ultra thick, black] (0,0) -- (4,0);
\draw[ultra thick, black] (4,0) -- (6,2);
\draw[ultra thick, black] (4,0) -- (6,-2);
\node at (-2.2,2.2) {$1$};
\node at (-2.2,-2.2) {$2$};
\node at (6.2,-2.2) {$3$};
\node at (6.2,2.2) {$4$};
\node at (0.2,-0.4) {$u_{1}$};
\node at (3.8,-0.4) {$u_{2}$};
\node at (2,-0.4) {$\gamma$};
\end{tikzpicture}
\end{center}
\caption{Four external scalars: Scalar exchange }
\label{diag:4sc}
\end{figure}
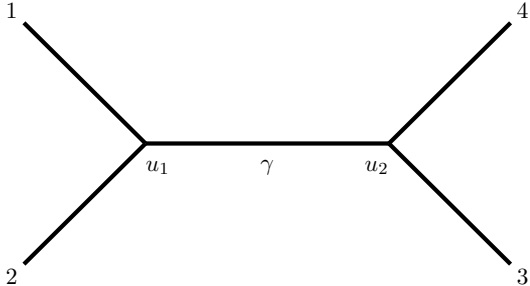

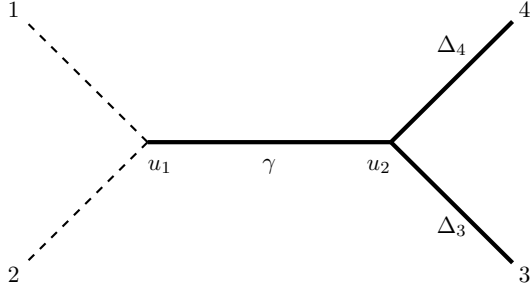
\begin{figure}[h]
\begin{center}
\begin{tikzpicture}[scale=.8,transform shape]
\draw[thick, dashed, black] (0,0) -- (-2,2);
\draw[thick, dashed, black] (0,0) -- (-2,-2);
\draw[ultra thick, black] (0,0) -- (4,0);
\draw[ultra thick, black] (4,0) -- (6,2);
\draw[ultra thick, black] (4,0) -- (6,-2);
\node at (-2.2,2.2) {$1$};
\node at (-2.2,-2.2) {$2$};
\node at (6.2,-2.2) {$3$};
\node at (6.2,2.2) {$4$};
\node at (0.2,-0.4) {$u_{1}$};
\node at (3.8,-0.4) {$u_{2}$};
\node at (2,-0.4) {$\gamma$};
\node at (5,1.6) {$\Delta_{4}$};
\node at (5,-1.4) {$\Delta_{3}$};
\end{tikzpicture}
\end{center}
\caption{Recursive method: First step }
\label{diag:recur1}
\end{figure}

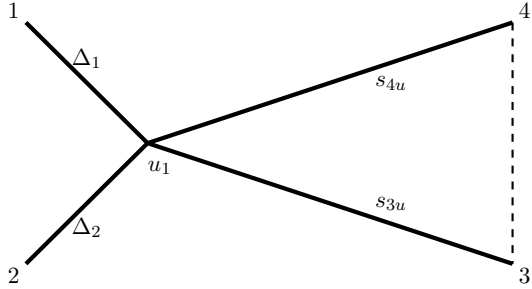
\begin{figure}[h]
\begin{center}
\begin{tikzpicture}[scale=.8,transform shape]
\draw[ultra thick, black] (0,0) -- (-2,2);
\draw[ultra thick, black] (0,0) -- (-2,-2);
\draw[ultra thick, black] (0,0) -- (6,2);
\draw[ultra thick, black] (0,0) -- (6,-2);
\draw[thick, dashed, black] (6,2) -- (6,-2);
\node at (-2.2,2.2) {$1$};
\node at (-2.2,-2.2) {$2$};
\node at (6.2,-2.2) {$3$};
\node at (6.2,2.2) {$4$};
\node at (0.2,-0.4) {$u_{1}$};
\node at (4,1) {$s_{4u}$};
\node at (4,-1) {$s_{3u}$};
\node at (-1,1.4) {$\Delta_{1}$};
\node at (-1,-1.4) {$\Delta_{2}$};
\end{tikzpicture}
\end{center}
\caption{Recursive method: Second step }
\label{diag:recur2}
\end{figure}

The calculation of Mellin amplitudes associated to conformal integrals presented in \cite{Nizami:2016jgt} involved the following steps: Each propagator (internal and external) in a given diagram was expressed in a Schwinger parametrized manner, and then the position space integrals over the interaction vertices were evaluated successively. Following this, we would be left with a Schwinger parameter integral that could be simplified drastically using the conformality of the overall integral, and the resulting integral could be evaluated exactly to give the Mellin amplitude as a product of beta functions. \\
\\
In the present case, when the position space conformal integral has fermionic legs, the simplications in the Schwinger parameter integral using the conformality condition are not as good, consequently the final Schwinger parameter integrals are complicated. Hence we shall apply a recursive method which allows us to reduce the calculation of any Feyman diagram to the calculation of a series of contact interaction diagrams \footnote{This technique was developed by Arnab Rudra for scalar conformal integrals while working on \cite{Nizami:2016jgt}}.\\
\\
To illustrate the procedure, we apply the recursive method to a simple example: a four point diagram of scalars with a scalar propagator as in Figure ~\eqref{diag:4sc}.
The conformal integral corresponding to this diagram is given by,
\be
I^{\phi_1 \phi_2}_{\phi_3 \phi_4}
	& = &\Cooint{u_1} \Cooint{u_2} \prod_{i=1}^{2} \Singprop{x_i}{-u_1}{\Delta_i} \prod_{i=3}^{4} \Singprop{x_i}{-u_2}{\Delta_i} \frac{1}{\abs{u_1-u_2}^{2 \gamma}}                  \label{confusc}
\ee
The conformality condition is $\Delta_{1}+\Delta_{2}=\Delta_{3}+\Delta_{4}=d-\gamma$. Now we shall treat the second interaction vertex $u_{2}$ like it existed indepedently as a contact interaction diagram with the ``external'' legs at $x_{1}$, $x_{2}$ and $u_{1}$. This is depicted pictorially in Figure \ref{diag:recur1}.\\
\\
We know the Mellin Barnes representation of the contact interaction conformal integral with scalar legs \cite{Symanzik:1972wj, Paulos:2012nu, Nizami:2016jgt}. 
\be
\Cooint{u_2}\prod_{i=3}^{4} \Singprop{x_i}{-u_2}{\Delta_i} \frac{1}{\abs{u_1-u_2}^{2 \gamma}}&=&\left(\Mellintu{3}{4} \right)\frac{1}{\Gamma\left(\gamma\right)} \mellintb{34}\Singprop{x_i}{-u_1}{s_{iu}} \nonumber \\
     && \Singprop{x_{34}}{}{\bar{s}_{34}} \prod_{i=3}^4 \mdelta{\Delta_i -\bar{s}_{34} - s_{iu}} \mdelta{\gamma -s_{3u}-s_{4u}}    \label{2ndinto}
\ee
We shall always assume that the contours of the Mellin-Barnes integrals are chosen such that the poles of the gamma functions are not separated and that the integrals converge (see \cite{Nizami:2016jgt}). Now we can plug the result ~\eqref{2ndinto} back in ~\eqref{confusc} to obtain the second contact interaction conformal integral that we need to evaluate. The legs are now given by $(x_{1},u_{1})$ with dimension $\Delta_{1}$, $(x_{2},u_{1})$ with dimension $\Delta_{2}$, $(x_{3},u_{1})$ with ``dimension'' $s_{3u}$ and $(x_{4},u_{4})$ with ``dimension'' $s_{4u}$. This is represented pictorially in Figure \ref{diag:recur2}. Using the $2 \pi i \; \delta(\gamma -s_{3u}-s_{4u}) =  \mdelta{\gamma -s_{3u}-s_{4u}}$ in ~\eqref{2ndinto}, we also get the required conformality condition for this integral $\Delta_{1}+\Delta_{2}+s_{3u}+s_{4u}=d$.\\
\\
Once again, we use the known result for the contact interaction of scalars, and plug it back into ~\eqref{2ndinto}, to obtain,
\be
I^{\phi_1 \phi_2}_{\phi_3 \phi_4} 
	& = &  \Mellintt{1}{4} \Singprop{x_{il}}{}{\tilde{s}_{il}} \mellintb{34} \Singprop{x_{34}}{}{\bar{s}_{34}}  \frac{1}{\Gamma\left(\gamma\right)}      \nonumber 
		\\
			\nonumber
	& &  \times \left(\Mellintu{3}{4} \right)  \mdelta{s_{3u} - \tilde{s}_{13} -\tilde{s}_{23}-\tilde{s_{34}}} \mdelta{s_{4u} - \tilde{s}_{14} -\tilde{s}_{24}-\tilde{s_{34}}}
		\\
			\nonumber
	& & \times \mdelta{\Delta_1 - \tilde{s}_{12} - \tilde{s}_{13} - \tilde{s}_{14}} \mdelta{\Delta_2 - \tilde{s}_{12} - \tilde{s}_{23} - \tilde{s}_{24}}	\prod_{i=3}^4 \mdelta{\Delta_i -\bar{s}_{34} - s_{iu}} \mdelta{\gamma -s_{3u}-s_{4u}}      
\ee
The Mellin variables introduced in the second step are indicated with the tilde. Next, we integrate out the $s_{iu}$ using the delta functions, rename $\tilde{s}_{ij}= s_{ij}$ for $(i,j)\neq (3,4)$ and take $\bar{s}_{34} = s_{34} - \tilde{s}_{34}$, such that we obtain,
\be
I^{\phi_1 \phi_2}_{\phi_3 \phi_4}
	& = &  \Mellint{1}{4} \Singprop{x_{il}}{}{s_{il}} \Melldel{1}{4}        \nonumber
		\\
	& & \times	 \frac{1}{\Gamma\left(\gamma \right)} \mellintt{34} \frac{\Ga{\tilde{s}_{34}} \Ga{s_{34} -\tilde{s}_{34}}}{\Ga{s_{34}}} \mdelta{\gamma - K_{12,34} -2\tilde{s}_{34}}          \label{4thinto}
\ee
We have introduced the notation $K_{ij,kl}=s_{ik}+s_{il}+s_{jk}+s_{jl}$. Now we can integrate over $\tilde{s_{34}}$ and simplify the result to obtain,
\be
 I^{\phi_1 \phi_2}_{\phi_3 \phi_4} &=& \Mellint{1}{4} \Singprop{x_{il}}{}{s_{il}} \Melldel{1}{4}        \nonumber
		\\ 
	&& \frac{1}{2 \Ga{\gamma}} \Be{\gamma - K_{12,34}}{d-2\gamma}            \label{5thinto}
\ee
We simplified the second argument of the beta function using the conformality condition and the constraints imposed by the delta functions:
	\be
		\nonumber
	d = \gamma +\Delta_3 + \Delta_4 \qquad \text{and} \qquad 
	\Delta_i = \sum_{k=i,i\neq k}^4 s_{ik}.
	\ee
~\eqref{5thinto} is the familiar result obtained for the scalar propagator in Mellin space as obtained in \cite{Paulos:2012nu, Nizami:2016jgt}. \\
\\
In general for more complicated Feynman diagrams, one can carry on this procedure and use the result for the contact interaction at each step. This would typically give a nested Mellin-Barnes integral over beta functions. All of the technicalities in the method presented in \cite{Nizami:2016jgt}, for example making a suitable choice for the order of integration over the vertices, still continue to hold. To summarize the differences between the two methods: we are trading some nested Schwinger parameter integrals for some nested Mellin-Barnes integrals. Thus, in the case of scalars, this technique does not offer any simplifications over the method presented in \cite{Nizami:2016jgt}. However for conformal integrals with legs with spin, the Schwinger parameter integrals are particularly difficult and therefore this method is very helpful. One has to do a set of Schwinger parameter integrals while calculating the Mellin amplitude associated with the contact interaction diagram, but for all other Feynman diagrams there are no further Schwinger parameter integrals to be evaluated.

\providecommand{\href}[2]{#2}\begingroup\raggedright\endgroup

\end{document}